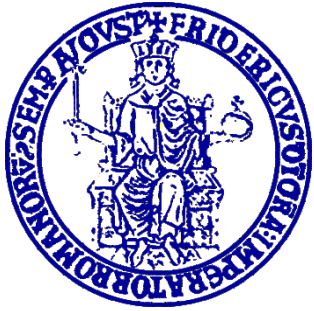

# Università degli Studi di Napoli
## *Federico II*

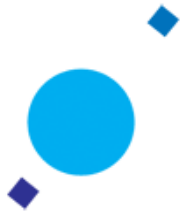

DOTTORATO DI RICERCA IN
**FISICA**

**Ciclo 30°**
Coordinatore: Prof. Salvatore Capozziello

Settore Scientifico Disciplinare FIS/05

# Characterisation of dust events on Earth and Mars
# the ExoMars/DREAMS experiment and the field campaigns in the Sahara desert

| **Dottorando** | **Tutore** |
|---|---|
| Gabriele Franzese | dr. Francesca Esposito |

Anni 2014/2018

A birbetta e giggione
che sono andati troppo veloci
e a patata
che invece adesso va piano piano

# Summary

# Introduction

Suspended dust can significantly affect the thermal profile of the atmosphere, absorbing and reflecting the solar radiation, influencing also the cloud formation and life-cycle. Once entrained at altitude, smallest grains can be transported all over the planet, mixing the soil composition and affecting the amount of mineral and nutrients of the different ecosystems. The average global effect on the terrestrial climate is still not well established. However, except peculiar and rare conditions, it is estimated as not predominant. Indeed, the presence on Earth of oceans, vegetation and other complex factors limits the abundance of dust sources and mitigates the aerosol total contribution.

The martian environment presents instead a different situation. Despite being much thinner than the terrestrial one, martian atmosphere can still sustain aeolian processes: they are common and diffuse all over the planet and able to set in motion sand and dust grains, as evidenced by the diffuse present of dunes. The presence of fewer competing factors, the arid environment and the widespread distribution of dust sources make the dust contribution one of the crucial factors that drive the planet climate. Indeed, martian dust storms can last several months, even covering the whole planet surface absorbing up to 80% of the incoming solar radiation.

Another important phenomenon associated with the dust clouds is the mutual contact electrification of the grains, that can generate electric fields up to several kV/m. On Mars, the triboelectricity associated with dust impacts is expected to be the main charging mechanism of the atmosphere and could lead to potential greater than the breakdown voltage, with the consequent formation of electrostatic discharges. These phenomena can potentially affect the atmospheric composition and the habitability of the planet.

Therefore, in the frame of the Martian exploration, the study of the dust lifting phenomena covers a key role and represents one of the goals of the present and future missions.

The ExoMars 2016-2020 programme aims to search for signs of past or present life on Mars, and investigates the Martian atmosphere and the long-term climate changes.

The first mission has been launched in 2016. It included the Schiapparelli lander, developed to demonstrate the technology for entry, descend and landing on the Martian surface, and the Trace Gases Orbiter, actually in orbit around the planet with the aim to map the sources of atmospheric trace gases.

The second mission will be launched in 2020 and it will include a surface platform and a rover. The aims of the mission are the search for life signs, the investigation of the boundary and subsurface layer to characterize the water and geochemical distribution, the climate evolution and the habitability of the planet.

In this optic, the missions also involve the deployment on the martian surface of instruments able to study the meteorological, electric and dust activity parameters of the planetary boundary layer.

The team I joined during my PhD work at the INAF astronomical observatory of Capodimonte is directly involved in the ExoMars programme, being in charge of the DREAMS meteorological station on board of the Schiapparelli lander of the ExoMars 2016 mission, and of the MicroMED optical particle counter on board of the surface platform of ExoMars 2020.

In the frame of the instruments developing and the acquisition of martian analogous data on Earth, our team has carried out various campaigns in the Sahara desert, during the peak activity season for the dust processes, to study the environment and the lifting phenomena that are expected on Mars.

The data acquired in the Sahara represent up to now the most comprehensive data set available for the dust phenomena, being also the only one that crosses synchronous acquisition of meteorological, electric, sand and dust activity parameters.

Currently, there are still many open questions regarding the dust processes physics, largely due to the lack of proper field surveys. Indeed, both on Earth and Mars, we still lack a proper estimation of the injection rate, global amount and physical proprieties of atmospheric dust. The wind speed threshold during dust storms and devils for the entrainment of sand and dust grains are still matter of controversy, due to the not precise knowledge of the magnitude of the different forces involved. In particular, the role of the electric forces is very unclear and, in general, the proprieties and the development of the electric field related to the dust cloud composition and environmental meteorological parameters heavily need further investigation to be understood.

Therefore, in order to answer at least partly to the aforementioned questions, my PhD research topics have been:
- the development of proper detection algorithms able to individuate the dust events acquired in the data, structuring the techniques in such a way that they can be used be used also for the analysis of the data collected by the ExoMars missions;
- the physical characterization of the dust storm and dust devils activity observed during the Saharan campaigns, studying in particular the electric proprieties of the events and the relations between the electric field, the amount of sand and dust lifted and the environmental conditions;
- the comparison of our terrestrial results with the martian data available in literature and the study of what we can expect and what we have to look for in the future ExoMars data, that may fill many of current lacks.

The work is structured as follows:
- In Chapter 1 I will discuss the role of airborne dust on the planet climate, both on Earth and on Mars, describing also what are the main lifting processes, such as dust storms and devils. In addition, I will briefly introduce the boundary layer physics, the basic laws of the surface-atmosphere interaction that lead to the grains lifting and the involved electrification process.
- In Chapter 2 I will present the ExoMars programme and the related field campaigns we performed in the Sahara. In addition, I will describe the detection algorithms we developed to individuate the active lifting processes acquired in the surveys.
- In Chapter 3 I will discuss the analysis of the dust lifting activity monitored during the terrestrial campaigns, focusing on the description of the observed electric proprieties.

- In Chapter 4, I will present the analysis of the Saharan campaign I personally planned and led in 2017, specifically aimed to the study of the dust devil processes.

Enjoy my thesis.

# Chapter 1
# Atmospheric dust on Earth and Mars

In this chapter we explain what is the dust and its role on the planetary atmosphere and climate. We introduce the main dust lifting phenomena and the physic at the base of the lifting process. In addition, we discuss about the triboelectrification process associated with windblown grains and the resulting induced E-field.

## 1.1 Mineral Dust

The mineral dust is the component of the atmospheric aerosol originated by the suspension of mineral grains ($\lesssim$ 60 μm in diameter) from the soil. The principal mechanism for the mobilization of the soil grains is the wind erosion, however, there are cases where other natural mechanisms like the thermophoresis or the electric forces can give a relevant contribution.
The dust is mainly originated in arid and semi-arid region like deserts and dry lake beds. On Earth, all the main dust sources are located in the north hemisphere, while the contribution of the south hemisphere is estimated under 10% of the global dust emission (Choobari et al., 2014). The Sahara Desert represents the first contributor to the global dust budget giving around 55% of the total (Rajot et al., 2008), and the highest concentration of the emissions comes from the Lake Chad Basin. Other important sources are spread in Middle East and Asia, like the Lut Desert, the Karakum Desert, the Taklimakan Desert and the Gobi Desert.
Dust is also produced during the grinding and abrasion by ice over bedrock in river beds and glaciers and hence it is abundant in cold climate regions and at high latitudes where it is emitted by strong glacier-driven or katabatic winds (Dagsson-Waldhauserova et al., 2013).
The abundance of the emissions depends on the season and it is related to the dust storms activity; some regions like the Africa and the Middle East are more active in the summer period (Goudie and Middleton, 2006), while the activity in the Chinese region peaks in spring (Shao and Dong, 2006; Sun et al., 2001). The dust originated from different sources can present substantial different composition and, even considering a single source, the composition could be not homogeneous variating with the size of the grains. The characteristics and concentration of dust in an area are not just related to the local sources. Indeed, once lifted the grains can remain in suspension for weeks, travelling for thousands of kilometers, driven by the global circulation.
Alongside the seasonality due to the atmospheric dust lifting phenomena, the dust concentration can be affected by sporadic phenomena, like the volcanic eruptions, that can tremendously increase the amount of dust in the atmosphere. In addition to the natural sources, a percentage between the 10 and 30% of the emission has anthropic origin (Tegen and Fung, 1995; Sokolik and Toon, 1996). The human contribution to the deforestation and the desertification of many area of the planet favors the increase of the dust emission. Moreover, dust grains are directly lifted by activities like the mining, agriculture and overgrazing (Moulin and Chiapello, 2006), e.g. in Australia the anthropogenic dust emission counts for about 75% of the total (Bryant, 2013).

## 1.1.1 Impact on the Terrestrial land-atmosphere-ocean system

The effects of the dust on the atmospheric balance are complex and heavily depend on the concentration, size and composition of the grains. Usually, the are divided into direct, semi-direct and indirect effects. The direct effects are related to the dust impact on the atmospheric radiative budget, while the semi-direct and indirect effects are both related to the interaction of the dust with the clouds. In particular, the semi-direct effects describe the role of the dust absorption of radiation on the cloud's lifecycle, while the indirect effects are connected to the role of the dust as ice nuclei and cloud condensation nuclei. In this way the dust is able to influence the weather and the hydrological cycle. In addition, it affects the biogeochemical cycle, working as a trans-continental vector of nutrients for land and ocean habitats.
Lastly, the atmospheric dust affects the quality of the air affecting the human care. Grains under 10 μm of size are easily inhalable. Coming in contact with wet airspace surfaces they can deposit inside the respiratory system, melting and releasing toxic constituents for the organism and serving also as a vector of biological components such as bacteria, endotoxins, and fungi.

### 1.1.1.1 *Direct effect*

The dust grains have a diameter of the order of tens of microns or less because the larger grains are too heavy to enter in suspension for a long period. The range of diameters makes the atmospheric dust able to interact with both the short and the long wave radiation. Hence, the dust plays the dual action of reflecting in space part of the solar radiation (short wave radiation) and trapping part of the radiation emitted by the surface (long wave radiation), heavily affecting the thermal gradient of the atmosphere and consequently the climate. In presence of a suspended dust layer the surface will be cooled during the day-time and warmed during the night-time (Hansell et al. 2010), resulting in an overall cooling, due to the preponderance of the interaction with the shortwave radiation (Yoshioka et al. 2007). Thus, the atmospheric dust plays on average a role opposite of the greenhouse gases, giving a cooling contribute that counterattacks the gases warming effect.

### 1.1.1.2 *Semi-direct and indirect effects on the cloud physics*

After the suspension, part of the dust is embedded within clouds. Here, it can lead to the "cloud burning effect", enhancing the evaporation through the absorption of the solar radiation (Hansen et al. 1997; Ackerman et al. 2000). In addition, the influence of the grains on the atmospheric temperature profile affects the vertical and horizontal development of the cloud coverage (Koch and Del Genio, 2010).

Due to their size, the dust grains work as ice cloud nuclei (Sassen et al., 2003), and, when they are coated with soluble materials or absorb water, they can growth until to act as efficient cloud condensation nuclei (Levin et al., 2005). In this way, the dust affects the amount of droplet inside the cloud, modifying their optical proprieties and the precipitation rate. These interactions are complex, and, depending on the circumstances, the net effect can be both a suppression or an enhancement of the rain precipitation. For this reason, the atmospheric dust can heavily affect the hydrological cycle.

### 1.1.1.3  *Indirect effects on the biogeochemical system*

Alongside these contributions, the dust re-deposition plays an indirect effect on biogeochemical system (Mahowald, 2011). Suspended dust is removed from the atmosphere by gravitational sedimentation and wet precipitation. Dry deposition dominates over the lands, while the wet deposition is preponderant on the oceans (Zender et al., 2003). The dust is vector for long range transport of crucial nutrients for the ecosystems, in particular nitrogen, phosphorus, and iron. For example, we know that the tropical forests are phosphorus-limited (Vitousek, 1984; Swap *et al.* 1992), while many ocean biota, such as phytoplankton, are iron limited (Martin et al., 1990). The atmospheric $CO_2$ is one of the stronger greenhouse gases and could inhibits the growth of the Ozone layer. The concentration of this gas is regulated by the plant respiration and by the "biological pump" mechanism of the oceanic phytoplankton. The dust cycle modulates on a long time scale (hundreds and thousands of years) the productivity of these ecosystems, thus strongly affecting the absorption of the atmospheric $CO_2$ (Okin et al., 2011). This effect is so pronounced that the variation in the suspended dust concentration has been proposed as one of the main factor that drives the interglacial-glacial variation of the $CO_2$ concentration and thus the climate change (Martin, 1990).

### 1.1.1.4  *Estimation of the total effect*

It has been estimated that both the dust indirect effect on the cloud development and on the biogeochemical cycles give a neat contribution on the global radiative forcing comparable to the direct effect (Mahowald, 2011; Foster et al., 2007). However, the impact of the dust on the global climate is far from well established, primarily due to the uncertainty that affects the evaluation of the global dust burden value, the not total comprehension of the chemical and physical properties of the lifted grains, as well the lack of knowledge of the temporal and spatial variability of the concentration.

The intergovernmental panel on climate change (IPCC) provides at regular intervals Assessment Reports of the state of knowledge on climate change. The latest Fifth Assessment Report, given in November 2014, highlighted how aerosols contribute the largest uncertainty to the total climate forcing estimate for the Earth (IPCC, 2014). The in situ observations provide direct information on dust composition, size distribution, chemical and physical proprieties; however, they are too sparse for adequate global coverage and we have to use satellite observations to monitor the emissions on a global scale. The planetary dust cycle is currently simulated using different models, but, depending on the assumptions and starting input parameters, the aerosol emission estimate ranges over more than a factor 2.

To improve these results, the acquisition of further field data, as the ones presented in this work, plays a crucial role. In addition, we have to take into account the dust lifting processes are not limited to the terrestrial environment, they are present also on other planet equipped with an atmosphere able to sustain aeolian process. Therefore, the understood of dust lifting physics on Earth is important also for the implications on the planetary exploration. In particular, in this work we will focus on the Martian case, in the frame of the ExoMars 2016-2020 programme that aims to investigate the possible presence of past or present life on Mars. One of the requirements necessary to achieve this objective is the study of the climatic conditions of the planet that, as we will see in the next section, are strongly influences by the dust processes.

## 1.2 Mars

Mars is the fourth planet of the solar system. Its radius (~3.396 km) is about one half of the terrestrial one, while its gravity (~3.71 m/s$^2$) is around one third. The solar longitude (Ls) is the parameter used to indicate the Martian season: it is the angle that the planet describes in its motion along the ecliptic. It ranges from 0° to 360°, where 0° correspond to the start of the year, while 360° is the end. The Martian year lasts 687 terrestrial days, while the rotational speed is similar to that of the Earth with a Martian day that lasts 24h and 39'. During its evolution the planet has lost most of its atmosphere and currently it maintains only a tiny one with a superficial pressure of ~6 mbar, around 6‰ of the terrestrial one. Despite this, the atmosphere is still able to sustain aeolian processes.

The environmental conditions of the two Martian hemispheres are strongly asymmetric. The Martian rotational axis is tilted of 25.2° and its orbit has a large eccentricity; moreover, the southern hemisphere is on average around 5 km higher than the northern one. For these reasons, the southern hemisphere summer is shorter than the northern one, but it is warmer, due to the stronger insolation, and it is characterized by a stronger Hadley circulation favored by the topography (Joshi et al, 1995). During the so called "dust storm season" (from $L_s$~180° to $L_s$~330°), these factors lead to the generation of strong boundary winds and consequently the planet exhibits the strongest dust lifting activity.

The Martian regolith forms a loosely packed, porous medium, which is the source of the ubiquitous airborne dust and allows the exchange of volatiles with the atmosphere.

The space missions for Mars exploration have been ongoing since the early 1960s, and nowadays we count more than twenty successful missions. Currently, there are six orbiters around the planet (Mars Reconnaissance Orbiter, Mars Express, Mars Odyssey, Mars Atmosphere and Volatile Evolution, Mars Orbiter Mission and ExoMars Trace Gas Orbiter) and two active rovers on the surface (Mars Exploration Rover Opportunity and Mars Science Laboratory Curiosity).

The presence of dust in the atmosphere and the high frequency of dust lifting events on the surface have been known since early Martian missions. Indeed, it has been observed how the near surface atmosphere is characterized by an ever-present dust haze. Despite this, the mobilization of the sand size grains has not been observed until recent years. Indeed, until 2008 it was still considered that Martian aeolian bedforms were relict features formed in the past when atmospheric densities or wind speeds were higher (Merrison et al., 2007). On Mars, it is supposed that, like on Earth, the dust lifting is initially triggered by the set in motion of the sand grains (see saltation process in par. 1.5.2). Therefore, the abundance of dust mobilization coupled with the lack of bedform migration has represented a paradox for several years (Sullivan et al., 2008). The acquisition of high-resolution images from both land and orbits has resolved this paradox, showing that the mobilization of sand actually takes place on the planet (Bourke et al., 2008; Bridges et al., 2007; M. Chojnacki et al. ,2010; Sullivan et al., 2008). Silvestro et al. 2010 has given the first evidence of widespread ripple and dunes migration on the planet, proving that the Martian surface is currently active, with frequent sand transport events.

### 1.2.1 Impact on the Martian land-atmosphere system

Overall, the dust cycle is a critical factor that drive the weather and climate of the planet. As we described for the Terrestrial case, Martian airborne dust affects the absorption and reflection of the solar radiation inside the atmosphere. Due to the enhanced presence of dust, this effect is stronger than on Earth. In addition, also on Mars dust grains serve as condensation nuclei for the clouds of water, $CO_2$ and ice present on the planet. Moreover, due to the of lack of vegetation and any other kind of coverage of the surface, the optical properties of the soil heavily affect the energetic balance between the amount of radiation absorbed and reflected by the planet. The measure of the reflectance of a surface is called albedo. The continuous transport and redeposition of dust clouds around the planet and the tracks left by dust devils passage can substantially change the local albedo of the Martian region. These variations affect the thermal structure and circulation of the boundary layer, causing a direct feedback on the momentum exchanged between soil and atmosphere and hence on the lifting processes.

In absence of precipitation and fluvial systems on the present Mars, the aeolian processes represent the principal erosion phenomena of the landscape, causing the migration of the ripple and dune systems and the creation of wind streaks on the surface.

We will discuss more deeply the dust lifting events active on the Martian surface in the next section.

Another important aspect to take into account is related to the induction of strong electric fields near the ground due to the presence of suspended dust; we will discuss more deeply this topic in par. 1.6.

# 1.3 Dust lifting phenomena

## 1.3.1 Dust Storms

On Earth, dust storms are local phenomena arising by strong unidirectional winds and extended gust fronts able to lift sand and dust particles from the surface, reducing the visibility under 1000m (Goudie 1978). They are common in arid and semi-arid environment and can growth up to scale of thousands of km, becoming visible from the Terrestrial orbit. Inside the larger events, the wind speed can exceed 30 km/h, while the visibility can decrease up to few tens of m.
On Earth, aside from the case of sporadic episodes like volcano eruptions, dust storms give the main contribution to the injection of dust into the atmosphere. The lifted dust can travel for thousands of km, evidence of this is the frequency with which the Europe is covered by the dust originated in the North Africa dust storms. In very rare cases, the same storm can extend over an area with a scale length of some thousands of km, e.g the great North American dust storm of 1933 that covered the region between Montana, Ohio, Missouri and the Lake Superior. However, on average, the Terrestrial storms are much smaller than their Martian counterpart.

### *1.3.1.1* Martian Dust storm

Martian dust storms have been observed from Earth since the nineteenth century: Antoniadi firstly reported large yellow clouds obscuring part of all of the surface.
With the beginning of the Martian space exploration the Mariner 9 probe in 1971 was the first spacecraft to orbit around the planet, encountering the largest global storm observed so far. Successively, the Viking missions have observed the global dust storm of 1977 both from land and orbit.
The rate of dust injection into the atmosphere is comparable on Mars and on Earth ($\sim 3 \cdot 10^{15} g/yr$, Greeley and Iversen, 1987) and local dust storms occur frequently and are probably dynamically similar to dust storms on Earth.
In period of high activity, there may be several tens of storms contemporaneously active on the planet. Typically, the activity peaks in late spring and early summer, when Mars is near perihelion. However, in some years, the storms do not develop at all. They can last from few weeks to few months, even various months in case of global dust storms. Usually, the major storms become visible as local bright clouds core, yellow or red, of few hundreds of km. After few days the core starts an expansion phase with the formation of secondary cores that join the main cloud. Under particular circumstances, that are still not completely clear, on average every three martian years, this expansion can lead to the formation of a planet encircling storm (see Fig. 1). There is no evidence that dust is raised simultaneously on a global scale, more likely different strong storms originated in particular geographical localities lift enough dust to cover the whole planet. About one-half of the dust clouds appeared near the edge of the southern polar caps, and the other half occurred mainly in the southern hemisphere, for the different elevation and topography, as we explained in par 1.2.

The greater intensity of the Martian dust storm compared to the Terrestrial ones is related not only to the major abundance of dust sources but also to the stronger orographic effects that drive the wind circulation. Indeed, the Martian surface has a height gradient that overcomes 30 km and many storms originate from slope winds generated off the flanks of the volcanoes in the Tharsis region (up to 22 km of height) (Lee et al., 1982).

Guzewich et al. 2015 have analyzed the dust storms active in dust lifting recognizable from the orbital images of the Mars Global Survey and the Mars Orbiter Camera, for Martian years (MY) from 24 to 28. They observed how the dust storm activity peaks in the equinox periods but is nearly absent near each solstice. Comparing theirs results with the ones obtained for the global dust opacity during MY24-27 (Montalbone et al 2014), Guzewich et al. 2015 concluded that the dust storm activity is strictly related to the global dust opacity, being responsible of around 50% of the total lofted dust. Smaller scale dust lifting phenomena, in particular dust devils, have to be taken into account and have to play a role on the optical depth comparable to the dust storms one. These kind of phenomena become predominant in the solstice and aphelion periods, due to the decrease of dust storm activity.

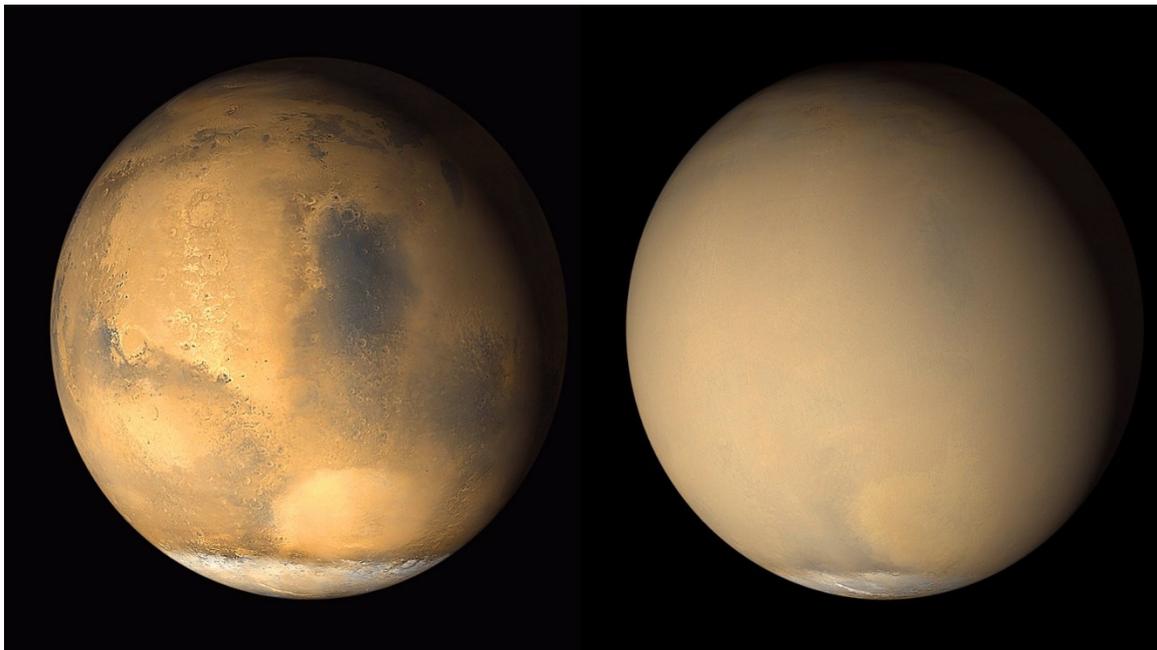

**Fig. 1** *Two images of Mars, taken by the Mars Orbiter Camera on NASA's Mars Global Surveyor orbiter, before and after the great dust storm of 2001. The first image is taken at the start of June, while the second one is relative to the end of July. Credit to: NASA/JPL-Caltech/Malin Space Sciences Systems https://mars.nasa.gov/resources/21448/*

## 1.3.2 Dust Devils

Dust devils are convective vortices, stable vertical columns of air in rotary motion, strong enough to entrain material from the surface.

Their formation is favored in conditions of strong insolation, low humidity environment, lack of vegetation and buildings or other high obstacles and gently sloping topography (Balme and Greeley, 2006). For these reasons, they are often observed in terrestrial deserts and are also very common on the surface of Mars. Fig. 2 shows a dust devil photographed during one of the field campaigns. The diameter of the dust column can remain roughly constant or increase with the height, the column can be perpendicular to the surface or slightly tilted in the direction of motion.

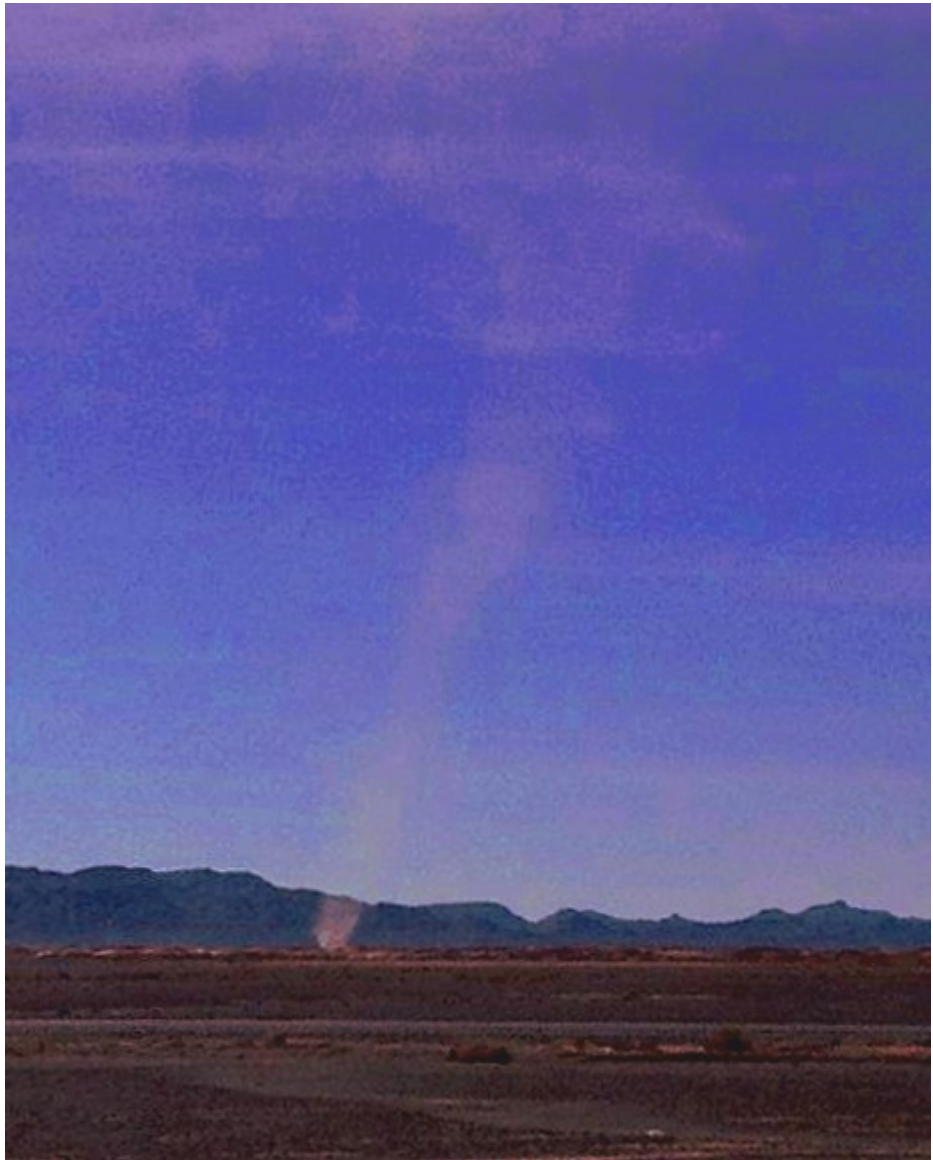
**Fig. 2** *A dust devil observed a few kilometers from the measurement site.*

At the center of the column there is a low pressure core, related to the vertical convection, that induces a horizontal rotary motion of the air flow. The rotational speed $v_r$ in function of the distance d from the center of the vortex, can be reasonably described using the Rankine vortex model (Giaiotti and Stel 2006):

$$v_r(d) = \begin{cases} V_r \dfrac{R}{d} & if\ d > R \quad (1a) \\ V_r \dfrac{d}{R} & if\ d < R \quad (1b) \end{cases}$$

where d is the distance from the vortex center and R is its radius, i.e. $d = R$ is the vortex region with maximum rotational speed, called vortex wall. The speed linearly increases with the distance d from the center of the vortex up to its wall, where it reaches its maximum $V_r$, then decreases as the reciprocal of d.

The pressure has its minimum value in the center of the vortex and its trend in time can be described by a Lorentzian curve (Ellehoj et al. 2010):

$$P(t) = \frac{-\Delta P_o}{1 + \left(\dfrac{t - t_o}{\frac{1}{2}\Gamma}\right)^2} + B \quad (2)$$

where $B$ is the background pressure value, $\Delta P_o$ is the magnitude of the pressure drop at the center of the vortex relative to $B$, $t_o$ is instant relative to the minimal pressure and $\Gamma$ is the full width at half maximum (FWHM) of the signal. The first term of on the right side of eq.(2) is the pressure drop $\Delta P(t)$ and it represents the theoretical signal that a vortex leaves in the pressure time series.

Dust devils with a single pressure core are the most common, however in some cases the vortex can present more than one core. The magnitude of $\Delta P$ is usually of the order of one thousandth of the atmospheric pressure (~1 mbar on Earth).

At vortex wall, the rotational speed is approximately in cyclostrophic balance with the magnitude of the pressure drop, namely, the force due to the pressure gradient is equal to the centrifugal force:

$$\frac{V_r^2}{R} = \frac{\Delta P_o}{\rho} \quad (3)$$

The rotational motion of the vortex overlaps the motion of the wind background. When the total wind speed reaches a sufficient magnitude, the whirlwind starts to lift sand and dust from the surface becoming a true dust devil. These vortices are one of the most efficient aeolian lifting mechanisms, thanks to the combined effect of horizontal and vertical wind, saltation and pressure-gradient force (Klose et al., 2016). The relative importance of the different mechanisms is still unclear, but, their sum makes the dust devil more effective in dust size grains lifting than the shear stress related to unidirectional wind (Greeley et al., 2003). Inside the wall of the column the lifted grains follow an upward helical pattern, forming a debris-laden annulus, while the central region of the vortex is relatively dust free. The vertical wind speed rapidly decreases outside the vertical column, and the dust devil is surrounded by a region characterized by a downward directed motion, where the lifted grains relapse toward the surface. In some cases, the passage of the dust devil can leave a visible track on the surface due to the removal of the first soil layer and the consequent albedo change.

On Earth, the whirlwinds affect the atmospheric dust concentration on local scale, but, they seem to be responsible of only some percent of the global dust budget (Jemmett-Smith et al. 2015).

Sinclair in 1966 has been the first to systematically study a large dust devil dataset in the Sonora Desert (Arizona). He observed that the best environmental wind speed to favor vortex formation is around 5 m/s and that the activity peaks around noon, when ground is at its warmest. Indeed, one of the main mechanism to generate the vorticity that lead to the whirlwind is the vertical air motion induced by surface region warmer than the neighbors. The horizontal gradient of surface temperature leads to a pressure gradient that induces the air convection and rotation.

### 1.3.2.1 *Dust Devils on Mars*

The presence of dust devils on Mars has been confirmed by several Mars Mission, using both landed and orbital instruments.
Ryan, 1964 has been the first to suggest the possible existence of dust devils on Mars. The possibility to find convective vortices on the surface was strongly suggested by the desertic conditions of the planet (Neubauer, 1966; Gierasch and Goody, 1973). The first evidences of their presence was given by the NASA Viking missions: Ryan and Lucich, 1983 and Ringrose et al., 2003 have identified many convective vortices in the Viking landers wind and temperature measurements. The complete characterization of the pressure signatures of these events was not possible yet, due to the infrequent sampling and the poor digital resolution of the data. The confirmation that the vortices could be actually dusty was later given by the analysis of the Viking Orbiters images, where a hundred of near-surface dust columns were identified (Thomas and Gierasch, 1985). During the Pathfinder mission, tens of dust devils have been identified both in the images of the lander (IMP camera) and in the data of the surface meteorology package (MET), allowing the first Martian study of the vortices pressure and temperature profiles (Schofield et al., 1997; Murphy et Nelly., 2002; Metzger et al.,1999; Smith and Lemmon, 1999; Ferri et al.,2003).
Afterwards, the NASA Mars Global Survey and the ESA Mars Express mission have allowed the observation from the orbiter of dozen dust columns and dust devils tracks through the high-resolution Mars Orbital Camera (Edgett and Malin, 2000; Malin and Edgett, 2001) and the High Resolution Stereo Camera (Greeley et al., 2005; Stanzel et al., 2008).
The NASA Mars Exploration Rover mission landed in 2004, has monitored three seasons of dust devil activity in Gusev crater using the Spirit rover (Greeley et al., 2006). Spirit was equipped with a composite cameras system without meteorological instruments and it collected images of hundreds of events. The NASA Phoenix Mars Lander, landed in 2008, collected meteorological data almost continuously for 151 Martian sols, measuring 502 pressure drops compatible with dust devil passages (Ellehoj et al., 2010). The NASA Martian Science Laboratory (MSL) Curiosity rover, landed in 2012, is still active on the surface, monitoring the boundary layer atmosphere through the Rover Environment Monitoring Station (REMS). So far it has collected 245 convective vortices signatures (Steakley and Murphy 2016).
The nearly global Mars coverage of the orbiter images show how dust devils are common and widespread on the surface of the planet. Dust columns have been identified in both hemispheres in every season, and dust devil tracks have been observed almost at every latitudes and elevation (Malin and Edgett, 2001; Greeley et al., 2004; Stanzel et al., 2008). The diameter of the observed Martian dust devils ranges from the meter to the kilometer, while the height goes from the hundreds of meters to various kilometers. The pressure drop is of the order of ten µbar, the rotatory speed ranges from few m/s to over the 50 m/s, while the translational speed usually goes from few m/s to about 15 m/s, with rare peaks up to 50 m/s. The temperature variation is of the order of the kelvin degree.

Pathfinder, Phoenix and Curiosity are the only missions that have collected usable pressure time series of the events. Despite the three mission have landed in very different sites, subtropical for Pathfinder, norther polar for Phoenix and equatorial for Curiosity, they observed very similar dust devil characteristics. In all the cases the cumulative distribution of the pressure drops is describable with a power law with exponent around 2-3, in the same range of variation. Both the Martian surface pressure P and the pressure drop of the events are around one thousand times smaller than the ones on Earth. Hence, the ratio $\Delta P/P$ is quite similar on the two planets. However, the Martian dust devils can be an order of magnitude larger in diameter and height than the terrestrial ones (Fenton et al., 2016). Overall, the meteorological characteristics of the Martian dust devils are really close to the terrestrial ones suggesting that the events could have a common formation mechanism and dynamics (Ringrose et al., 2003).

Usually the whirlwinds are observed between 8:00 and 17:00 (Local True Solar Time LTST) and most of the daily occurrence is concentrated around noon: between 11:00 and 13:00 for Pathfinder and Curiosity, and between 11:00 and 15:00 for Phoenix. The seasonal occurrence usually peaks in the summer season.

As already mentioned, the role and importance of dust devils in the Martian climate is a highly studied and debated subject, however they seem to count for around 50% of the global dust budget (Guzewich et al., 2015). They represent a continuous source of lifted dust, active even outside the dust storms season. For these reasons they have been proposed as the main mechanisms able to sustain the ever present dust haze of the Martian atmosphere (Neubauer, 1966; Thomas and Gierasch, 1985; Klose et al., 2016; Murphy et Nelly., 2002; Ferri et al.,2003; Fisher et al.,2005; Stanzel et al., 2008).

The study of dusty vortices is one of the key scientific points to be pursued by the next Mars space missions: the NASA InSight 2018 and the ESA/ROSCOSMOS ExoMars 2020.

## 1.4 Physics of the boundary layer

The planetary boundary layer (PBL) is the region of the atmosphere closest to the surface, where the principal factors that influence the air flow are the friction with the surface and the vertical temperature gradient (Kaimal and Finnigan 1994). The height of the PBL on Earth is usually under the km and above this layer the air flow is in near-geostrophic balance. Indeed, increasing the height, the effects of the surface become less and less relevant.

The boundary layer can be in turn sub-divided in two regions:

- a surface layer of about 50/100 m where the quantity of momentum $\tau$, heat $H$ and moisture $E$ exchanged between the soil and the atmosphere are independent from the height;
- and a region above up to a Km, where $\tau$, $H$ and $E$ can slightly depend on the altitude and the Earth rotation.

Our main interest will be focused on the surface layer, where the dust lifting happens.

### 1.4.1 Atmospheric Stability

The atmospheric pressure P decreases with the altitude z due to the lesser weight of the overlying air column. When the pressure gradient is affected only by the gravity force we say that the atmosphere is in hydrostatic equilibrium and we have that:
$$dP = -g\rho dz \qquad (4)$$
where g is the gravity acceleration and $\rho$ is the air density.

Hence, even in adiabatic conditions, the temperature T of the air layers is not constant with the altitude z, but it changes according to the pressure variation. We can define the potential temperature $\theta$ as the temperature that the air layer would have if it was at the reference pressure $P_o$:
$$\theta(z) = T(z)\left(\frac{P_o}{P(z)}\right)^{\frac{R}{c_p}} \qquad (5)$$
where $R$ is the gas constant, and $c_p$ is the specific heat at constant pressure. Usually, the reference $P_o$ is chosen equal to the pressure at surface level.

Using the eq.(4) and the ideal gas law, we can connect the potential temperature gradient to the temperature gradient by the equation:
$$\frac{\partial \theta(z)}{\partial z} = \frac{\partial T(z)}{\partial z} + \frac{g}{c_p} \qquad (6)$$
That is more useful to rewrite in the form:
$$\frac{1}{\theta}\frac{\partial \theta}{\partial z} = \frac{1}{T}(\Gamma_d - \Gamma) \qquad (7)$$
where $\Gamma_d = \frac{g}{c_p}$ is the dry adiabatic lapse rate and $\Gamma = \frac{\partial T(z)}{\partial z}$ is the environmental lapse rate.

There are three different conditions of atmospheric stability:

- stable, when $\frac{\partial \theta(z)}{\partial z} > 0$ and $\Gamma > \Gamma_d$. Any kind of vertical mixing of the atmosphere is inhibited by the buoyancy forces.
- neutral, when $\frac{\partial \theta(z)}{\partial z} = 0$ and $\Gamma_d = \Gamma$. An air layer does not suffer any buoyancy forces ascending or descending.
- unstable, when $\frac{\partial \theta(z)}{\partial z} < 0$ and $\Gamma < \Gamma_d$. The vertical motion of the air layers is favoured by the buoyancy forces.

The unstable condition is usually observed during the day-time, when the surface is heated by the solar radiation, raising the potential temperature of the lowest air layers. During the night-time, the situation is inverted because the superficial soil layer cools faster than the atmosphere, lowering the temperature of the near ground air. The neutral condition is observed at the sunset and sunrise, when there is the transition between the stability and instability of the atmosphere. The amount of momentum exchanged between the surface and the air flow is affected by the atmospheric stability.

## 1.4.2 Neutral case

### *1.4.2.1 Wind Speed Profile*

We start splitting the wind speed $v$ in two components, the main component (denoted by an overbar) and the fluctuating component (denoted by the prime):
$$v = \bar{v} + v' \qquad (8)$$
We can define the same subdivision for each atmospheric parameter, like pressure, temperature and humidity.

The main component is related to the large scale motion of the flow, while the fluctuations are linked to the behavior of the atmospheric turbulence. Hence, we have to mediate on a time interval much longer than the turbulence time scale, so that $\overline{v'} = 0$. More generally, the mean has to be evaluated over a time interval where the Reynolds averaging conditions are verified (Kaimal and Finnigan, 1994).

Let's us call $(u,v,w)$ the tridimensional components of the wind. The air flow is in a laminar regime if $\bar{w} \cong 0$, otherwise we are in presence of a turbulent regime. In this work, we will deal only with laminar flows.

Above the boundary layer, in particular at middle latitudes, the wind flow is usually in geostrophic balance: the two main forces that affect the flow, the Coriolis force and the Pressure gradient force, are nearly equal. In this condition, if the atmosphere is in hydrostatic equilibrium, the wind speed does not directly depend on the altitude and the wind blows at the so called geostrophic speed.

Instead, in the boundary layer the air flow is mainly affected by friction with the soil surface and the effect increases by decreasing the height. The wind is slowed down by friction and hence transfers energy to the soil. We define $\tau$ the flux of momentum that flows through the horizontal plane. Using the so called K-theory, $\tau$ can be expressed as a function of the wind speed vertical gradient as:
$$\tau = K_m \rho \frac{\partial \bar{u}}{\partial z} \qquad (9)$$
where $K_m$ is the turbulent exchange coefficient for momentum, $\rho$ is the air density.

Since it is a flux, $\tau$ can also be defined in terms of the covariance between the horizontal and vertical wind component. For simplicity, if we consider the coordinate system with the axis x parallel to the wind propagation direction $(u, v) = (u, 0)$, we have:
$$\tau = -\rho \, \overline{u'w'} \qquad (10)$$
It is convenient to introduce a parameter $u^*$, called friction velocity, to parametrize $\tau$ and the wind profile. The friction velocity has indeed the dimension of a speed and all the estimates on the lifting process efficiency can be brought back to the evaluation of this parameter. It is defined by the relation:

$$\tau_o = \rho\, u^{*2} \qquad (11)$$

the subscript symbol 0 indicate that the flux is evaluated at the surface level. In the general case where $(u, v) \neq (u, 0)$, we obtain:

$$u^* = \sqrt[4]{\overline{u'w'}_o^{\,2} + \overline{(v'w')}_o^{\,2}} \qquad (12)$$

Even if all the quantities are evaluated right at the surface level, the value of $\tau$ is practically constant up to few tens of meters above the surface. Hence, it is possible to evaluate the covariance also at some meters from the soil (Monin and Obukhov, 1954; Haugen et al., 1971).

$K_m$ has the dimension of a speed times a length, hence we can introduce a constant of proportionality $k$ and rewrite the coefficient as:

$$K_m = k\, u^* z \qquad (13)$$

k takes the name of Karman constant, its value is empirically derived and ranges between 0.35 and 0.43. In this work we will assume $k = 0.4$.

Joining the equations (9), (11) and (13), we obtain the following equation for the horizontal speed gradient:

$$\frac{\partial u}{\partial z} = \frac{u^*}{k\, z} \qquad (14)$$

By integrating, we obtain the so called "law of the wall" (Priestley, 1959):

$$u(z) = \frac{u^*}{k} \log \frac{z}{z_o} \qquad (15)$$

where $z_o$ is the roughness length that represents the altitude where the wind speed becomes zero. The value of $z_o$ is related to the height of the soil coarseness, and it is roughly equal to 1/30 of the mean grains diameter.

To summarize, increasing the altitude the value of the wind speed logarithmically increases from the value 0 to the value of the geostrophic wind.

### 1.4.2.2 Temperature Profile

Starting from the temperature gradient, we can define the amount of heat $H$ exchanged between the surface and atmosphere using the K-theory:

$$H = -K_h\, \rho\, c_p \frac{\partial \bar{\theta}(z)}{\partial z} \qquad (16)$$

where $K_h$ is turbulence heat exchange coefficient.

More specifically $H$ represents the heat flux that passes through the horizontal plane. In analogy with what we did for the flux of momentum, we can also define $H$ in term of the covariance between the vertical turbulence and temperature fluctuation:

$$H = -\rho\, c_p \overline{w'\theta'} \qquad (17)$$

Usually, it is introduced the temperature scale $\theta^*$ as analogous of $u^*$:

$$\theta^* = \frac{\overline{w'\theta'}}{u^*} \qquad (18)$$

In theory, also this quantity has to be evaluated right over the surface, but, as we said, the fluxes variations are negligible in the surface layer.

If no heat is exchanged ($\theta^* = 0$), we are in the neutral case and using the definition of potential temperature we will have simply:
$$\bar{\theta}(z) = \theta_o \qquad (19)$$
that is, the potential temperature is constant with height.
Whereby, the temperature change with height is given only by the pressure variation:
$$\bar{T}(z) = T_o - \frac{g}{c_p} z \qquad (20)$$
where $T_o$ is the temperature of the air layer immediately over the surface.

### 1.4.3 Not neutral case

#### 1.4.3.1 Stability parameters L, $R_i$, $R_b$

Different parameters can be used to evaluate the atmospheric stability. The choice of one over another depends on the quantities that are known for the system that we want to study. The most used parameter has been introduced by Monin and Obukhov and has the form:
$$L = -\frac{u^{*2}}{k \frac{g}{\bar{\theta}} \theta^*} \qquad (21)$$
$L$ has the dimension of a length and the ratio $z/L$ (where z is the altitude) expresses the stability of the atmosphere:
- stable, if $z/L>0$;
- neutral, if $z/L=0$;
- unstable, if $z/L<0$.

Generally speaking, the Monin-Obukhov similar theory introduces the concepts that the fluxes $\tau$ and $H$ can be evaluated using only the parameters $u^*$, $\theta^*$ and $\frac{z}{L}$, as we will see in the next two sections.

One of the practical difficulties in evaluation of $L$ is its directly dependence on $u^*$ and $\theta^*$ and hence the requirement to measure the wind turbulence components in all the three dimensions. In most of the cases, the wind is measured only in the horizontal plane and with a rate of acquisition too low to evaluate the turbulence term.

For this reason, it can be more practical to utilize the Richardson number $R_i$:
$$R_i = \frac{g}{\bar{\theta}} \frac{\partial \bar{\theta}}{\partial z} \left(\frac{\partial \bar{u}}{\partial z}\right)^{-2} \qquad (22)$$
where, usually, the gradient terms are evaluated directly by using the difference of temperature and wind speed measurements taken at two different heights:
$$R_i = \frac{g}{\bar{\theta}} \frac{\Delta \bar{\theta}}{\Delta \bar{u}^2} \Delta z \qquad (23)$$
whereby, $R_i$ does not require the measure of the turbulence terms, and it doesn't depend directly by the vertical wind speed.

Using the Monin-Obukhov similarity function $\phi$, that we will describe in the next section, we can directly link $L$ to $R_i$:

$$Ri = \begin{cases} \dfrac{z}{L}\dfrac{1}{\left(1+\dfrac{5z}{L}\right)} & stable \\ \dfrac{z}{L} & unstable \end{cases}$$

Lastly, it has been introduced the Richardson bulk number $R_b$ for the cases when the wind speed and the air temperature are known at only one height $z$, but the surface temperature is also available:

$$Ri_b = \frac{gz\,\Delta\bar{\theta}}{\bar{\theta}\,\bar{u}^2} \tag{24}$$

where $\Delta\bar{\theta}$ indicates the difference between the surface and air temperature.

Both $R_i$ than $R_b$ are defined with the same sign of $L$, hence if the parameter is positive it indicates stable atmosphere, unstable if it is negative and neutral if it is zero.

### 1.4.3.2 Wind Speed Profile

We can introduce the effects of not neutrality of the atmosphere on the wind profile, incorporating a function that dependent from the stability parameter in eq.(14):

$$\frac{\partial \bar{u}}{\partial z} = \frac{u^*}{k\,z}\,\phi_m\left(\frac{z}{L}\right) \tag{25}$$

$\phi_m$ is called momentum similarity function and its trend has to be derived empirically from the experiments. Currently, the most used expression is the Businger-Dyer form (Dyer, 1974; Businger, 1988; Högström, 1988):

$$\phi_m\left(\frac{z}{L}\right) = \begin{cases} 1 + 5\dfrac{z}{L} & stable \\ \dfrac{1}{\sqrt[4]{1 - 16\dfrac{z}{L}}} & unstable \end{cases} \tag{26}$$

anyway, these expressions are valid in conditions not too far from the neutrality.
The general form of the integrated eq. (25) is:

$$u(z) = \frac{u^*}{k}\log\frac{z}{z_o} - \varphi_m|_{z_o}^{z} \tag{27}$$

where $\varphi_m$ is the primitive of $\phi_m$ and it represents the corrective term to the law of the wall. However, it is clear that the introduction of $\phi$ makes the integration much more complex than the neutral case and the eq. (27), as it is, can be very unpractical to use in a context of real data. Depending on the case, various simplification can be assumed, as rewrite the eq. (26) in function of $R_i$ and consider $\phi_m$ not depending on z during the integration, reintroducing the altitude only in eq. (27).

### 1.4.3.3 Temperature Profile

We can construct the analogous of eq. (25) for the potential temperature:

$$\frac{\partial \bar{\theta}}{\partial z} = \frac{\theta^*}{k\,z}\,\phi_h\left(\frac{z}{L}\right) \quad (28)$$

where $\phi_h$ is the heat similarity function, whose expression is similar to that of $\phi_m$:

$$\phi_h\left(\frac{z}{L}\right) = \begin{cases} 1 + 5\frac{z}{L} & stable \\ \dfrac{1}{\sqrt[2]{1 - 16\frac{z}{L}}} & unstable \end{cases} \quad (29)$$

In the limit of neutral atmosphere, $\phi_h(0) = 1$, while $\theta^* = 0$ and the right term of eq. (28) simply reduces to 0.

### 1.4.4 Dust feedback

The emission of dust can establish a feedback on the local dust lifting processes (Yue et al. 2010). As discussed earlier, the dust layer absorbs the solar radiation, heating itself and the atmosphere, cooling the surface at the same time. This increases the stability of the boundary layer decreasing the value of $u^*$ and the shear stress (eq. (27)): the wind speed is increased over the dust layer and decreased below it. For this reason, the feedback is usually negative, leading to the partial slowdown of further lifting phenomena. For example, in presence of aerosol clouds, a wind speed reduction up to 8% and 4% has been measured over the China and the Sahara, respectively (Jacobson and Kaufman, 2006; Heinold et al., 2008).

On the other hand, this effect can be inverted during the night time, where the emission due to the dust cloud can warm the surface, leading to an increase of the wind shear stress. Moreover, in some cases the feedback can result positive even during the daytime. Indeed, a dense near ground dust haze capturing the outward thermal radiation of the soil, can lead to an overall heating of the lower air layers. This leads to the increase of the atmospheric instability, favoring the convective motions and hence a further injection of grains in atmosphere.

## 1.5 Physic of dust lifting

R. A. Bagnold has been one of the most important early researcher of the physics of grains lifting. During his studies he individuated three main modes for the wind to set in motion the grains: the suspension, the saltation and the creep.

In the following we will analyze these different processes, explaining their range of effectiveness, both on Earth and on Mars.

### 1.5.1 Suspension

Let us consider firstly a grain that is already in atmosphere. The particle is only affected by the wind and the gravity (*G*). When the air flow encounters an obstacle, in this case the particle, it exerts a dynamic pressure *P*. Following the Bernoulli theory P assumes the form:

$$P = \frac{1}{2}\rho v^2 \tag{30}$$

where $\rho$ is the air density and *v* is the relative speed between the flow and the particle motion. Let's define *A* as the cross sectional area of the grain, i.e. the projection of the particle surface on a plane orthogonal to the relative motion direction. The force that affects the grain depends also on the shape of the object, texture, viscosity and other proprieties of the particle surface and fluid turbulence. Hence, we define a generic drag coefficient $C_d$ that takes into account these factors and that is usually empirically estimated. We obtain that the wind drag force *D* over the particle is:

$$D = \frac{1}{2}\rho v^2 C_d A \tag{31}$$

We can define the drag as the opposition force exerted by the wind on a body that tries to penetrate the flow. Hence, *D* has the same direction but opposite verse with respect to the motion of the body in the rest frame of the flow.

Considering only a vertical motion, the particles will fall accelerated by the gravity until reaching a maximum speed called free fall value $v_F$, for witch *D* equals *G*. Smaller is the grains, and hence less is *G*, less will be the $v_F$ needed to obtain the equality $D = G$.

As seen in par.1.4.2.1, even in case of laminar flow, the air has a vertical turbulence component *w'*. The motion of the grain relative to this vertical component is downward directed with a speed equal to $v_F + w'$. Hence, on average, the turbulence will exercise a drag *D'* upward directed. *D'* tends to lift the grains with a speed *w'*, hence if we have *w'* > $v_F$, the particles will stop falling and will be maintained in suspension by the turbulence. Obviously, this process is effective only for smaller particles, the ones of dust size that have a low $v_F$, while sand size and bigger particles will tend to return to the surface. As we mentioned, dust in suspension can be lifted up in the atmosphere and travel for thousands of km before to re-falling on the ground.

### 1.5.2 Saltation

Until now, we faced the case of a particle that is already injected in the atmosphere. To consider the case of a particle that is still on the surface we have to introduce other important forces. Let consider a spherical grain placed on a bed of other similar particles under the influence of a horizontal wind flow. The drag tends to set in motion the grain in the horizontal direction, however this is not the only force exerted by the wind. There is another aerodynamic force directed orthogonally to the direction of relative motion that take the name of *lift force*. Indeed, when the wind blows around the particle, the stream moves faster over the particles than below it, where it encounters an increased friction and number of obstacles. The Bernoulli principle states that the velocity is inversely proportional to the pressure, hence a region of low pressure is created over the particle and a high one below, leading to a net upward pointing pressure gradient force, also called Saffman force, that generates the lift.

In order to set in motion the soil grains, the aerodynamic drag and lift have to overcome the gravity and the friction that link them to the bed below. This friction is made by the interparticle forces of chemical and electrical nature that coalesce the grains, as Van der Walls, electrostatic and water absorption forces.

As we said in the last section, the magnitude of the drag force depends on the factor $C_d$ that has to be evaluated empirically. However, the evaluation of the lift force is even more complicated, as it is the evaluation of the interparticle cohesion. The factors that regulate these forces are both macroscopic (e.g. composition of the soil, shape of the grains, fluid turbulence and environmental condition as the humidity and temperature) than microscopic (e.g. the distribution of the electron on the grains surface and around the contact points) and a proper grain by grain modelling is currently unachievable.

Instead to resolve the singular forces for each particle, we can consider on average the interaction between the air flow and the soil. As we introduced in par 1.4.2.1, the amount of momentum transferred by the wind to the soil is $\tau$ and it can be parametrized with the parameter $u^*$, eq. (11). The friction velocity is not a real speed of the wind, however, it approximatively gives the amount of vertical turbulence $w'$ near the surface. As particles of different sizes will need a different amount of momentum $\tau$ to be set in motion, we can define this threshold as a function of $u^*$.

Bagnold defined two kinds of threshold: a static or fluid threshold $u_t^*$ needed to set in motion the particles when they are still at rest, and an impact threshold needed to sustain the motion once this has already started. We can define $u_t^*$ as (Bagnold, 1941):

$$u_t^* = A_f \sqrt{\frac{\rho_p - \rho}{\rho} g\, D_p} \tag{32}$$

where $D_p$ and $\rho_p$ are the diameter and density of the grain, while $A_t$ is a factor that takes into account the contribute of cohesion, flow turbulence level and lift force.

For the smaller particles the cohesive forces strongly overcome the gravity contribution becoming the main resistance to motion; this results in an enhanced difficult in lifting dust size grains than sand ones. Indeed, on Earth, it has been observed that the optimum grain size for the lifting by wind is around 75 μm, with a value $u_t^*$ around 0.2-0.3 m/s (see Fig. 3). The threshold rapidly increases for particle sizes greater and smaller than the optimum. On Mars, where the atmosphere is more rarefied, the threshold is one order of magnitude higher: ~2 m/s for the optimum size ~ 115 μm (Greeley et al., 1976).

This means that the first particles will be set in motion by wind speed of around 50 m/s, a value that is quite uncommon on the planet. Moreover, if we consider dust size particles, the necessary wind speed value grows up to over 200 m/s, a condition extremely rare on the planet. Hence, the direct aerodynamic wind lift cannot explain the abundance of dust lifting phenomena on Mars and the mean distribution of the dust haze, that peaks around diameter of ~ 1-2 μm (Pollack et al., 1979). Indeed, even if a particle is too heavy to enter in suspension, when the drag is high enough to break the cohesion, the lift force could imprint enough impulse to raise it even slightly off the surface.

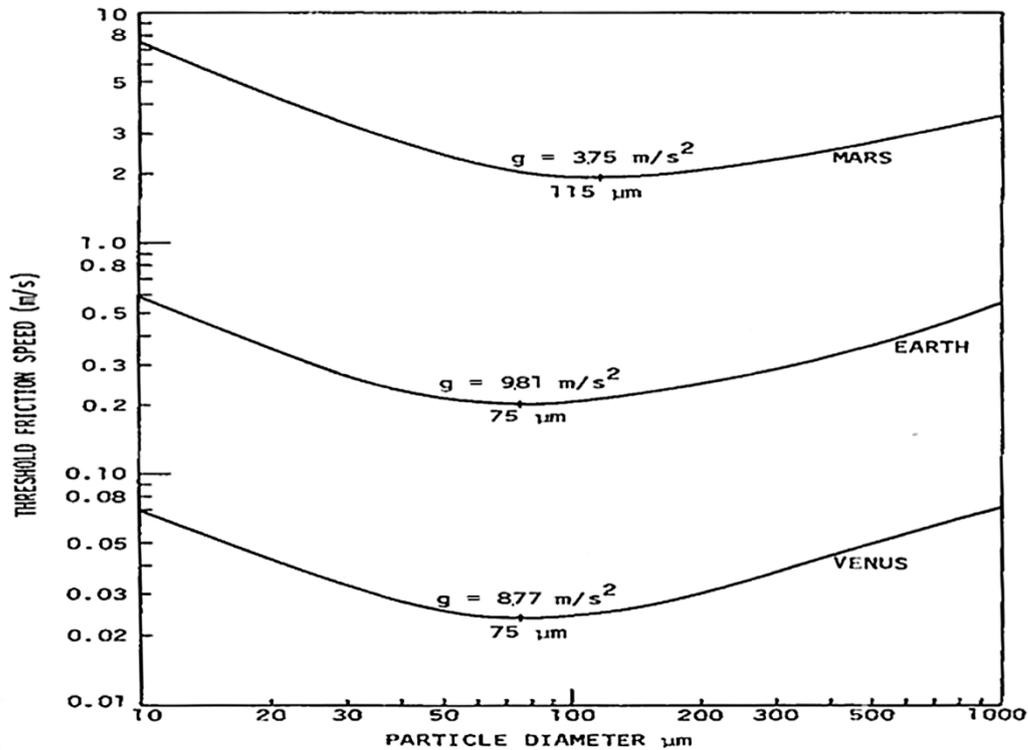

Fig. 3 *The fluid threshold in function of the grain size for Earth, Mars and Venus. (modified, from Iversen et al., 1976)*

After the detachment from ground, the sand grains trajectories are mainly affected by the wind drag and gravity, while, the lift force due to horizontal wind that raised the object rapidly becomes negligible due to re-balancing of the flux speed above and below it. Also other contributions, as turbulence lift and the Magnus force due to grains spin, seem to be negligible. Hence, the grains will move in ballistic trajectories gaining horizontal momentum by drag before to re-fall. By impacting on the ground, these bullets partially convert a fraction of their horizontal momentum in vertical one, transferring energy to other grains. The energy can be enough not only to raise other sand grains but also to break the agglomerates of dust with a sandblasting process (the impacts on the ground of the re-falling sand grains at high velocity), making them much easier to lift (Alfaro et al., 1997; Shao, 2008; Greeley, 2002). The whole process is known as "saltation" and, starting from the particles of optimum size, it can set in motion a range of grain from large sand to small dust. Once started, the saltation can be sustained by wind less intense than the static threshold, whereby, it is possible that a single strong gust begins the process and that this is able to propagate even in a low wind speed background (Almeida et al., 2008; Kok, 2010).

The saltation process is generated by the wind, but has a feedback effect on its profile. Indeed, the particle motion increases the roughness length $z_0$ with a mechanism called "Owen effect" (Owen, 1964; Raupach, 1991).

Some authors have also argued that electrostatic forces can affect the motion of saltators (the grains in saltation) (Schmidt *et al.* 1998; Zheng *et al.* 2003; Kok and Renno, 2008) increasing also the number of grains involved (Zheng *et al.* 2006; Rasmussen *et al.* 2009). Finally, if the particle is too heavy to be lifted, but still light enough to be set in motion by the drag, it will start to creep along the surface displacing other particle on its trajectory. The displaced grains could in turn enter in suspension or saltation depending on their size.

## 1.6 Electrification

The terrestrial atmosphere is characterized by a global electric circuit that connect the ionosphere to the surface. Inside the ionosphere, the electric conductivity is increased by the ionization by solar and cosmic rays and its value is approximatively constant with height. Under the ionosphere there is the so called electrosphere where the conductivity decreases toward the surface. A general positive current flows from the upper atmosphere and to the soil, tending to positively charge the electrosphere and the surface. The value of the electric field is inversely proportional to the conductivity, hence, its maximum is reached in a layer of roughly 1 km above the surface with a fair weather field of ~50 V/m downward directed.
Sand and dust lifting phenomena can lead to the induction of electric field (E-field) up to three orders of magnitude greater than this background. For example, dust devils with a vertical electric field of over 4000 V/m have been reported, firstly by Freier (1960) and Crozier (1964, 1970) and more recently by Farrell et al. 2004.
Inside the dust cloud, the grains mobilized by the wind collide with each other and with the surface, acquiring charge by triboelectricity, leaving the whole system approximatively neutral (Kunkel, 1950; Eden and Vonnegut, 1973).
The exact mechanism that leads to the grains electrification is still debated and not completely understood (see Harrison et al. 2016 for a review on the topic). For a heterogeneous system the charge acquired during the impacts depends primarily on the composition of the colliders and their superficial proprieties (McCarty and Whitesides 2008). In the simplified case that the whole cloud is approximatively homogeneous, the charging process is size-dependent: the smaller grains tend to acquire a charge opposite to the larger ones (Inculet et al., 2006; Duff and Lacks, 2008; Esposito et al., 2016a; Harrison et al., 2016; Neakrase et al., 2016). The smaller particles are lighter and are then driven upwards by the flow, while the larger ones remain closer to the ground, thus producing the charge separation.
The presence of suspended dust negatively charged over a cloud of sand of the opposite sign (upward directed E-field) is consistent with the majority of field and laboratory measurements reported in literature (Bo and Zheng 2013; Frier 1960; Croizer 1964 and 1970; Farrell et al. 2004, Jackson et al. 2006). However, there are cases where this electric configuration is not reproduced (Trigwell et al. 2003, Sowinski et al. 2010, Kunkel 1950) and there are also reported cases where the electric field is downward directed (Demon et al. 1953; Esposito et al. 2016).

Overall, the contact electrification is still poorly understood. Laboratory experiments confirm that composition and the diameter distribution of grains play a fundamental role; however, the quantification of their effect is still not clear. The problem is even more complex when we want to study the triboelectrification inside the dust lifting events, because the dynamic of the phenomena and the environmental atmospheric conditions can heavily affect the final result.

These reasons have motivated the execution of more detailed field surveys, as the ones reported in this work. Indeed, the main purpose of this thesis is exactly the study of the proprieties of the E-field induced by dust lifting events, in order to deepen our knowledge on a field that is still really poor understood and apply the results to the Martian atmosphere, that recent studies hypothesize considerably affected by the triboelectrification.

## 1.6.1 Electrification on Mars

A global atmospheric electrical circuit is likely to exist also on Mars, but, due to the lack of an adequate in situ instrumentation, its presence has not yet been confirmed. However, the abundance of entrained dust, the generally favorable condition for lifted grains to acquire and hold charge and laboratory experiments performed in Martian like condition suggest the existence of an atmospheric electric field, widespread across the planet and highly variable in relation to the dust lifting activity (Eden and Vonnegut 1973, Forward et al. 2009, Barth et al 2016). The triboelectricity associated with dust impacts is expected to be the main charging mechanism of the atmosphere, differently than on Earth where the atmospheric ionization dominates.

The electric field induced by the lifting processes may significantly affect the composition of the Martian atmosphere and the habitability of the planet, locally enhancing by up to 200 times the chemical formation of oxidants able to scavenge organic material from the surface (Atreya et al. 2006).

If the potential difference equals the breakdown voltage of the system an electrostatic discharge can occur. According to the Paschen's law, this voltage decreases with the gas density, that on Mars is two order of magnitude lower than the terrestrial one. Hence, the Martian atmospheric electric breakdown field strength is only ~20 kV/m (Melnik and Parrot 1998) and the concentration of suspended charges can lead to the formation of electric discharges, potentially able to interfere and cause damage to the landed instrumentation, representing also an issue for the human exploration.

# Chapter 2
# The ExoMars missions and the Moroccan field campaigns: detection and study of dust lifting events

Airborne Dust is one of the main factors that drives the short and long term variation of the thermal structure and global circulation of the atmosphere on Mars. Therefore, in order to study the martian climate it is fundamental to understand the physics of the dust lifting processes. In this chapter, we give a brief introduction to the ExoMars programme, whose principal mission will land on Mars in 2021. We will focus in particular on the pieces of the scientific payload under the direct responsibility of our team: the DREMS station of ExoMars 2016 and the MicroMED sensor of ExoMars 2020.
In the frame of the development of the DREAMS instruments and their test in harsh environment, our team performed various field campaigns in the Sahara desert. We will present the different campaigns and the characteristics of the meteorological station deployed. One of the main purposes of terrestrial surveys was the acquisition of a martian analogous data set, in order to study the physics of the dust lifting phenomena. Hence, we will discuss the algorithms we developed to detect dust storm and dust devil events occurred during the campaigns and how we intend to apply these techniques in martian environment.

## 2.1 The ExoMars program

ExoMars (Exobiology on Mars) is a joint program of the European Space Agency (ESA) and the Russian federal Space Agency (Roscosmos). The aim of the project is to search for evidences of life on Mars, extant or extinct, and investigate the Martian atmosphere and long-term climate changes, monitoring the atmospheric trace gases and the boundary layer meteorology in both clear and dusty environments. The program foresees two separate mission. The first mission launched on 14$^{th}$ March 2016 included the Trace Gas Orbiter (TGO) and the Schiaparelli Entry descent and landing Demonstrator Module. The TGO is actually in orbit around the planet to map the sources of various atmospheric trace gases characterizing their spatial and temporal variation. In particular, the TGO will monitor the methane, in order to study the biological or geological origin of the gas.
The scientific payload of Schiaparelli included the DREAMS (Dust Characterization, Risk Assessment, and Environment Analyser on the Martian Surface) meteorological station, we will describe the DREAMS experiment in the next subsection.

One of the aims of the Schiaparelli lander was to test the entry, descent and landing of a payload on the surface of Mars, in preparation of the next ExoMars mission, planned in 2020.
The ExoMars 2020 mission will deliver to Mars a surface platform and a rover.

The rover aims to investigate the martian subsurface to search for sign of organic material, particularly from the early planet stages. The martian atmosphere is not able to shield the surface from the ultraviolet radiation, therefore the biological traces could be host only in the underground. The rover is powered by solar panel and, through its stereo and close up camera (PanCam and CLUPI), it is able to navigate fully autonomously. In this way it could travel up to 100 m per sol, in order to reach the sites more suitable for finding organic material, selected by using a ground penetrating radar (WISDOM) and a water searcher instrument (Adron). The rover is also able to automatically exact sample up to 2 meters deep with its drill, characterizing at the same time the borehole with an infrared spectrometer (ISEM) and a camera to establish the mineralogy (Ma_MISS). The collected samples will be delivered to the internal instruments and crushed in powder in order to be analysed by the infrared (MicrOmega) and raman (RLS) spectrometer and by the organic molecule analyser (MOMA). The study of the sub surface composition coupled with the analysis of the trace gases performed by the TGO will allow to retrieve the past climatic condition of the planet.
The landing platform instead will characterize the atmospheric environment of the landing site, acquiring data of an entire martian year. Overall, its main scientific goals are:
- the study of the vertical structure of the atmosphere, from the orbit to the surface;
- the determination of the present climatic conditions;
- the analysis of the aeolian processes and of their causes in both cases of clear and dusty environment;
- the proof of the existence of an atmospheric electric field;
- the first direct measure of atmospheric dust concentration;
- the providence of ground data to validate the observations made by the orbiters;
- the study of the amount of atmospheric humidity and its daily and seasonal variation;
- the study of the internal structure of the planet, its rotation and orientation, using X-band radio measurement between the surface platform and the Earth
- the investigation of the amount of volatiles exchanged between the atmosphere and the surface, in order to study the possible presence of subsurface water;
- the monitoring and characterization of the radiation environment, including the ultraviolet one.

Among the other instruments, the platform will carry also a METEO package, equipped by a temperature, pressure, humidity and wind speed and direction sensor. It will study the vertical atmospheric profile starting from the descending, monitoring the global circulation and the cycles of carbon dioxide and water vapor. Moreover, the lander hosts also the Dust Complex, a suite of instruments able to monitor the activity and electric proprieties of lifted dust and saltation. It is a suite of 4 sensors:
- an impact sensor, for the detection of the saltating sand grains and the measurement of their charge;

- an electric probe for the study of the atmospheric E-field;
- an electromagnetic activity sensor (EMA) for the observation of the possible electromagnetic discharges;
- an optical particle counter (MicroMED) able to monitor the airborne dust concentration and size distribution.

The synergy between the meteorological package and the Dust Complex will be able to fully characterize the aeolian processes and the environmental conditions where they are generated. The station deployed by our team in the Sahara desert has acquired measurements analogous to the ones expected by the Meteo and the Dust Complex.

## 2.2 The DREAMS Experiment

DREAMS was the only scientific payload that equipped the Schiaparelli lander, it is a meteorological station comprehensive of the first electric field sensor ever sent on Mars.

The purpose of the experiment was the characterization of the Martian boundary layer in both clear and dusty weather, providing in particular the first investigation of the atmospheric electric proprieties, through the crossing of meteorological and electric measurements. This data can also help to study the hazards for the landed instruments and human exploration, quantifying the velocity of the wind and the magnitude of possible electrical discharges. The meteorological station consists of the following sensors: a thermometer (MarsTEM), a pressure sensor (DREAMS-P), an air humidity sensor (DREAMS-H), a 2D anemometer (MetWind), an electric field sensor (MicroARES) and a Solar Irradiance Sensor (SIS). The data are collected by an autonomous Central Electronics Unit (CEU) and all the systems are powered by an internal battery.

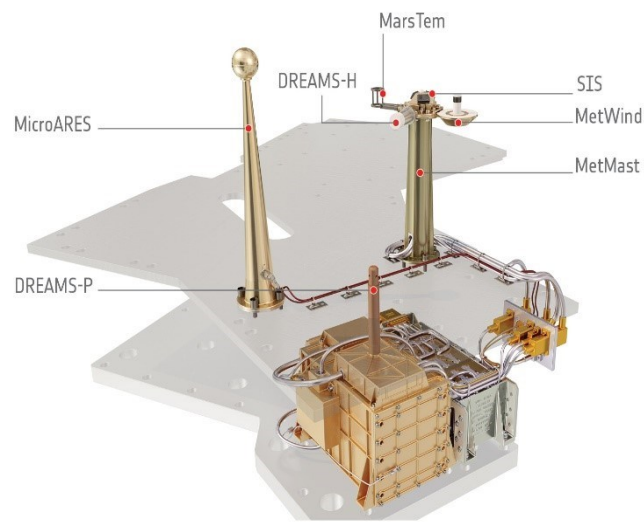

**Fig. 4** *The DREAMS suite*

DREAMS has been developed through a cooperation of six European Countries (Italy, France, Spain, Netherlands, Finland, United Kingdom). The lead of the project is in Italy, at the "Osservatorio Astronomico di Capodimonte", under the supervision of the P.I dr. Francesca Esposito. During my PhD I joined the team of dr. Esposito, working on the analysis of terrestrial data analogous to the foreseen DREAMS and ExoMars 2020 ones, in preparation to the analysis of the Martian data.

The ExoMars 2016 mission arrived at Mars on $19^{th}$ October 2016. Unfortunately, Schiaparelli failed the last phase of the landing. DREAMS was supposed to start to operate after the touchdown, and few seconds before crashing it switched on for the sequence of operations scheduled after landing, proving to be healthy and ready to start measurements.

## 2.2.1 MicroARES

On Mars, the effective triboelectrification of the grains during the lifting processes and the existence of an atmospheric E-field and the has still to be proven by measurements.

MicroARES has been designed to acquire the first evidence of martian electric activity by measuring the amplitude of the vertical atmospheric E-field component, using the lander potential as a reference. In order to check its capacity to monitor the dust lifting events, the instrument has been tested in the Sahara desert during one of the campaign performed by our team, see par. 2.4.

MicroARES is constituted by a spherical electrode installed on a stiff metallic support with an electronics board housed in the common electronics box (DREAMS CEU). In order to be in condition of floating potential, namely in this case the condition necessary to obtain a uniform potential on the electrode surface when it is immersed into the atmosphere, we need the internal instrument impedance to be an order of magnitude greater than the surrounding one.

The MicroARES experiment was design to obtain the first observation of:

- the electric conductivity, its diurnal and day by day variations and its perturbations following solar events;
- the quasi DC (direct current) E-field;
- the ELF/VLF (extreme and very low frequency) radio-electric emissions linked to the AC (alternating current) component of the signals.

The instrument can indeed work in different modes: "a relaxation probe mode" to measure the conductivity during fair weather, and a "high voltage mode" in case of strong electric activity.

Working in relaxation probe mode, when the potential ranges between the ±90 V, the sensor measures the DC component of the signal with a sensitivity of 10 mV/m.

When the potential exceeds the value of 90 V the instrument enters in the high voltage mode to sense potential up to the order of kV. However, this mode can be maintained only for a limited amount time of time due to the increased power output required.

The acquired signal is separated in two components in the analog portion of the electronics. The first is related to the large amplitude and low frequency of the DC channel and the second is related to the small amplitude and high frequency of the AC channel.
The AC channel can be used to detect the impacts of charged dust particles on the antenna.

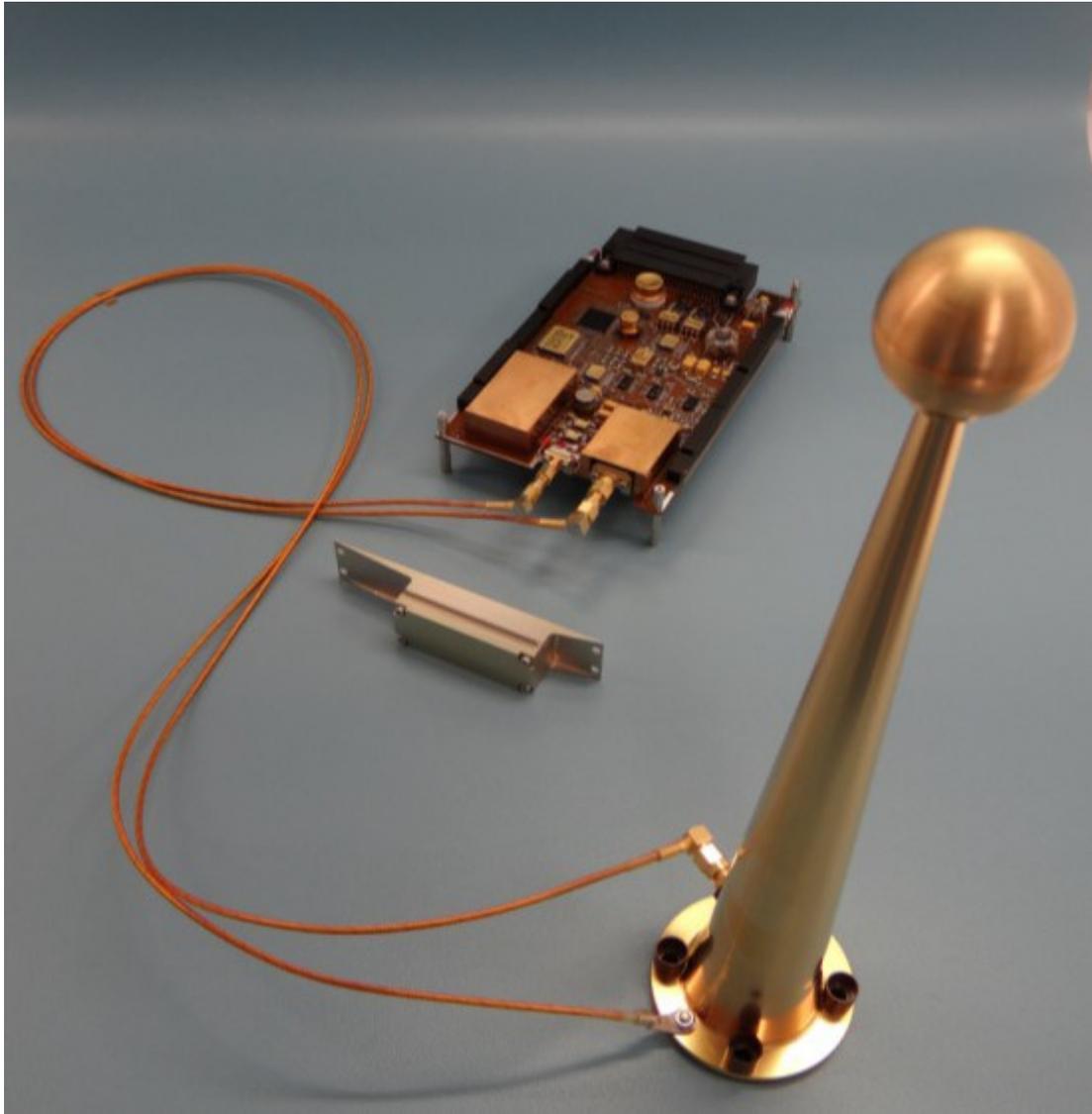

**Fig. 5** *Flight version of Micro-ARES. (source: The Micro-ARES Experiment as Part of the DREAMS, The Sixth International Workshop on the Mars Atmosphere)*

### 2.2.2 MicroMED

As we explained in Chapter 1, the atmospheric dust heavily influences the thermal structure and dynamics of the atmosphere affecting the circulation at all the scales. This is especially true in the martian case, where the stronger dust storms can lead to an absorption of over 80% of the solar radiation. However, the exact mechanisms of raising and settling of dust as its temporal and geographical variability are still not well understood, in particular due to the lack of proper data. Indeed, the estimation of the suspended dust concentration and size distribution is indirectly inferred from the opacity obtained using land or orbit images. The indirect evaluation of these quantities need several a-priori assumptions and a lot of the information related to near surface layers cannot be extracted. Moreover, the surface dust and sand flux and their granulometry represent key input parameters for the Mars climate models.

MicroMED (Fig. 6) has been developed to fill this lack of measure, directly monitoring for the first time the lifted dust concentration and size distribution, as well as its temporal variability. As for the DREAMS station, the lead of the MicroMED sensor is at the Osservatorio Astronomico di Capodimonte with the dr. Francesca Esposito as P.I.. The instrument is an updated and miniaturized version of the sensor MEDUSA that was selected for an early version of the ExoMars payload.

The sensor consists of a single particle optical counter that analyses the grains in the 0.4-20 µm range. A pump will inject the air through an inlet and the flow will cross a collimated laser beam emitted by a diode laser. The light scattered by the passing grain will be detected by a photodiode producing a signal related to the particle size.

The instrument can work in a range from the clear weather case up to the heavy load scenario.

Working in tandem with the other instruments of the Dust Complex and METEO package, MicroMED will be able to fully characterize the lower layer's dust processes, from a meteorological, electric and lifting activity point of view.

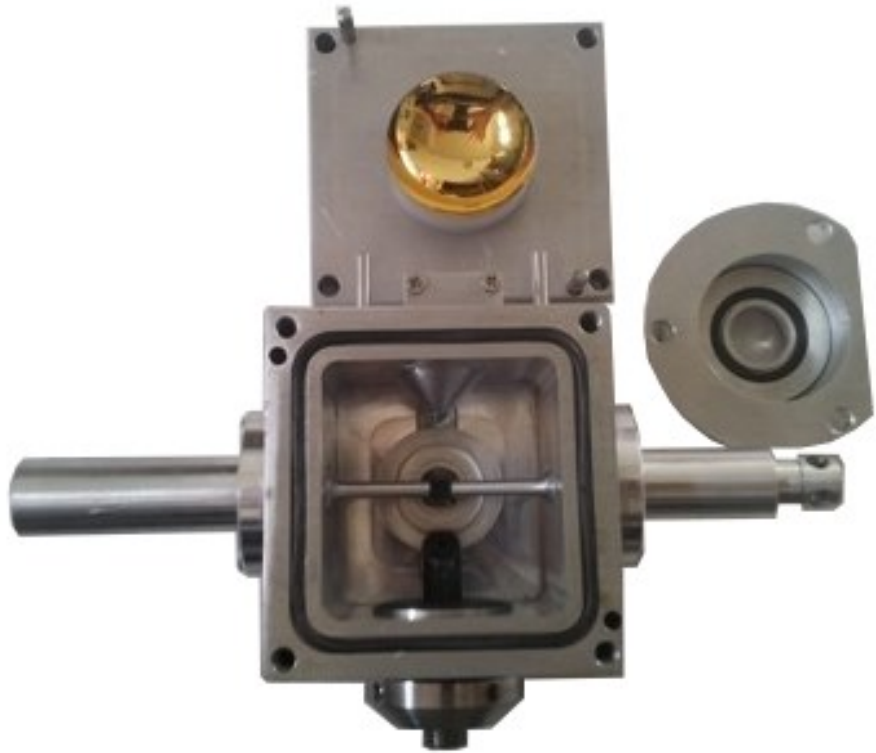

Fig. 6 *Terrestrial test version of MicroMED*

## 2.3 Saharan field campaigns

The Sahara Desert is the biggest source of dust on our planet. Hence, it represents an optimum environment to study the dust lifting phenomena and the dust proprieties. The Terrestrial results, when appropriately tuned to take into account the differences in the atmospheric pressure, composition and gravity, can be very useful to understand the physics of dust processes on Mars.

Our team has carried out various field campaigns in the North West Sahara Desert, in the Moroccan region of Tafilalt, deploying a meteorological station equipped in a manner similar to DREAMS. The aims of the missions were the acquisition of measurements useful to prepare for Martian data analysis and the test of DREAMS sensors in a harsh environment. Indeed, some of the sensors on board of DREAMS, like MarsTEM, SIS and MicroARES have been tested and tuned in the Moroccan expeditions.

Moreover, the data set acquired in the Sahara is in many aspects specular to the ones that will be acquired during the ExoMars 2020 mission. Both the surveys foresee not only the analysis of the atmospheric meteorological parameters as pressure, temperature, humidity and wind (Meteo Suite), but also the monitoring of the saltation activity and lifted dust concentration and distribution, coupled with the study of the induced electric field (Dust Complex).

The first campaign has been carried out in the 2012. However, the mission represented mostly a test useful for the subsequent campaigns. Afterward, three other missions have been performed in 2013, 2014 and 2017 during the local dust storm season, allowing a deep analysis of the local dust lifting activity. In this thesis, we will focus on the data collected in these last three missions.

## 2.3.1 2013 field campaign

The field campaign took place between July and September 2013 at geographical coordinates 4.113° W, 31.161° N, elevation of 797 m a.s.l. The area is characterized by an arid environment, rich in both sand and dust and very active from an aeolian point of view. The site chosen to deploy the meteorological station is near the center of a flat Quaternary lake sediment bed few kilometers away from the Erg Chebbi, an agglomeration of sand dunes height ~150m. The composition of the sediment (sand, silt, and clay fractions) is the result of the erosion of the regional bedrock of late Paleozoic sedimentary rocks outcropping in the area and is chiefly constituted of detrital shale, quartz, and carbonates grains. The soil clay fraction (< 2 µm), is represented by Fe oxides and carbonates, whilst it is depleted in clay minerals. The position near the center of the lake made the site rich in hygroscopic and soluble minerals. For this reason, most of the soil grains are aggregated in an extended saline crust, as Fig. 7 shows.

It is likely that the absorbed water produces deliquescence, forming local brine pockets/films that influence the overall physical properties of the dust (e.g., electrical conductivity, electrical charge/discharge).

The deployed meteorological station consisted of:
- soil temperature (CS thermistor) and moisture (CS616-C) sensors,
- three 2D sonic anemometers (Gill WindSonic) placed at 0.5, 1.41, 4 m,
- one temperature and humidity sensor (Vaisala HMP155) at 4.5 m and one thermometer (Campbell Sci. (CS)) placed at 2.5 m,
- pressure sensor (Vaisala Barocap PTB110) at 2 m,
- solar irradiance sensor (LI-COR LI-200 Pyranometer) at 4 m,
- atmospheric electric field sensor (CS110) faced down at 2 m.

In addition, to monitor the sand and grain motion were deployed also:
- size-resolved airborne dust concentration sensor at 1.5 m (Grimm EDM 164-E) that analyses dust in 31 channels in the range 0.265- 34 µm,
- two sand impact sensors (Sensit Inc.) for the detection of saltating sand grains,
- three sand catchers (BSNE) at different heights (12, 25 and 40 cm) for daily collection of sand in saltation.

The station was set to operate 24 hours/day at a sampling rate of 1 Hz. A solar panel system powered the station.

In this site, as well for the sites of the following campaigns, the surrounding soil was the main source of the lifted dust observed.

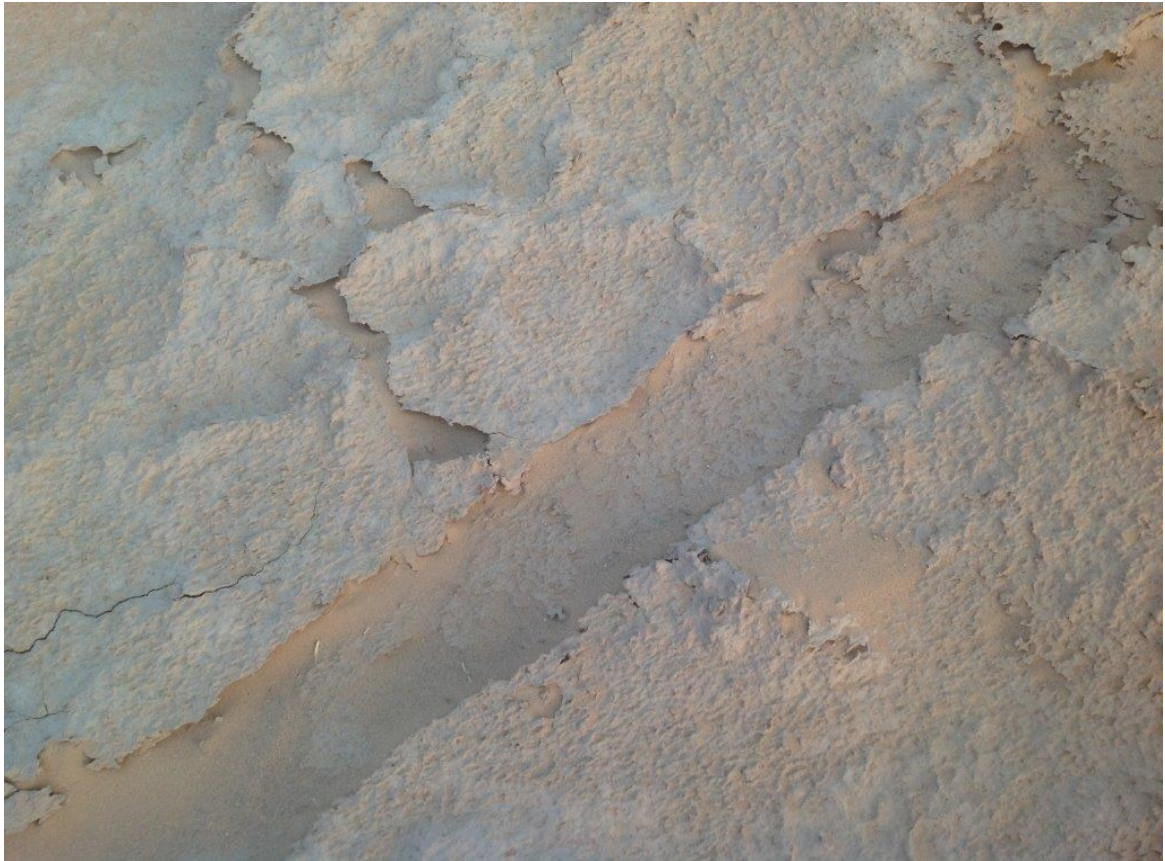

*Fig. 7 The 2013 soil site, where it is possible to notice the abundance of saline crusts.*

## 2.3.2 2014 field campaign

The 2014 field campaign lasted 83 days from June 15[th] to September 5[th] at geographical coordinates 4.110° W, 31.193° N. Unlike the 2013 mission, the team deployed the station on the edge of the Quaternary lake sediment, where the decrease of soluble minerals makes the soil more easily erodible, as shown in Fig. 8. The sand, silt and clay fractions of the soil have a composition similar to the 2013 one, consisting of detrital shale grains, quartz and carbonates.
At the beginning of the campaign, we deployed the same instruments of 2013 placing the three 2D sonic anemometers at 0.5, 1.41, and 4m heights above the ground, one thermometer and humidity sensor at 4.5 m and the other thermometer at 2.5 m, the pressure sensor at 2 m, the solar irradiance sensor at 4 m, and the atmospheric electric field sensor at 2 m. The atmospheric dust concentration sensor and the sand catchers have been placed at same height as in 2013. In addition to this set up, from July 14[th] we added a second mast with two 2D sonic anemometers (Gill WindSonic) placed at 7 and 10 m and one 3D sonic anemometer (Cambell CSAT3) placed at 4.5 m.

The station operated 24 hours/day at a sampling rate of 1 Hz, despite the 3D sonic anemometers that acquired at 20 Hz. The final set up is shown in Fig. 9.

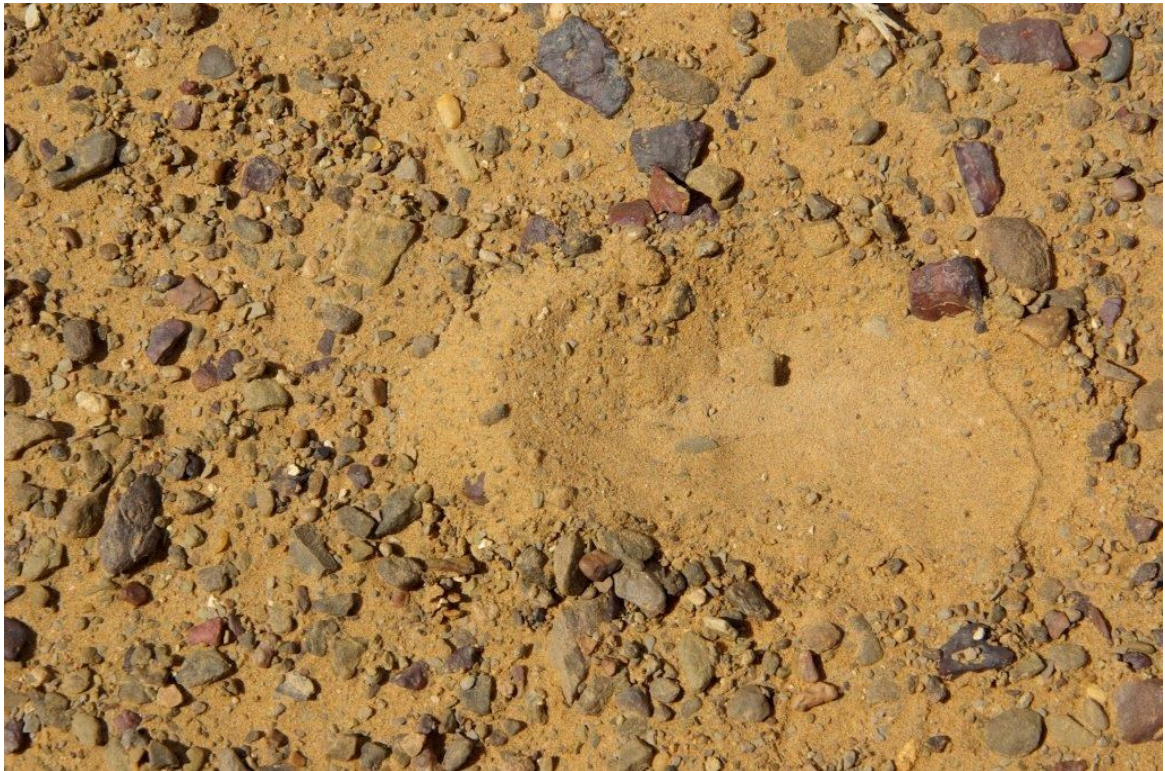

**Fig. 8** *The 2014 soil site.*

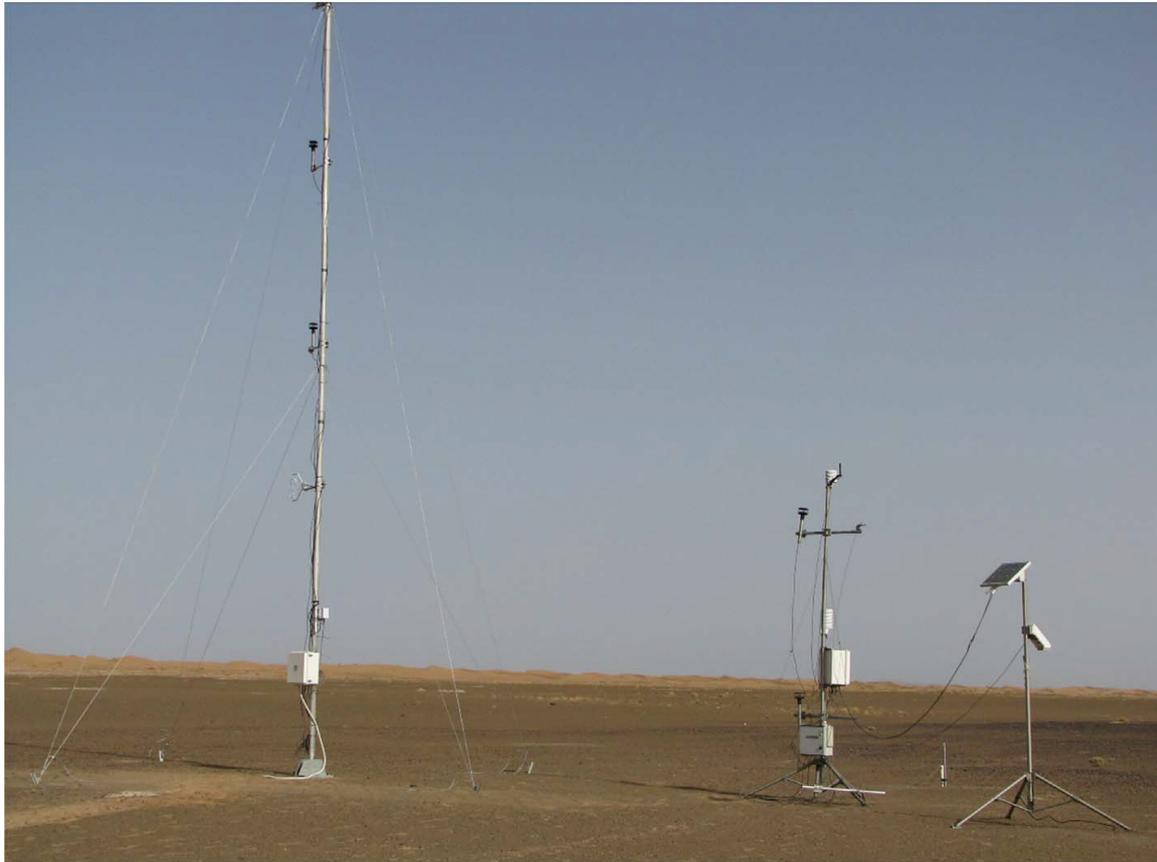

**Fig. 9** *The final set up of the 2014 meteorological station. From the right to the left: the first mast hosts the electric field sensor and the solar panel; the second mast hosts two thermometers, three 2D anemometers, the solar irradiance sensor and the dust concentration sensor; the third mast hosts two 2D anemometers and one 3D anemometers. At the surface level there are the two sensits and the tree sand catchers. Under the surface there are the thermometer and the soil humidity sensors.*

### 2.3.3 2017 field campaign

The mission lasted from the 13$^{th}$ July to the 27$^{th}$ July, for a total of eleven scheduled days of measurements, in a site near the previous ones, at coordinates: 4. 0274° W, 31. 2018° N. The purpose of the campaign was the deployment of a station equipped in a similar manner to the 2013 survey, with the addition of a high rate acquisition camera to catch on pictures the encountered dust lifting phenomena.
The final set up envisaged the use of two masts:
- the first one equipped by three anemometers, three thermometers, a humidity sensor for air and another one for soil, a sand impact sensor and a night lightning system;
- the second mast placed about ten meters from the first one with the solar panel system and the camera.

The camera had to be oriented in order to catch the first mast and the passing dust event in the same image with a field of view of about 180°. All the instruments were powered by a battery, recharged by the solar panel. After the installation, the entire station was planned to be totally independent with the need to visit the measuring site just to download the acquired data or to fix possible problem. The station had to acquire day and night with a sample rate of 1 Hz, while, the camera had to take a picture every 2 seconds during the day, and to enter in a sleep mode during the night, to preserve the battery. The camera would be turned on by a trigger system that we developed on the base of the dust devil research software used to analyse the data of the 2014 field campaign: when the meteorological station acquires a signals compatible with the passage of a dust devil, the trigger turns on the camera and the night lighting system in order to acquire a 30 seconds video.

Unfortunately, due to various issues faced during the campaign we had to reduce the acquisition time, partially changing also the station set up. We will deeply discuss on the plan, organization and on field correction of the campaign in Chapter 4.

Fig. 10 shows a zoom of the Tafilalt region near the border between Morocco and Algeria, where the three Sahara campaigns took place, the sites are there indicated.

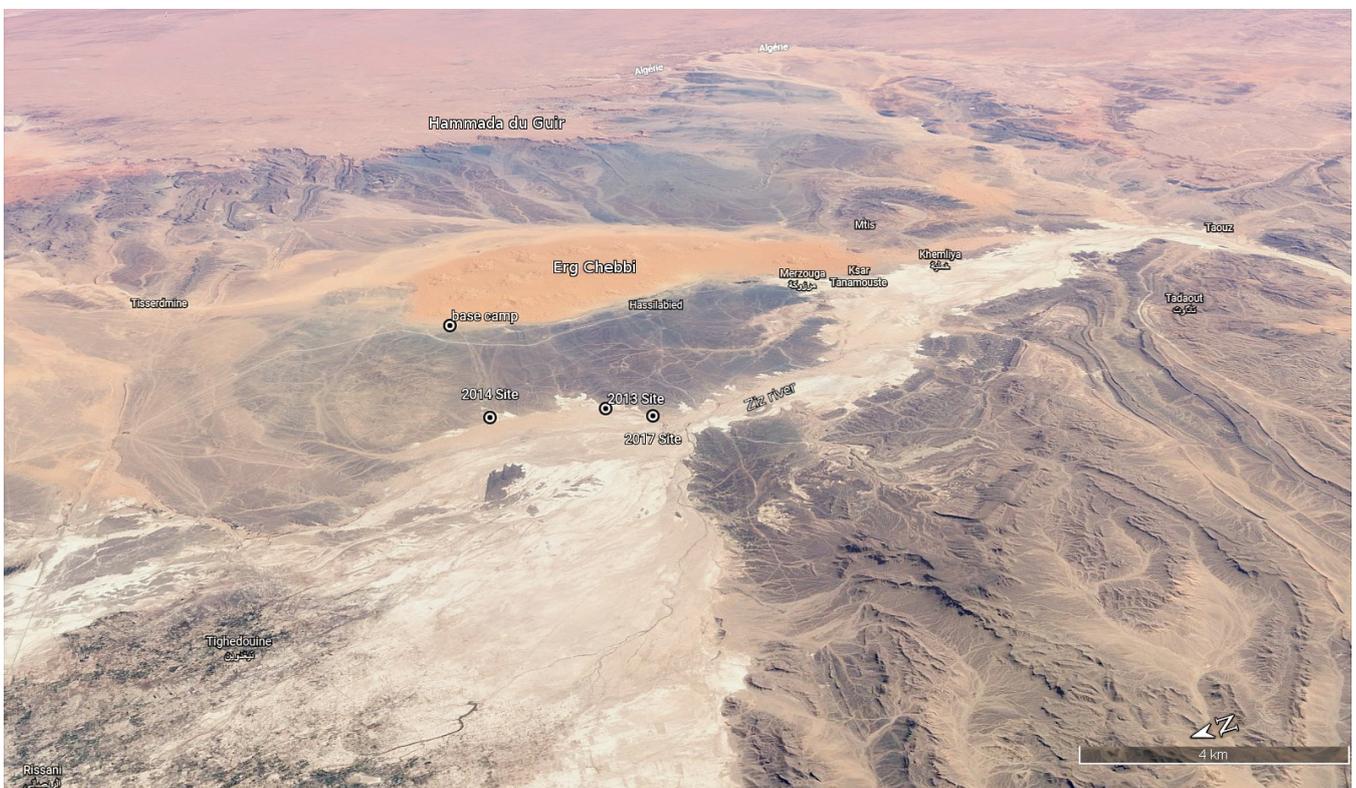

Fig. 10 *The field campaign sites in the Tafilalt region (Morocco). (source: Google Earth)*

## 2.4 DREAMS as dust devils probe on Mars

A terrestrial version of MicroARES has been tested during the 2014 field campaign. MicroARES has to work in condition of floating potential, hence its internal impedance has to be one order greater than the environment. The terrestrial conductivity is two order of magnitude smaller than the martian one; for these reasons we had to use an electrode much larger than the martian version (see Fig. 11). The purpose of test was the study of the instrument behavior in dusty condition and its capability to monitor also the transient events, such as dust devils. We compared the results with the ones obtain through the deployed meteorological station.

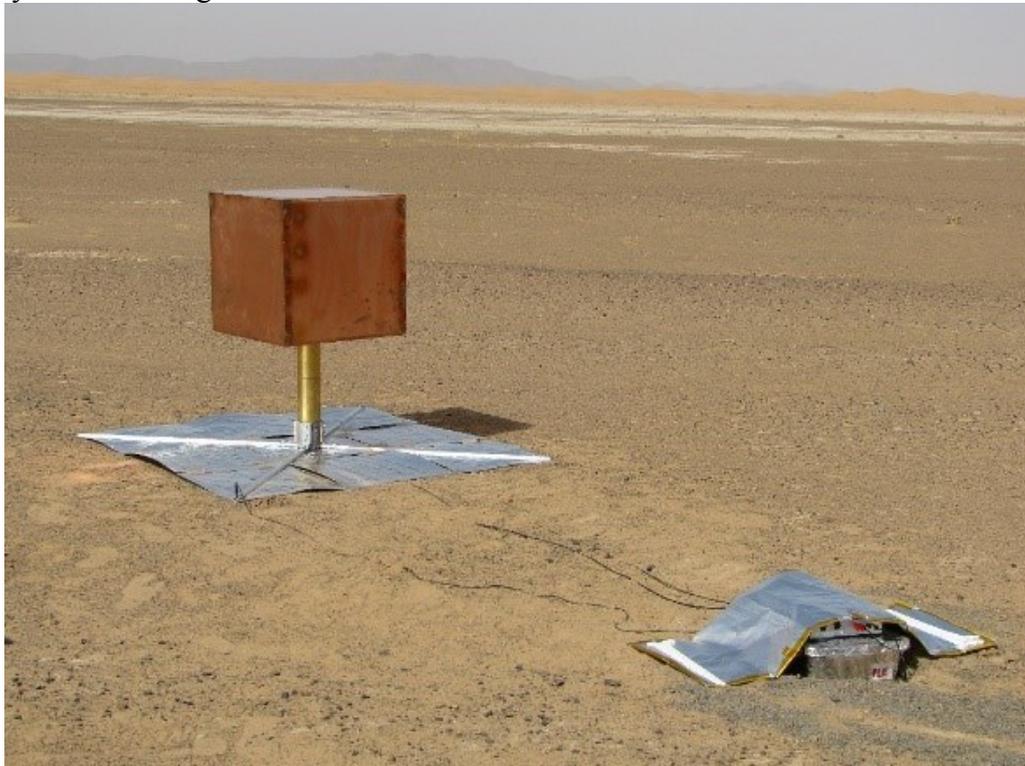

Fig. 11 The terrestrial version of MicroARES tested in the Moroccan campaign

The CS110 is the commercial instrument that we mounted on the Moroccan station to monitor the atmospheric E-field. This sensor can measure only the signals with a frequency below 0.5Hz, hence, only the MicroARES DC channel data can be compared.

Overall, there is a good level of agreement between the measurements of the two instrument, taking also into account that a punctually comparison of the data is not possible due to their different positions. Indeed, on average MicroARES measures the E-field at 0.5 meters from the ground in contrast with the 2 meters of the CS110, moreover there are 50 meters between the two locations.

We also observed how, in most of the occasions, dust devils detected by the CS100 are identifiable also in the MicroARES data. Fig. 12 shows the plots of one of these detection as seen by the whole set of instruments of the station and by MicroARES.

The Morroccan test reasonably demonstrate how DREAMS had the concrete possibility to monitor the dust lifting activity, not only for fairly long events such as the dust storms, but also for the short duration events such as the dust devils. The vortices detection techniques used for the analysis of the Moroccan campaigns has been developed in view of their forthcoming use on the DREAMS data.

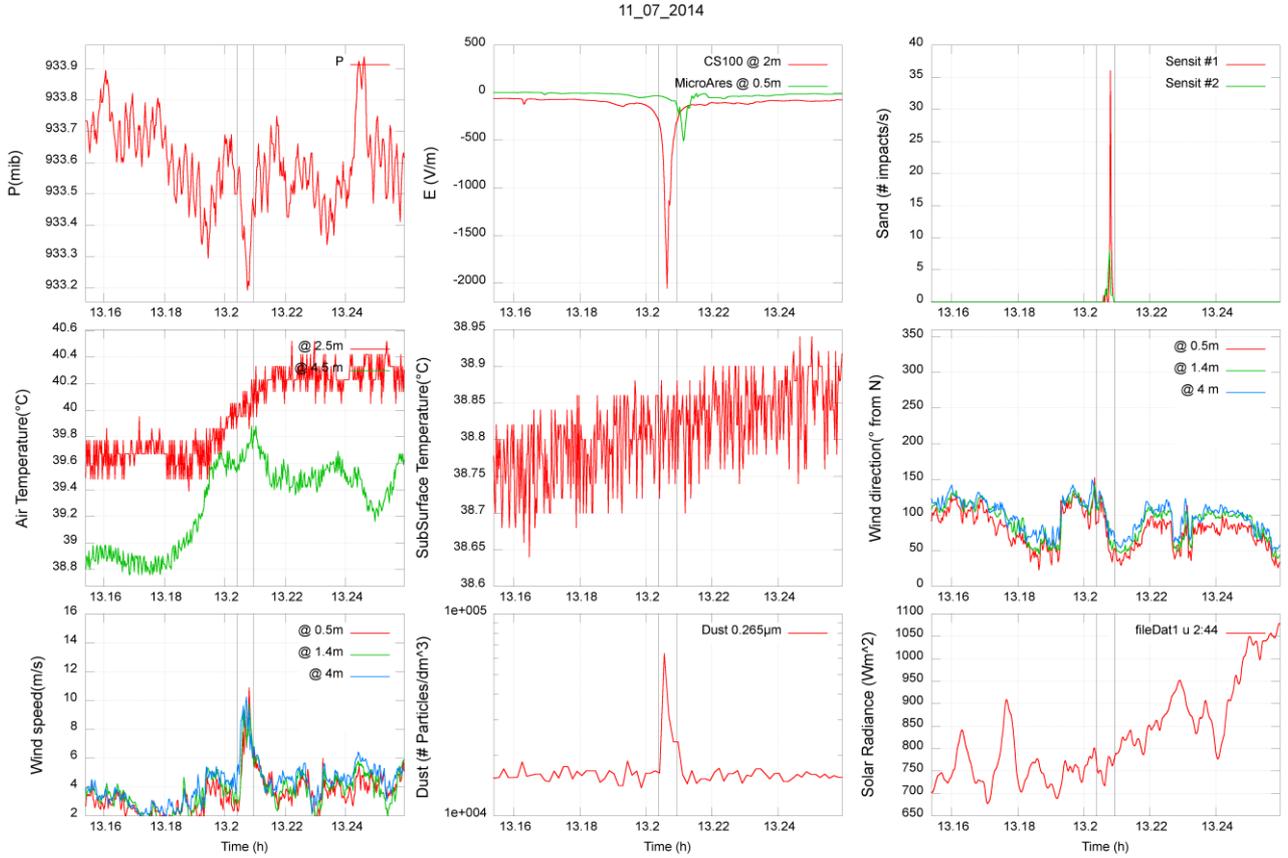

**Fig. 12** *One of the 2014 dust devil events caught both by the deployed meteorological station and the terrestrial version of MicroARES (in green in the second plot).*

## 2.5 Dust storm detection

We used an instrument called SENSIT' to monitor saltation activity. The SENSIT monitors the number of saltating grains measuring the number of impacts per seconds through a piezoelectric crystal (Stockton & Gillette, 1990). In theory, it is possible to use the SENSIT output signal to estimate the sand mass and momentum flux, however we'll use the instrument principally to individuate the time intervals characterized by saltation. The sensor is calibrated to detect the sand size grains and it is unlikely to respond to particles with a diameter less than 150µm (Stout and Zobeck 1997).

The detection of the time intervals characterized by active saltation processes has been performed by using a custom automatic routine. The routine identifies periods where activity is almost continuous, following a procedure similar to that described in Stout and Zoebeck (1997). For each instant, we have considered a time window $T$=3600s around the measurement and we have defined the saltation activity function $S_f$ as the ratio between the number of non-null sensit measures and the total number of measures within $T$:

$$S_f = 100 \frac{N_{Sensit \neq 0}}{N_{Tot}} \qquad (33)$$

When the saltation frequency is higher than a chosen threshold value, the data interval is selected as an active lifting event. The analysis of the detected saltation events is discussed in the Chapter 3.

## 2.6 Dust Devil detection

The main signatures of the passage of a dust devil are (see also par.1.3.2):
- a peak in wind speed,
- a change in wind direction,
- a drop in pressure,
- a peak in the electric field,
- a peak in concentration of the lifted dust and sand,
- a raise in atmospheric temperature.

Depending on the dust devil distance and its magnitude, these features can be more or less evident and some may be totally hidden. Clearly, the simultaneous occurrence of all of them strongly indicates the passage of a dusty vortex.

The detection of dust devils recorded into the meteorological data starts from the search for one of these features. However, generally the study of a single parameter is not sufficient to avoid the false-positive detection. Indeed, on a first approximation, a wind gust can give a signal of duration and wind speed trend similar to a vortex; the gust could also be strong enough to lift grain, perturbing also the dust concentration and E-field value in a manner similar to a dust devil.

Regarding the pressure, the enhanced speed of the gust can also be related to a horizontal pressure gradient, inducing a variation in the acquired measures. In addition, also the air convective motions related to solar heating can be measured as transient variations of atmospheric pressure. However, it is less likely that gusts and convection pressure signals could be confused with the pressure drop induced by a vortex passage.

For these reasons, usually the pressure is chosen as the main probe parameter to start the events detection, with the addition of the other parameters analysis as a next step, in order to eliminate the false positive detections.

For example, the dust devils have been individuated in the Pathfinder, Phoenix and Curiosity data using the atmospheric pressure as the principal identification parameter, performing an analysis called "phase picker" to recognize the passage of the vortices low pressure core (Murphy and Nelli, 2002; Ellehoj et al., 2010; Steakley and Murphy, 2016).

In order to analyze the data collected during our field mission, we developed three different detection algorithms. Two of them are based on the phase picker technique, that works on the time domain of the signal. For the first we have used the pressure as commonly done in literature, while for the second one we tested the possibility to utilize the E-field. For the last method, we developed a code based on a bilinear transform of the acquired signal, analyzing it in a domain constructed on purpose. The different techniques and their advantage and disadvantage are described in detail in the following paragraphs.

## 2.6.1 Time domain research

Phase picker is a term that comes from the seismology: it involves the search for a significant excursion between a short and a long-period average of the signal. When the difference between short and long values exceeds a chosen threshold the event is selected. The threshold depends on the fluctuations around the long-mean value, namely, on the variability and noisiness of the signal. The signatures of the dust devils passage last on average ~20 seconds, hence the short term interval has to be as brief as possible, while the long term one has to be long enough to marginalize the turbulence contribution (~ 10/20 minutes). A phase picker detection applied on the pressure signal is the most used procedure not only in the Martian, but also in the terrestrial measurements (Jackson and Lorenz, 2015). We developed two similar phase picker algorithms to individuate the dust devil events occurred during the 2013 and 2014 campaign. As we said, the vortex pressure drop is around only one thousandth of the atmospheric background, hence, its revelation can be challenging in case of a noisy signal. For this reason, we had to pre-process the acquired pressure time series before to proceed to the vortex detection.

### *2.6.1.1 Pre-processing of the data*

As the passage of the vortex lasts only a few seconds in the data, in order to perform a phase picker analysis, both the long-term time interval and the short one have to be as short as possible. We chose a long-term mean of 12 minutes, similarly to what is commonly done in literature (e.g. Jackson and Lorenz 2015). The standard deviation of our pressure measurements in this interval is ~0.3 mbar. This noise level is too high to allow a clear detection of the medium-small vortex signals and it could totally cover the weaker encounters.

In order to use the standard phase picker method on the pressure time series we need to reduce the noise. For this purpose, we have used a running average of the signal on a time window of 11 seconds. The extension of the window would lead to further cut the noise but also to reduce the pressure drop magnitude, until the complete elimination of the dust devils signals.

Hence, instead of increasing the running average time windows we decided to couple another kind of filter. The dust devils formation is not a strictly periodic phenomenon, hence the individuation of characteristic frequency in the variation of the pressure signal cannot be related to vortex passage. We performed a fast Fourier transform (FFT) of the pressure signal dumping the characteristic frequencies comparable or inferior the duration of the vortex passages.

Both the duration of the running average window and the dumping level of the FFT have been chosen to maximize the signal to noise ratio limiting the loss of information on the dust devil passages.

Fig. 13 shows the pressure signal in the three phases of the process: as acquired, after the running average and after the running average and the Fourier filtering.

After the application of the filters the standard deviation around the long-term mean is inferior to 0.1 mbar and we used the 1 Hz rate as the short term value of the phase piker analysis.

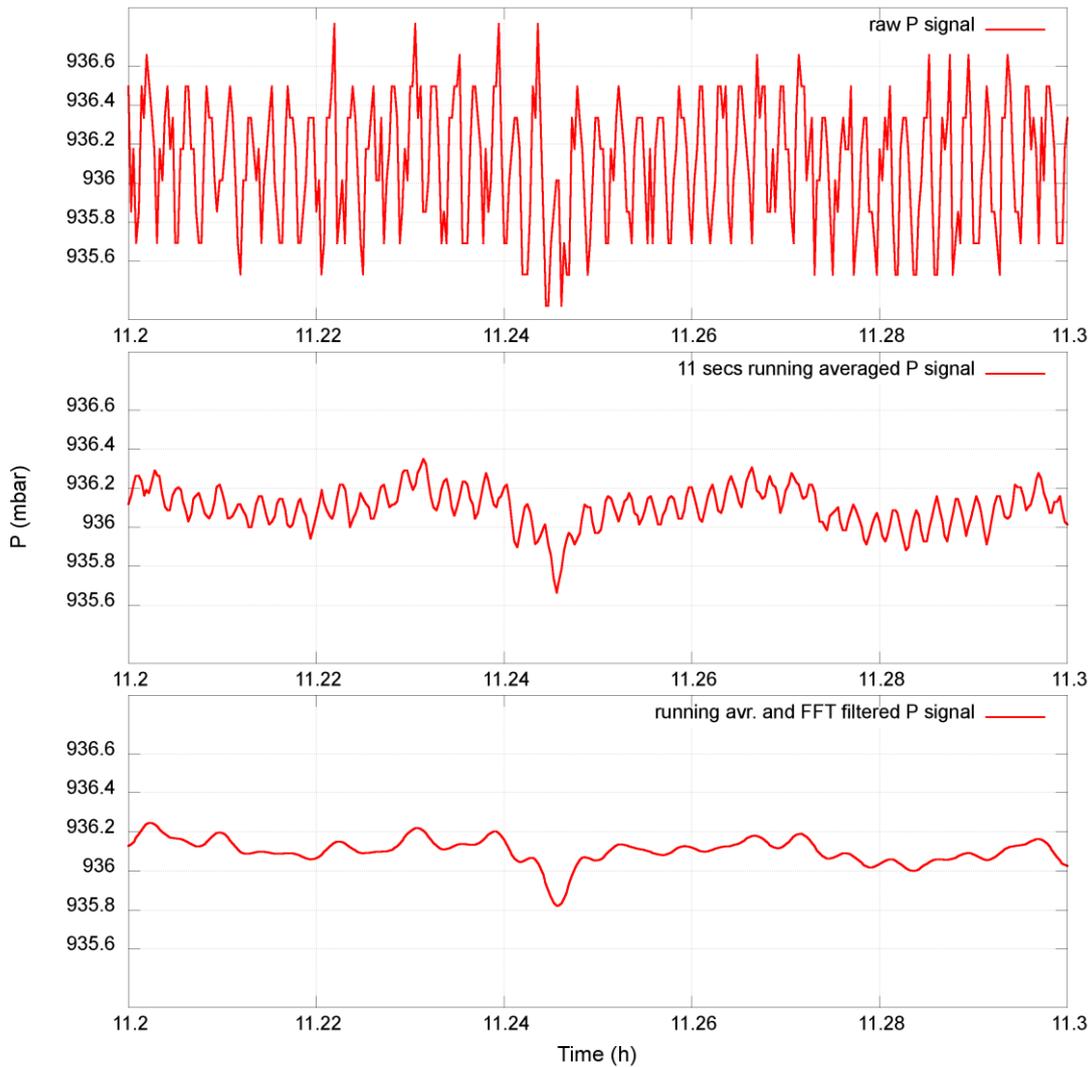

**Fig. 13** *The pressure time series relative to a dust devil observed during 27$^{th}$ July 2014. The first plot is relative to the raw acquired signal. The second and the third are relative to the two level of filtering, the running average and the fast Fourier transform respectively.*

### 2.6.1.2  2013 detection algorithm

We have developed a software that analyses the filtered data, dividing the whole day in time intervals of 12 minutes. For each one, it evaluates the median value of the atmospheric pressure. When the instantaneous and median pressure values differ more than a given limit ($\Delta P_l$), the time interval is selected. For these selections, the software analyses also the variations from mean of wind direction and electric field. If both these variations overcome the chosen thresholds, $\Delta W d_l$ and $\Delta E_l$, the event is identified as a possible dust devil. We have used the following limit values: $\Delta P_l$=0.18 mbar, $\Delta W_l$=30°, $\Delta E_l$= 50 eV. Indeed, we verified that these values give a good compromise between the possibility of detecting even the small dust devils and the ability to cut off the main part of the non-significant events. After this automatic selection, we crosschecked the selections using the whole set of measured parameter and we classified the events, from A to D, using the guidelines described in Table 1.

Fig. 14 shows how a Class A event appears for each measured parameters.

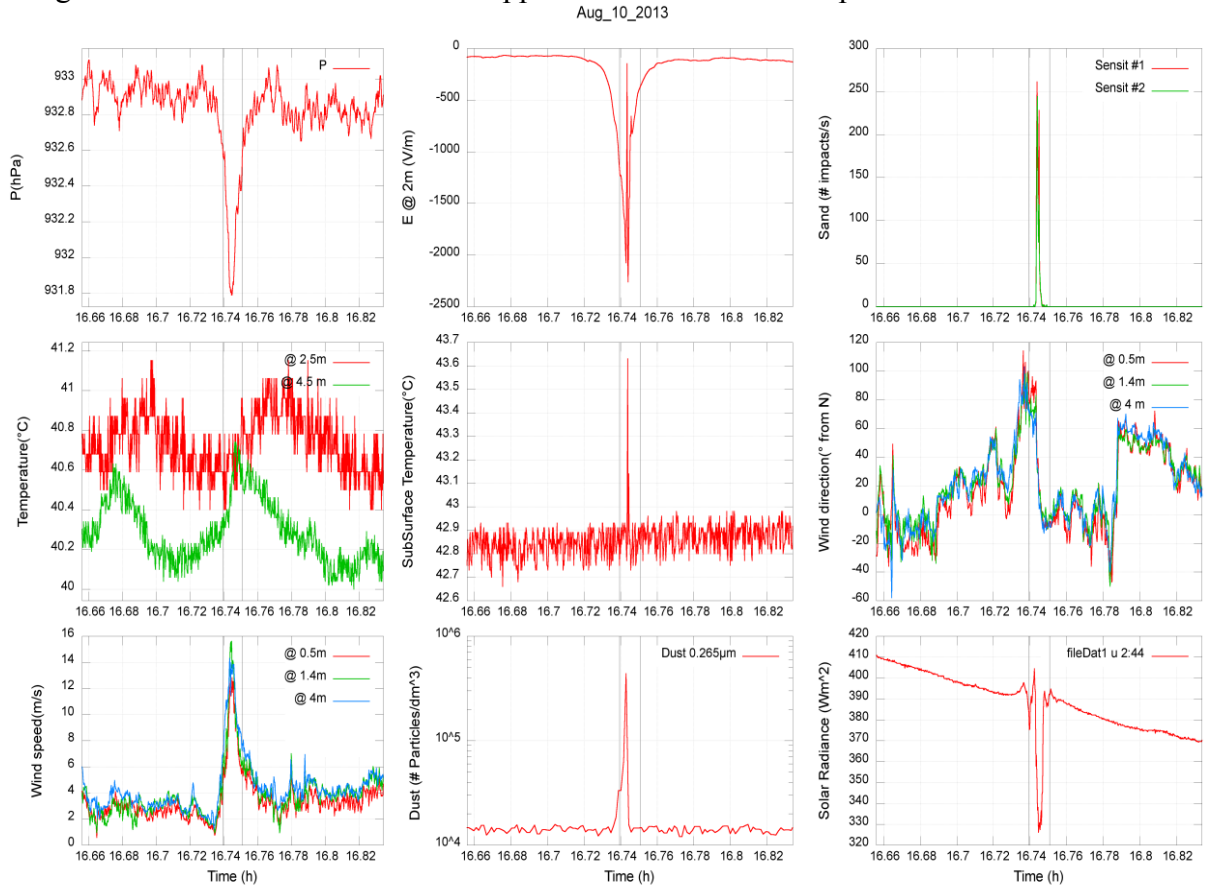

**Fig. 14** *A dust devil occurred on 8$^{th}$ August around a quarter to five p.m., identified by the meteorological instruments.*

| Classes | Main Characteristics |
|---|---|
| D | The event is a false positive, where the dust devil features are certainly not recognizable. |
| C | The pressure drop is barely observable and there are weak variations in electric field, wind speed and direction. The signal usually shows also features hardly compatible with a dust devil, i.e., a peculiar shape or |
|   | anomalous time duration. The event probably is not a dust devil. |
| B | The magnitude of the pressure drop is comparable with the noise level so could be partially hidden. However, the event shows a clear peak for each of the other main parameters. The event is probably a dust devil |
| A | The event shows a clear peak for each of the main parameters, it is clearly recognizable as a dust devil |

Table 1 *Classes and main characteristics regarding the division of the events identified by our phase picker research technique.*

### 2.6.1.3  2014 Detection algorithm

Unlike what we did for the 2013 data and the common behavior of literature, we have chosen to use the electric field instead of the pressure as the principal detection parameter. Indeed, the vortex pressure drop ranges from few tents of mbar to a couple of mbar, while the standard deviation along the twelve minutes' filtered pressure median is ~0.1 mbar. Instead, the vortex induced E-field ranges from few hundreds of V/m up to twenty thousand V/m, while the deviation around the twelve minutes' median is only few V/m. For this reason, we individuated in the E-field variation the most easily recognizable feature of the dust devil passage. Hence, we developed a code that follows the same passages of the 2013 one, but starting from the analysis of the E-field time series.

Again, our algorithm splits the whole day in time intervals of 12 minutes each and selects the events for which: $|E - E_m| \geq \Delta E_l$, where $E_m$ is the median E-field, $E$ is the 1 Hz measurement and $\Delta E_l$ is the chosen threshold. As we said, the detection of a single dust devil feature is not enough to unambiguously recognize the event. Hence, the next step of the algorithm analyses the wind direction and the pressure time series. When the pressure drop ($\Delta P$) and wind direction variation ($\Delta Wd$) overcome the chosen thresholds $\Delta P_l$ and $\Delta Wd_l$, the events are categorized as a possible dust devil.

We used the following thresholds: $\Delta E_l$=250V/m, $\Delta P_l$=0.18 hPa, $\Delta Wd_l$=25°. The threshold values have been chosen analyzing a subset of the whole dust devils sample, studying the typical magnitude of the induced variation and the noise level.

The value of $\Delta E_l$ is higher than the corresponding one used for the 2013 routine. This is due to the different characteristic of the two sites and in particular to the increased amount of soil humidity and the consequent decreased amount of dust lifted in 2013. Indeed, in the 2014 campaign, we observed that dust lifting events induced on average a stronger E-field. In order to eliminate the false-positive detections (mainly gusts and pieces of dust storms), the selections have been cross-checked using the other measured parameters as the horizontal and vertical wind speed, lifted sand and dust concentration, solar radiance and atmospheric temperature. We used the same categorization described in Table 2.

## 2.6.2 Tomography technique

As we mentioned (par. 2.6), isolated wind gusts could give a variation in wind speed and direction similar to the dust devil one, while, the pressure drop is more unlikely to be reproduced by not vortex phenomena. Hence, the pressure is usually chosen as principal detection parameter, coupling the analysis with the study of other key parameters, as the wind speed and direction, etc..
However, it is not uncommon that the monitoring of the dust devil's activity is affected by complications that prevent the acquisition of a complete set of meteorological parameters. This is true in particular for the extra-terrestrial missions. For example, the Viking Meteorology Instrument System on board of the Viking Lander 1, as well as Meteorology Package on board of the Pathfinder lander and the Rover Environment Monitoring Station on board of the Curiosity rover have suffered anomalies with the wind speed and direction detectors, making the wind data totally unavailable in some cases. The lack of these key parameters represents a serious issue for the unambiguous identification of the vortices.
This has led us to study the eventuality to perform dust devils detection using a single probe parameter. For this purpose, we decided to analyze the signal not directly in the time domain, but to use another domain built on purpose.
The most utilized integral transforms are the Fourier transform (Fourier, 1888) and the Wavelet transform (Daubechies, 1990). However, the Fourier transform does not provide information on the transient behavior of the signal, while, Wavelet transform provides some localization but it presents problems in coefficients interpretation and it is not appropriate for our signals. Localized transforms, such as the Windowed Fourier transform, for shorter window sizes allow a good localization in time but reduce the capacity of detection of low frequency components; on the other hand, longer window sizes reduce the capacity of time localization of the transformed signal.
We decided to adopt a bilinear transforms and among these, the Wigner-Ville quasi-distribution (Wigner, 1932) is the most commonly used. However, it does not guarantee the absence of spurious or even negative terms that can also appear in areas where there is no signal at all, presenting also the same issues of the Windowed Fourier Transform.
These problems in the bilinear transforms arise from the fact that time and frequency are two noncommutative operators and therefore a joint probability distribution cannot be defined, even in the case of positive quasi probabilities (Husimi, 1940; Kano, 1965).
For these reasons, we opted for the Tomograms: strictly positive bilinear transforms that provide a full characterization of the signal (Man'ko and Mendes, 1999; Man'ko et al., 2001). The transforms are obtained from the projections on the eigenstates of self-adjoint operators B obtained as a linear combination of a pair of commuting or non-commuting operators $O_1$ and $O_2$.

$$B(\mu,\nu) = \mu O_1 + \nu O_2 \qquad (34)$$

For the time ($t$) – frequency ($\omega$) operator we obtain:

$$B_{tf}(\mu, \nu) = \mu t + \nu \omega = \mu t + \nu \left(-i \frac{\partial}{\partial t}\right) \quad (35)$$

Taking $\mu = \cos(\theta)$ and $\nu = \sin(\theta)$ we have an operator that depends on a single value $\theta \in (0, \pi/2)$ interpolating between the time and frequency operators:

$$B_{tf}(\theta) = \cos(\theta) t + \sin(\theta) \left(-i \frac{\partial}{\partial t}\right) \quad (36)$$

When $\theta = 0$ we are in the time domain and when $\theta = \pi/2$ we are in the frequency domain. The construction of the time-frequency tomogram reduces to the calculation of the generalized eigenvectors of the operator $B_{tf}$. The projection $M_f(\theta, X)$ for a finite time signal $f(t)$ defined in an interval $t_0$ to $t_0+T$ is:

$$M_f(\theta, X) = \left|\int_{t_0}^{t_0+T} f^*(t)\, \psi_{\theta,X}(t) dt\right|^2 = |\langle f, \psi \rangle|^2 \quad (37)$$

where $\psi_{\theta,X}(t)$ are the eigenfunctions of operator $B_{tf}$, namely:

$$\psi_{\theta,X}(t) = \frac{1}{\sqrt{T}} \exp\left(\frac{i \cos\theta}{2 \sin\theta} t^2 + \frac{iX}{\sin\theta} t\right) \quad (38)$$

The dust devils pressure drop has a clear time behavior (see eq.(2)), but this trend could be totally or partially hidden by noise. Moreover, in the frequency domain, dust devils pressure does not possess a characteristic behavior. This fact suggests that a different kind of tomograms should be used. In this new tomogram, one of the operators should be adapted to the characteristics of the component we want to separate.

With the detection of dust devils in mind, a new type of signal-adapted tomogram has been recently proposed (Aguirre and Vilela Mendes, 2014; Gimenez-Bravo et al., 2013). The signal-adapted tomogram is a linear combination of a standard operator, such as time or frequency, with an operator $S$ that is specially tuned for the component that we want to extract:

$$B(\mu, \nu) = \mu t + \nu S \quad (39)$$

The construction of signal-adapted operator follows the same technique used in the bi-orthogonal decomposition of signals (Aubry et al., 1991; Dente et al., 1996).

We can consider a set of $k$ $N$-dimensional time sequences $\{\vec{x_1}, \vec{x_2}, \ldots, \vec{x_k}\}$ that are typical representations of the component one wants to detect. In our case, we will use $k$ signals of dust devils pressure drops of $N$-secs duration. This set of time sequences can be represented by means of a $k \times N$ matrix $U$, with usually $k < N$:

$$U = \begin{pmatrix} x_1(1\Delta t) & x_1(2\Delta t) & \cdots & x_1(N\Delta t) \\ \vdots & & \ddots & \vdots \\ x_k(1\Delta t) & x_k(2\Delta t) & \cdots & x_{1k}(N\Delta t) \end{pmatrix} \quad (40)$$

to construct the square matrix:

$$A = U^T U \in \mathcal{M}_{N \times N} \quad (41)$$

The diagonalization of $A$ provides k non-zero eigenvalues $(\alpha_1, \alpha_2, \ldots \alpha_k)$ and the corresponding $k$ $N$-dimensional eigenvectors $(\Phi_1, \Phi_2, \ldots \Phi_k)$.

The linear operator $S$ can be constructed from the previous set of eigenvectors in the following way:

$$S = \sum_{i=1}^{k} \alpha_i \, \Phi_i \Phi_i^T \in \mathcal{M}_{N \times N} \tag{42}$$

While the time operator for discrete time is built in the following way:

$$t = \begin{pmatrix} 1\Delta t & & & \\ & 2\Delta t & & \\ & & \ddots & \\ & & & N\Delta t \end{pmatrix} \in \mathcal{M}_{N \times N} \tag{43}$$

Hence, $B(\mu, \nu)$ assumes the form:

$$B(\mu, \nu) = \mu t + \nu S = \mu \begin{pmatrix} 1\Delta t & & & \\ & 2\Delta t & & \\ & & \ddots & \\ & & & N\Delta t \end{pmatrix} + \nu \sum_{i=1}^{k} \alpha_i \, \Phi_i \Phi_i^T \in \mathcal{M}_{N \times N} \tag{44}$$

In analogy to the time-frequency tomogram, a particular set of $(\mu, \nu)$ pairs can be selected by a single parameter $\theta$, with $\mu = \cos\theta$, $\nu = \sin\theta$. We can proceed in a way similar to the time-frequency operator, looking for the $N$ eigenvectors $\{\vec{\psi}_\theta^1, \vec{\psi}_\theta^2, \ldots, \vec{\psi}_\theta^N\}$ of operator $B(\theta)$. The projections of the signal $\vec{X}$ on these eigenvectors are obtained by:

$$c_\theta^i = \langle \vec{X}, \vec{\psi}_\theta^i \rangle \quad for\ i = 1, 2, \ldots, N \tag{45}$$

These projections construct a tomogram adapted to the operator pair $t$, $S$.

For the way in which $S$ has been built, it is possible to separate the signal components we are interested in by looking for particular values of $\theta$ where noise or undesired components vanish or becomes small. To do this, we have to search for high energy concentration in particular coefficients that indicates the presence of the component we are looking for. Therefore, we selected a threshold $\epsilon$ and we considered the set of values that contain a given amount of the total energy $c_\theta^i \geq \epsilon$.

We chose $\epsilon$ depending on the spectrum average $\left(\frac{1}{N}\sum_{j=1}^{N}|c_\theta^i|\right)$:

$$\epsilon = k_\epsilon \frac{1}{N} \sum_{j=1}^{N} |c_\theta^i| \tag{46}$$

where $k_\epsilon$ is only a constant.

In order to reconstruct the denoised signal $\vec{x}^f$, retaining its temporal structure information, we can consider only the indexes $i = i_1, i_2, \ldots, i_h$ for which $c_\theta^i \geq \epsilon$. We obtain the subset of $h$ coefficients $C = \{c_\theta^{i_1}, c_\theta^{i_2}, \ldots, c_\theta^{i_h}\}$. Considering only the eigenvectors $\{\vec{\psi}_\theta^{i_1}, \vec{\psi}_\theta^{i_2}, \ldots, \vec{\psi}_\theta^{i_h}\}$ of the subset $C$, we obtain the filtered signal:

$$\vec{x}^f = \sum_{j=1}^{h} c_\theta^{i_j} \vec{\psi}_\theta^{i_j} \tag{47}$$

### 2.6.2.1 Application to dust devils and categorization of the detections

We have built a 277x1000 matrix $U$ containing a set of 277 typical vortex pressure drop signals of 1000 second duration. The drop is around 15% of the background with durations ranging from few seconds to a minute; the entire signal is normalized to zero mean. The set of signals can be tuned for cases of low atmospheric pressure, to be adapted to the detection of dust devils on Mars.

As described in the previous section, we have built *A* and *S* from *U*. Fig. 15 shows the signal-adapted operator *S*, it is visible that it is symmetric and definite positive.

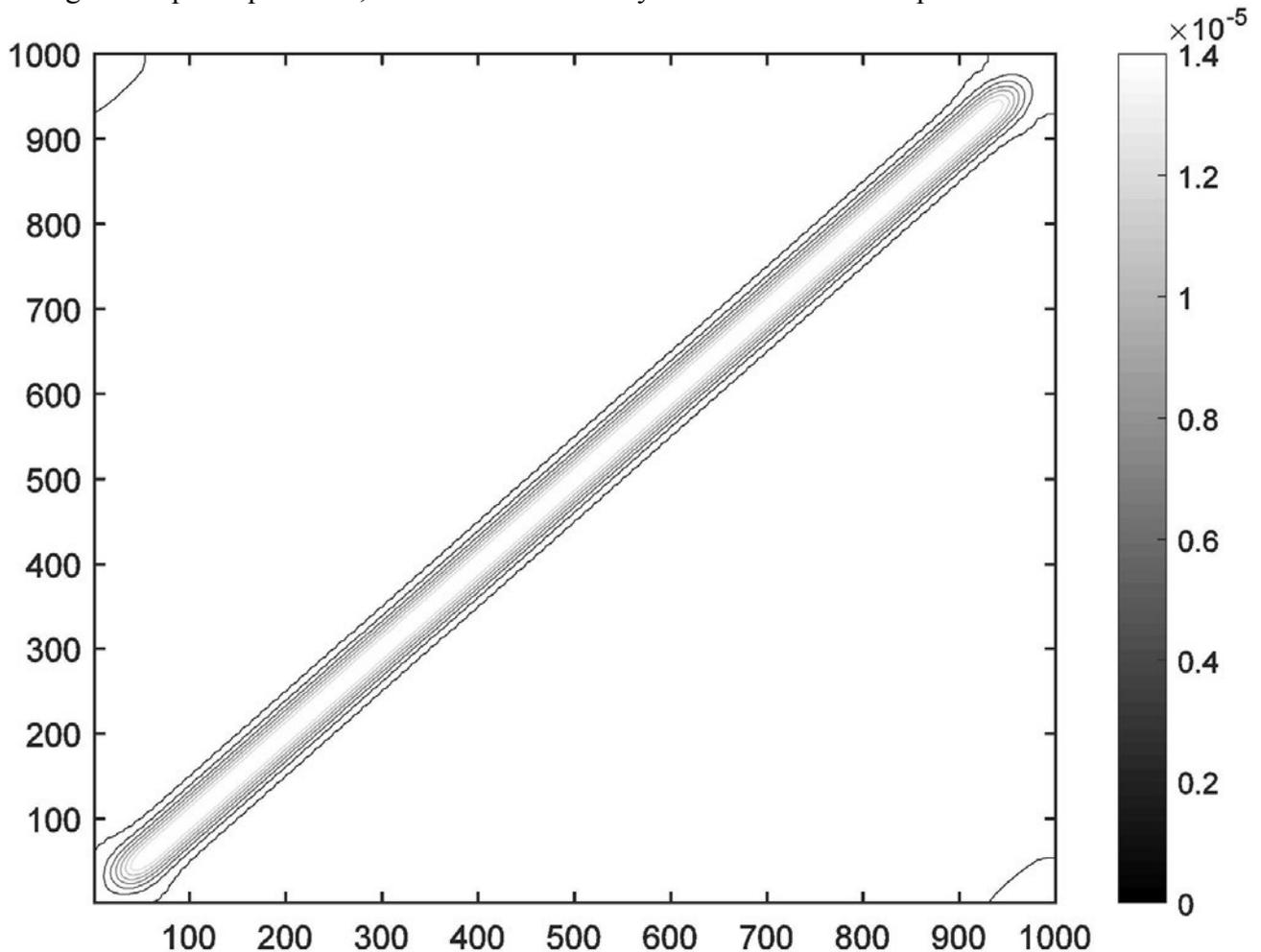

**Fig. 15** *Matrix form of the signal-adapted operator S.*

Finally, we built the tomogram with the linear combination $B(\theta) = cos(\theta)\, t + sin(\theta)\, S$ for the values $\theta = \frac{\pi l}{40}$ $l = 1,2 \ldots 20$.

To analyze each day of data we divided the signal in 1000-second samples with a 200-second overlapping margin between consecutive interval, to avoid loss of events close to the borders. We have normalized the signals to zero mean.

Fig. 16 shows one of the pressure samples of one thousand seconds containing a dust devil event relative to 10$^{th}$ August 2014.

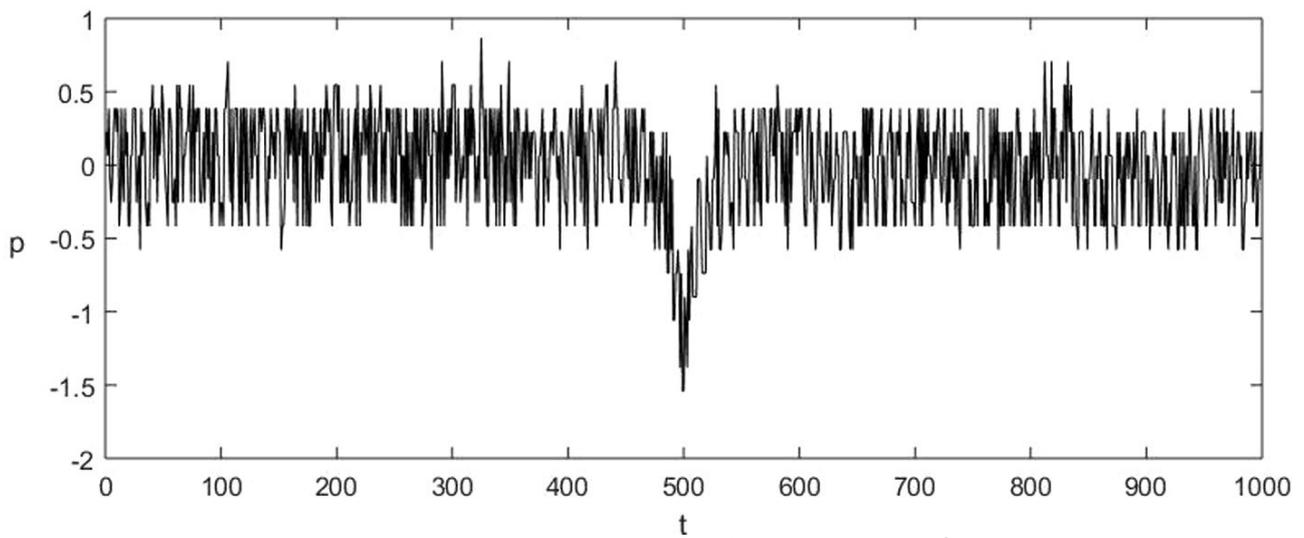

*Fig. 16 Raw pressure signal of a dust devil observed during the 2013 campaign, the 10th August around 4:45 pm, corresponding to ~500th second of the sample.*

For this event, Fig. 17 shows the value of the coefficients $c_\theta^i$ in function of $i$ and $\theta$. A clear peak in the tomogram is visible for every $\theta$ value when $i \sim 500$, that is the time instants relative to the dust devil passage.

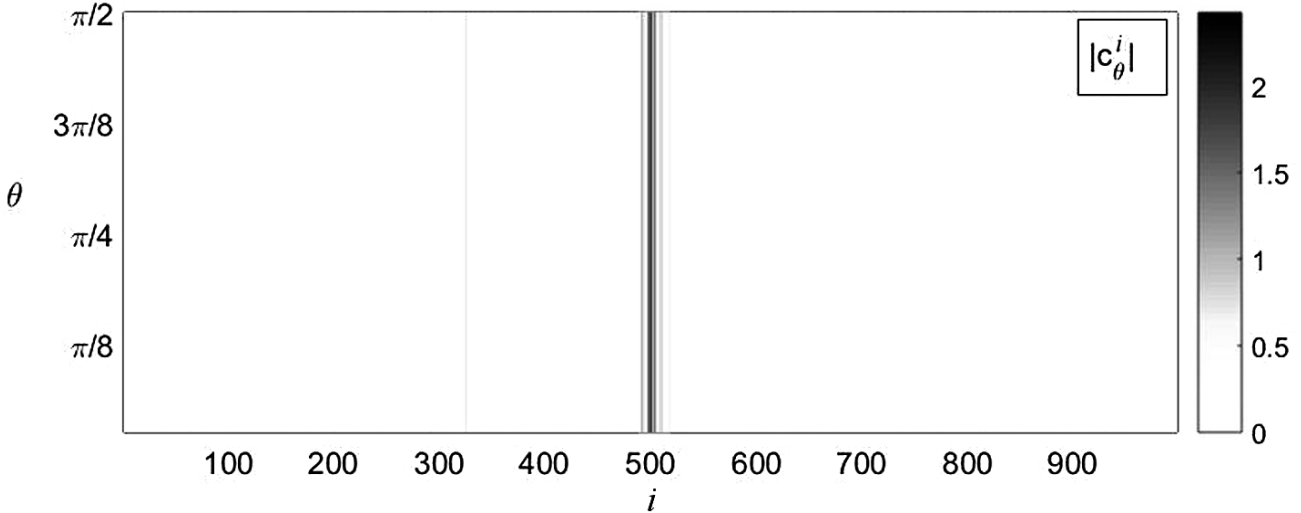

*Fig. 17 Tomogram of the event depicted in Fig. 16. A clear peak of the coefficients $c_\theta^i$ is visible for $i \sim 500$.*

Fig. 18 shows the projection of the tomogram (Fig. 17) for $\theta = \frac{\pi}{4}$. In order to avoid border effects, the first and last coefficient of the projection are discarded, as these coefficients tend to concentrate the energy of the signal that does not correspond to dust devil events.

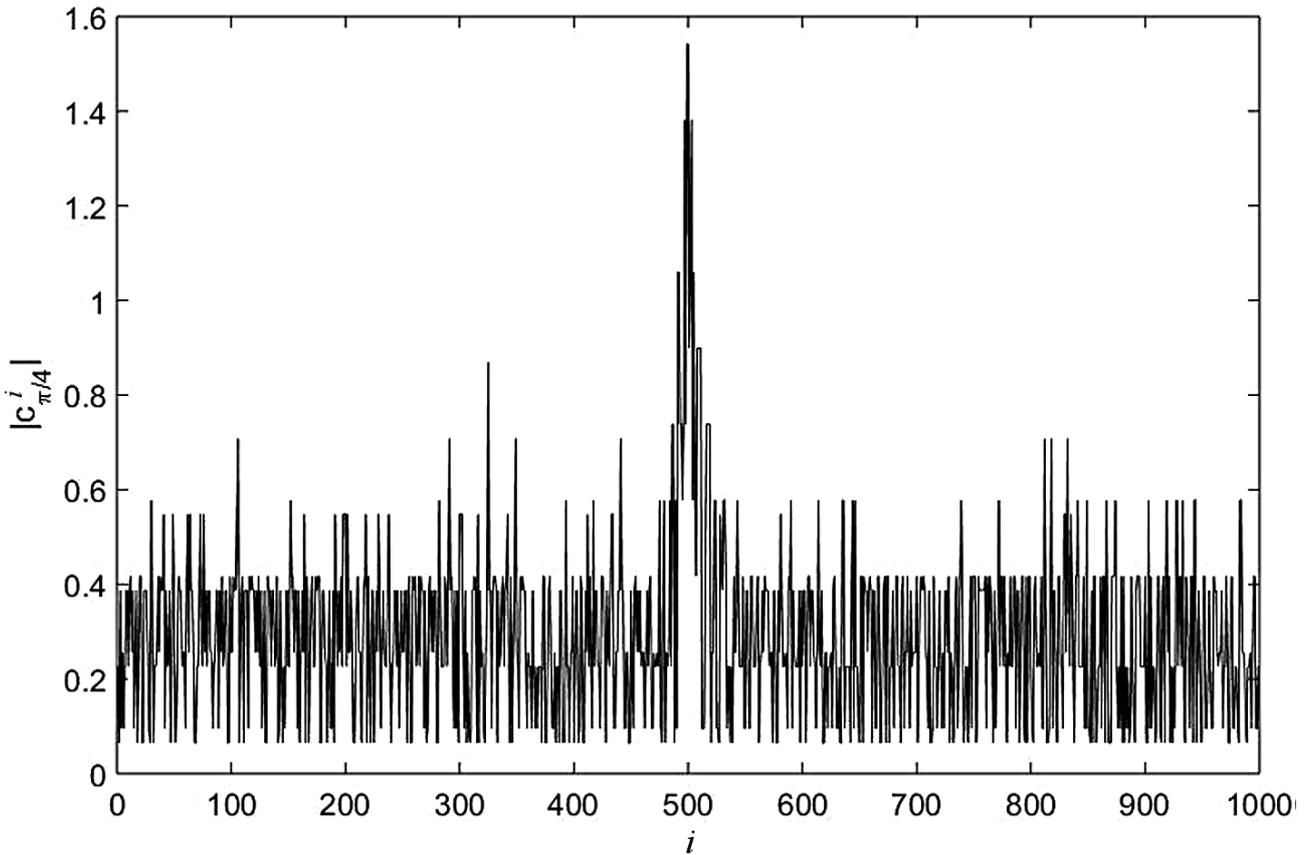

Fig. 18 *Projection of the pressure data tomogram for $\theta = \frac{\pi}{4}$.*

As explained in the previous section, we make use of the spectrum average $\left(\frac{1}{N}\sum_{j=1}^{N}|c_\theta^i|\right)$ to identify the dust devil components of the signal. The clearer the dust devil event is, the bigger is the value of the corresponding coefficients $c_\theta^i$, hence, we classify the dust devil events depending of the relative size with respect to the spectrum average. As can be seen in Fig. 17, the event can be clearly detected for each value of $\theta$, therefore, we opted to simply consider $\theta = \frac{\pi}{4}$.

The detected events have been divided into classes from the least to most probable, depending on the relative magnitude of $c_\theta^i$. The classification used is shown in Table 2.

| Classes | Main Characteristics |
|---------|----------------------|
| E | Transform coefficient > 6 · spectrum average |
| D | Transform coefficient > 6,5 · spectrum average |
| C | Transform coefficient > 7 · spectrum average |
| B | Transform coefficient > 7.5 · spectrum average |
| A | Transform coefficient > 8 · spectrum average |

Table 2 *Classes and main characteristics regarding the division of the events identified by Tomography technique.*

## 2.6.3 Validation of Tomography technique

Unlike the phase piker, the tomography is a completely new method for the dust devils detection. For this reason, we have to validate the technique evaluating its reliability and effectiveness.

In order to do so, we analyzed 5 days of data acquired in 2013 campaign: July 17[th], 21[th], 24[th] and August 10[th] and 11[th]. The days have been chosen randomly along the entire campaign, varying between the ones of high and low dust activity.

Overall, tomography has identified 47 dust devils candidates: 12 class E, 21 class D, 3 class C, 7 class B and 4 class A events. We have crosschecked all the selections, by analyzing the entire set of measured meteorological parameters, in order to confirm if they are truly identifiable or not as real dust devils. The complete list of the detected events and the results of the cross check are given in Table 3, they are labelled with the letter T and a progressively increasing number.

| ID | Date | ti(h) | Tomograms Class | Full parameters Crosscheck |
|---|---|---|---|---|
| T1 | 17_07_2013 | 4.7944445 | E | No |
| T2 | 17_07_2013 | 4.8316667 | E | No |
| T3 | 17_07_2013 | 4.8650000 | E | No |
| T4 | 17_07_2013 | 5.0838889 | E | No |
| T5 | 17_07_2013 | 5.1352778 | D | No |
| T6 | 17_07_2013 | 5.4038889 | D | No |
| T7 | 17_07_2013 | 5.4352778 | D | No |
| T8 | 17_07_2013 | 7.0977778 | E | No |
| T9 | 17_07_2013 | 9.6669445 | E | Not dusty Vortex |
| T10 | 17_07_2013 | 14.0333333 | E | No |
| T11 | 17_07_2013 | 15.5966666 | D | Yes |
| T12 | 17_07_2013 | 15.8788889 | D | Not dusty Vortex |
| T13 | 17_07_2013 | 16.0333333 | D | No |
| T14 | 17_07_2013 | 18.9736111 | A | Yes |
| T15 | 17_07_2013 | 19.5730556 | D | No |

| T16 | 17_07_2013 | 20.0319444 | D | No |
| T17 | 17_07_2013 | 20.5786111 | B | No |
| T18 | 17_07_2013 | 20.6494444 | B | Yes |
| T19 | 17_07_2013 | 21.5641667 | A | Yes |
| T20 | 21_07_2013 | 8.4327778 | D | No |
| T21 | 21_07_2013 | 12.8000000 | A | Yes |
| T22 | 21_07_2013 | 13.3691667 | B | Yes |
| T23 | 21_07_2013 | 14.9480556 | D | No |
| T24 | 21_07_2013 | 15.7708333 | D | No |
| T25 | 21_07_2013 | 18.6238889 | D | No |
| T26 | 24_07_2013 | 8.7080555 | D | No |
| T27 | 24_07_2013 | 9.0041667 | D | Not dusty Vortex |
| T28 | 24_07_2013 | 9.6888889 | B | Yes |
| T29 | 24_07_2013 | 10.9250000 | B | Yes |
| T30 | 24_07_2013 | 11.0016667 | D | Possible |
| T31 | 24_07_2013 | 11.5891666 | D | Yes |
| T32 | 24_07_2013 | 16.8805556 | B | Yes |
| T33 | 10_08_2013 | 0.0497222 | E | No |
| T34 | 10_08_2013 | 4.8011111 | E | No |
| T35 | 10_08_2013 | 11.9986111 | D | No |
| T36 | 10_08_2013 | 12.2894444 | D | No |
| T37 | 10_08_2013 | 13.9894444 | D | Not dusty Vortex |
| T38 | 10_08_2013 | 16.7441667 | A | Yes |
| T39 | 10_08_2013 | 17.9986111 | D | No |
| T40 | 11_08_2013 | 1.9922222 | E | No |
| T41 | 11_08_2013 | 3.9872222 | E | No |
| T42 | 11_08_2013 | 5.9919444 | E | No |
| T43 | 11_08_2013 | 9.9916667 | D | Not dusty Vortex |
| T44 | 11_08_2013 | 11.4319444 | B | Yes |
| T45 | 11_08_2013 | 13.0125000 | C | Yes |
| T46 | 11_08_2013 | 13.9905556 | C | No |
| T47 | 11_08_2013 | 19.9902778 | C | No |

**Table 3** *List of the events identified with the tomography technique. The date, the initial instant, the tomogram class and the result of the manual crosscheck are reported. The results of the crosscheck are simply given in term of yes and no, except one case for which the meteorological data is not conclusive and the event is catalogued as possible. We also indicate the events recognizable as convective not dust loaded vortices.*

As it is possible to see, all class A events are recognizable as dust devils, while all the class E ones are not. Regarding the class B events, just one does not seem to be a dust devil, while, there are two other "not dust devil" events in the class C. Three of the class D events are recognizable as dust devils, 4 do not seem to be a dusty convective vortex, while the remaining ones appear to be wind gusts or related to dust storms. Some of the events selected by the tomography are identifiable as vortex not strong enough to lift soil grains; however, these cases fall in lower classes D and E. The percentage of true dust devils in every class rapidly grows towards the A, how it is possible to notice in Fig. 19, where we showed the number of true dust devils in every class normalized by the number of events in the class.

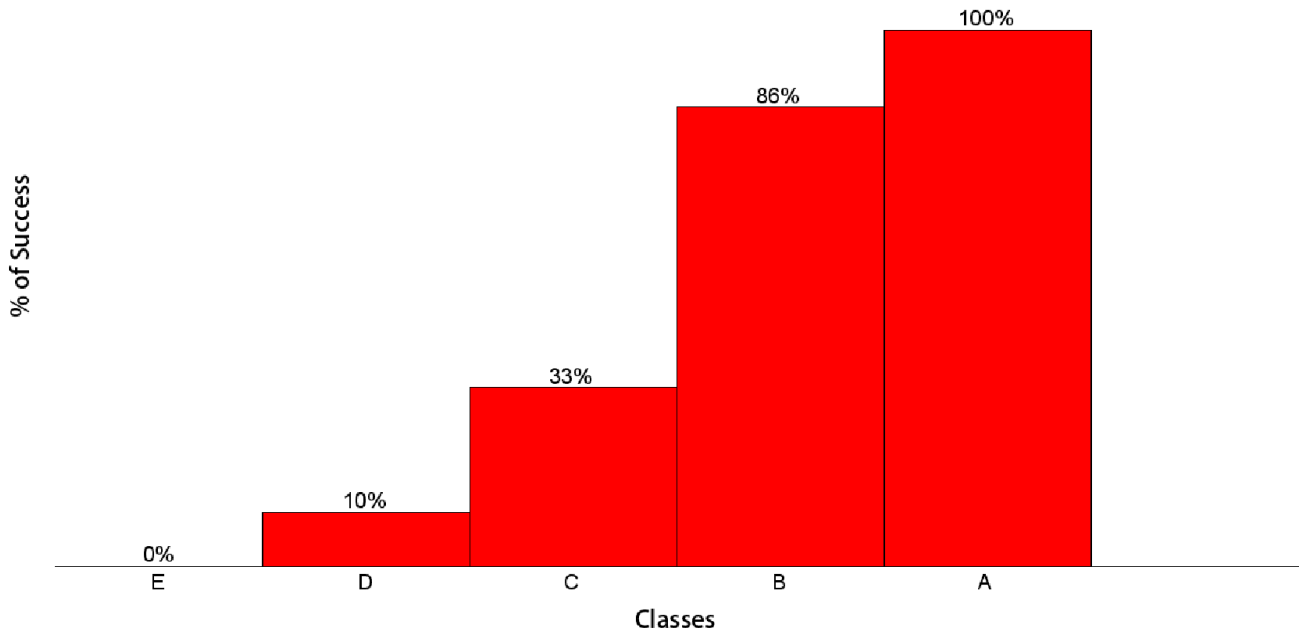

Fig. 19 *The percentage of true dust devils, recognized by the full parameters crosscheck, for every class of the tomographic analysis.*

The affability of the tomography technique and the reliability of the chosen classification is proven by the fact that 100% of the class A events are indeed dust devils, and overall it is highly probable that the events belonging to the higher class A and B are not false positive detections.

### 2.6.3.1 Comparison of the different methods

In order to test the effectiveness of the tomographic technique we decided to perform on the same 5 data days also the phase picker detection described in 2.6.1.2, comparing the two results.
Both algorithms use the pressure as main probe parameter, however, the tomogram works on a single parameter, while phase picker couples also the analysis of the wind direction and E-field.
Table 4 reports the class A, B and C identified by the time domain technique, the events have been labelled with the letter P and a progressively increasing number.

| ID | Date | $t_i$ (h) | Δt(s) | S (counts/s) | $ΔW_{dir}$ (°) | ΔP (mbar) | Classes |
|---|---|---|---|---|---|---|---|
| P1 | 17_07_2013 | 12.115 | 6.1 | 0.2 | 45 | 0.4 | A |

| P2 | 17_07_2013 | 13.0614 | 9.0 | 0.0 | 31 | 0.3 | B |
| P3 | 17_07_2013 | 13.7617 | 18.7 | 0.0 | 175 | 0.3 | A |
| P4 | 17_07_2013 | 15.595 | 23.0 | 1.4 | 31 | 0.5 | A |
| P5 | 17_07_2013 | 17.1817 | 4.0 | 2.1 | 34 | 0.3 | C |
| P6 | 17_07_2013 | 17.8178 | 4.0 | 3.1 | 41 | 0.3 | C |
| P7 | 17_07_2013 | 18.9661 | 45.0 | 3.6 | 94 | 0.8 | A |
| P8 | 17_07_2013 | 20.5736 | 37.1 | 0.9 | 52 | 0.7 | C |
| P9 | 17_07_2013 | 20.6439 | 36.0 | 12.8 | 42 | 0.7 | A |
| P10 | 17_07_2013 | 21.3003 | 38.9 | 152.5 | 60 | 0.6 | C |
| P11 | 17_07_2013 | 21.5617 | 34.9 | 35.9 | 164 | 1.3 | A |
| P12 | 17_07_2013 | 22.0197 | 12.2 | 5.3 | 36 | 0.3 | C |
| P13 | 17_07_2013 | 22.0675 | 29.9 | 8.0 | 51 | 0.5 | B |
| P14 | 21_07_2013 | 12.7928 | 41.8 | 0.0 | 179 | 0.8 | A |
| P15 | 21_07_2013 | 13.3683 | 13.0 | 5.9 | 80 | 0.5 | A |
| P16 | 21_07_2013 | 18.5372 | 13.0 | 0.4 | 37 | 0.4 | C |
| P17 | 24_07_2013 | 9.68722 | 12.0 | 0.0 | 53 | 0.4 | A |
| P18 | 24_07_2013 | 10.9225 | 24.8 | 0.1 | 92 | 0.7 | A |
| P19 | 24_07_2013 | 11.5881 | 20.9 | 0.0 | 94 | 0.5 | A |
| P20 | 24_07_2013 | 16.8775 | 51.1 | 0.0 | 124 | 0.9 | A |
| P21 | 10_08_2013 | 16.7394 | 41.0 | 24.4 | 116 | 1.0 | A |
| P22 | 11_08_2013 | 11.4311 | 13.0 | 2.9 | 120 | 0.6 | A |
| P23 | 11_08_2013 | 13.0119 | 40.3 | 0.0 | 174 | 0.5 | A |

**Table 4** *List of the events identified by the time domain research technique. For each events we report the date, the initial instant ($t_i$), the time duration ($\Delta t$), the mean values inside the event of Sensit counts, the maximum wind speed direction change ($\Delta W_{dir}$), the pressure drop ($\Delta P$) magnitude and the class.*

We focused on the best candidates detected by the phase picker research (class A and B), crossing the results with the ones obtained by tomography. As it can be noted in Table 5, the events detected are in good agreement for all days. There are only 4 events not detected by tomography: two class B and two class A, and they all happened during July, 17[th]. Overall, there is a match of 12 events over 16. Moreover, there is an event detected only by tomography during July 24[th], recognized by the full parameters crosscheck as a possible dust devil.

The first step of the time domain analysis performed on the pressure parameter only has identified a total of 6611 events. Such large number of detections shows that a simple pressure phase picker analysis is not sufficient to strictly constrain the identification of dust devils, especially when the noise level is relevant. As we said, to reduce the number of non-significant detected events, we had to couple the analysis of other two parameters: the wind direction and the E-field.

On the other hand, the tomography is specifically calibrated to search for the dust devil signature by analyzing one single parameter. It has reached a good efficiency in the detection, providing a clear classification of the events, allowing to individuate the best candidates. In addition, the tomographic analysis can be performed directly on the raw data, despite the presence of high noise level because the tomogram automatically filters the signal.

| Time domain research | | | | | Tomography research | | |
|---|---|---|---|---|---|---|---|
| ID | Date | $t_i$ (h) | Class | Match | ID | $t_i$ (h) | Class |
| P2 | 17_07_2013 | 13.0614 | B | No | | | |
| P13 | 17_07_2013 | 22.0675 | B | No | | | |
| P1 | 17_07_2013 | 12.115 | A | No | | | |
| P3 | 17_07_2013 | 13.7617 | A | No | | | |
| P4 | 17_07_2013 | 15.595 | A | Yes | T11 | 15.5966 | D |
| P7 | 17_07_2013 | 18.9661 | A | Yes | T14 | 18.9736 | A |
| P9 | 17_07_2013 | 20.6439 | A | Yes | T18 | 20.6494 | B |
| P11 | 17_07_2013 | 21.5617 | A | Yes | T19 | 21.5641 | A |
| P15 | 21_07_2013 | 13.3683 | A | Yes | T22 | 13.3691 | B |
| P17 | 24_07_2013 | 9.68722 | A | Yes | T28 | 9.6888 | B |
| P18 | 24_07_2013 | 10.9225 | A | Yes | T29 | 10.9250 | B |
| P19 | 24_07_2013 | 11.5881 | A | Yes | T31 | 11.5891 | D |
| P20 | 24_07_2013 | 16.8775 | A | Yes | T32 | 16.8805 | B |
| P22 | 11_08_2013 | 11.4311 | A | Yes | T44 | 11.4319 | B |
| P23 | 11_08_2013 | 13.0119 | A | Yes | T45 | 13.0125 | C |
| P14 | 21_07_2013 | 12.7928 | A | Yes | T21 | 12.8000 | A |
| P21 | 10_08_2013 | 16.7394 | A | Yes | T38 | 16.7441 | A |
| | 24_07_2013 | | | No | T30 | 11.0016 | D |

Table 5 *The match between the events identified by time domain research technique and by the tomography technique.*

On the base of these results, we can conclude that the tomography appears to be a very promising technique for dust devil detection, in particular in those cases when it is possible to count only on a single acquired parameter. It can be easily tuned for the Martian environment, providing good reliability and detection efficiency, even in case of strong noise.

## 2.7 Dust Devil impact parameter

We described how to individuate the dust devils signals recorded by a fixed meteorological station and what is the trend that expected rotational wind speed $v_r$ eq.(1) and pressure drop $\Delta P$ eq.(2) trend.
Now we want to briefly describe what are the difficulties and the limitation arising by studying the vortex signals using the measurements acquired by a fixed meteorological station, as the one we used in the Sahara or the ExoMars 2020 surface platform.
In both pressure and wind rotation equations, we can interchange between the variable space *d* (the distance between the vortex center and the station) and the variable time *t*. To do so, we have to make explicit the relation between these two variables. For simplicity, let us define $t_o$, the instant of minimum approach of the vortex to the station, as zero and let us call this minimum distance $d_o$. Therefore, we have simply:

$$d^2(t) = (s\,t)^2 + d_o^2 \qquad (48)$$

where $s$ is the translational speed of the vortex and we can call $d_o$ the impact parameter of the encounter vortex-station.

In general, the recorded signature of the dust devil passage in a specific meteorological quantity $q$ can be factorized as the product of quantity $q_o$, intrinsically depending on the dust devil characteristics as its sizes, for a function $f(d)$, that expresses how the signal decreases with distance:

$$q(d) = q_o\,f(d) + B \qquad (49)$$

where $B$ represent the background value of the signal.

If we are considering the pressure, $q_o$ will simply be the pressure drop at the vortex core ($\Delta P_o$), while if we consider the rotational wind speed $q_o$ will be $V_r$, the rotation at the vortex wall. The quantity $q_o$ represents the maximum variation of the parameter $q$ that the vortex could induce, but it is directly measurable only when the impact parameter $d_o$ is equal to zero. When $d_o$ is greater than zero, the variation observed $\Delta q = |B - q|$, will reach a maximum value smaller than $q_o$. The most easily observable is actually this maximum variation of the signal from the background. We can call $\Delta q_{Max}$ this quantity and rewrite the eq.(49) as:

$$q(d) = q_o\,f(d_o)f(d - d_o) + B = \Delta q_{Max}\,f(d - d_o) \qquad (50)$$

We defined $q_o$ as intrinsically dependent on the dust devil characteristics, that variates with the environmental condition of the dust devil formation, like temperature, wind regime, pressure, humidity, soils composition etc.. However, we expect that the different dependences of $q_o$ could be mainly summarized as a function of vortex radius $R$: the larger is the value of $R$, the larger will be the intrinsic induced variation $q_o$.

Hence, we expect:

$$\Delta q_{Max}(R, d_o) = q_o(R)\,f(d_o) \qquad (51)$$

As we said, $\Delta q_{Max}$ is directly measurable, however, we cannot resolve the degeneration radius-distance: i.e. a tiny dust devils passing near the station (low $R$ and $d_o$) or a huge one passing farther (larger $R$ and $d_o$) could give very similar signals $q(t)$ and then measured variation $\Delta q_{Max}$.

Varying the considered parameter $q$, we could have different forms for the functions $q_o(R)$ and $f(d)$. Indeed, we have already seen that rotatory speed and pressure signals follow two different decreasing trends with distance, being the first $f(d)$ linear, while the second is Lorentzian ( eq.s (1) and (2) ). Overall, also the other observed parameters, as f.i. the induced E-field or the concentration of lifted dust, can in theory follow different trends in time and intrinsic dependence on the vortex size.

Depending on the magnitude of $d_o$ we can divide the events in two categories:
- dust devils passing outside the station, when $d_o \geq R$ ($R$ is the vortex radius);
- and dust devils crossing over the station, when $d_o < R$.

Despite the difficulties in the evaluation of the $d_o$ from the measurement of the acquired $q(t)$, we can still recognize if an event passes over the meteorological station by some peculiar features of the signal. Fig. 20 schematically shows the passage of two dust devils: the first one does not intercept the meteorological station (Fig. 20a), the second crosses the instruments (Fig. 20b). In the case depicted in Fig. 20a we measure a single maximum in the variation of all the monitored parameters in the instant $t_o$, corresponding to the maximum approach of the vortex to the station $d = d_o$. Instead, when the impact parameter is smaller than the radius of the dust devil, the vortex wall passes two times over the station (time instants $t_1$ and $t_2$ of Fig. 20b). In these instants, the instruments measure the maximum values of horizontal and vertical wind speed, lifted sand and dust concentration and hence E-field variation. Between $t_1$ and $t_2$ the dust-free inner part of the vortex crosses the station resulting in a decreasing of the magnitude of the wind speed and grains concentration, that reach the local minimum in the instant $t_0$. For the pressure signal, taking into account that the magnitude of the pressure drop ΔP is maximum in the center of the vortex and not on the wall, we measure a single maximum of ΔP in the instant $t_o$ in both cases of Fig. 20a and Fig. 20b. We are then able to distinguish between the case of Fig. 20a and Fig. 20b by looking for a single or a double peak trend in the measured data of wind speed, grains concentration and E-field. The double peak feature is usually particularly marked in the horizontal wind speed data, becoming clearer also in the other affected parameters when the impact parameter approaches zero.

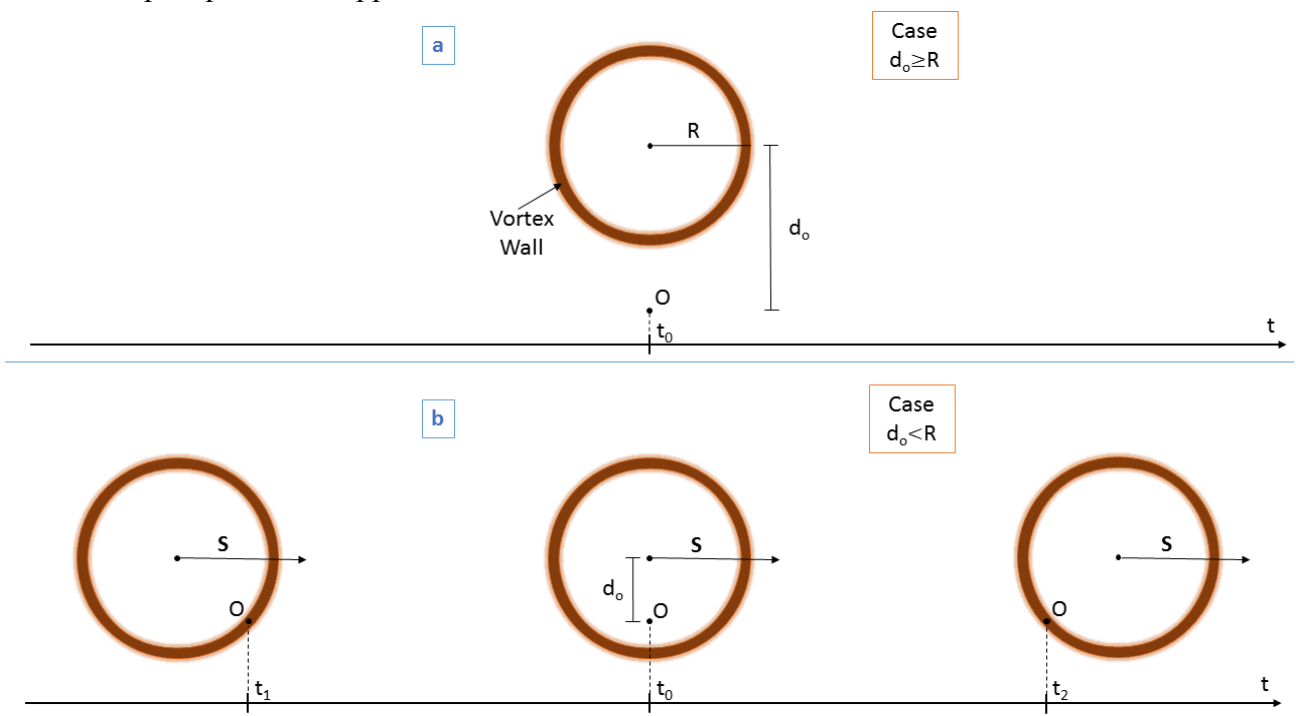

**Fig. 20 (a)** *Case $d_o \geq R$: dust devil of radius R that passes out of the meteorological station placed at the point O. $t_0$ represents the instant of maximum approach $d_o$ to the station. **(b)** Case $d_o < R$: dust devil that intercepts the meteorological station seen in three different instants. The vector S represents the translational velocity of the vortex. $t_1$ represents the instant when the vortex wall passes over the station the first time, $t_0$ represents the instant when the vortex center reaches its maximum approach to the station and $t_2$ represents the instant when the dust devil moves away crossing for the second time the station with its wall.*

Fig. 21 shows one of the dust devils that crossed the instruments. In contrast to the events depicted in Fig. 14 that passed outside the station, it is visible how the measured wind speed (seventh plot) shows a clear two peaks trend.

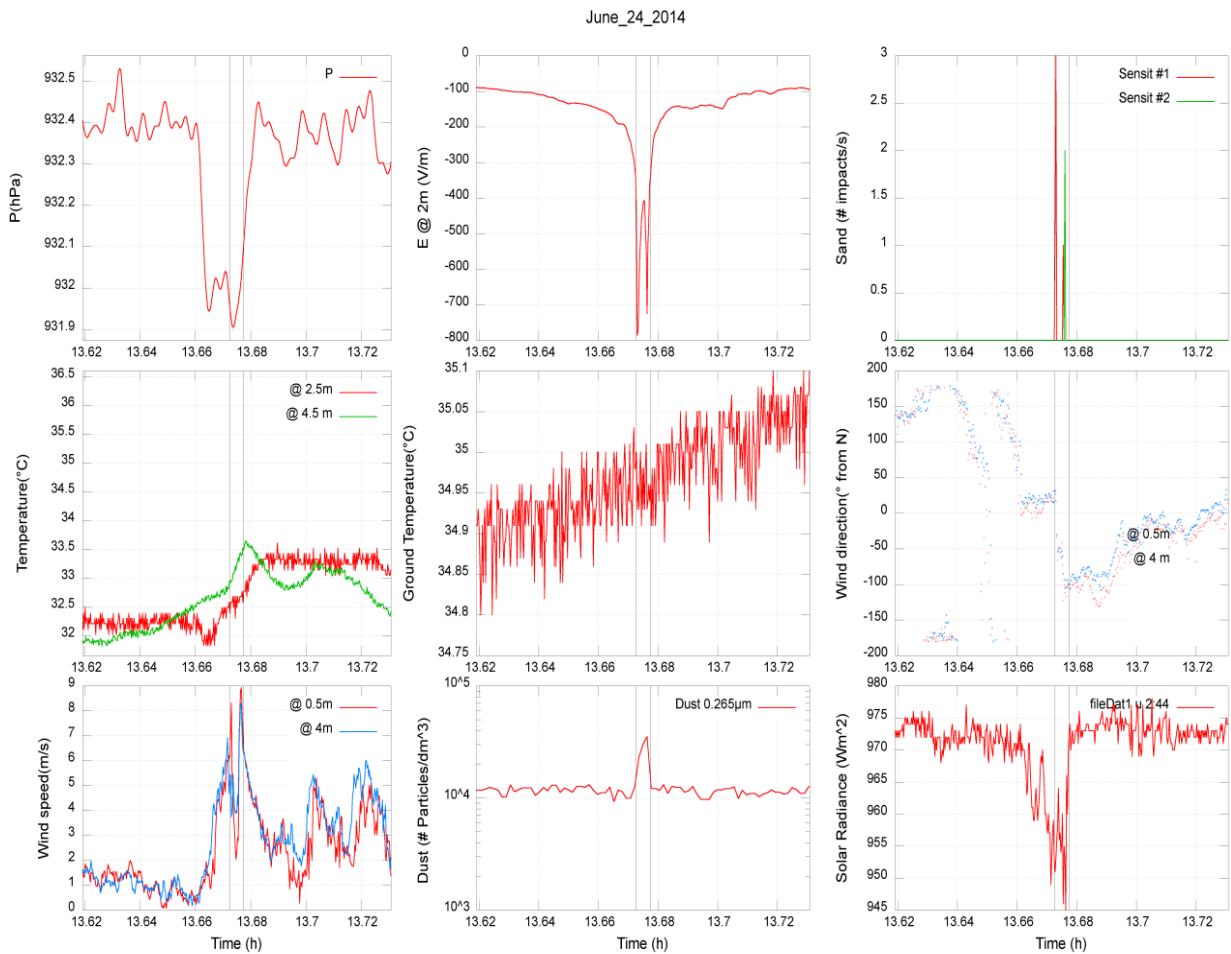

**Fig. 21** *A dust devil encounter of an event that passed over the meteorological station. The double peak trend in the horizontal wind speed (seventh panel) is clearly recognizable.*

# Chapter 3
# Proprieties of the dust induced E-field

Dust storms and dust devils are the main lifting phenomena able to mobilize and entrain into the atmosphere the dust on Mars. In order to understand the martian climate it is fundamental to study the dust cycle starting from the lifting phenomena that drive it. This is one of the main purposes of the present and future martian missions, as NASA InSight 2018 and Roscosmos/ESA ExoMars 2020. A key role in this process is represented by the acquisition and analysis of terrestrial analogous data, in order to understand the physics of the lifting processes and allow a proper preparation and planning of the Martian data surveys.

In this optic, we performed various field campaigns in the Sahara desert, deploying a set of meteorological instruments similar to the ones that will be on board of ExoMars 2020. In the last Chapter we described the detection techniques we developed and used to detect the dusty events. In this Chapter we will present the data analysis regarding the dust storms and devils activity observed during the first two Sahara campaigns. We will focus in particular on the electric proprieties in dusty environment, that currently represent one of the most unknown factor of the dust lifting phenomena. The generation and the behavior of the E-field and the contribution of the electrical forces to the lifting process are indeed still poorly understood, and our field data can significantly help to shed light on this subject. We will also draw a parallel to the martian case to see if our results are extendible or not in that environment. The results here presented have been published by our team in Esposito et al. 2016 and Franzese et al. 2018.

## 3.1 Dust Storms

In this section we will focus our attention on the dust storm activity observed during the 2013 and 2014 campaigns. Measurements were performed during the peak of the dust storm season in Morocco. By using the procedure described in par 2.5, we detected 83 events with a saltation level over 1%: 13 during 2013 campaign and 70 during 2014.

To the best of our knowledge, this is the first data set comprehensive of simultaneous measurements of the dust concentration and related E-field variation. We studied how the behavior of the dust storms is connected to the environmental conditions, focusing in particular on the dependence of the induced E-field on the humidity level. Moreover, we investigate if this E-field generated by the grains electrification can in turn affect their lifting.

### 3.1.1 Bimodality of the observed dust storm distribution

Fig. 22 shows the daily saltation activity observed during the two campaigns: a peak around the 4 p.m. is clearly visible, however we observed a sustained activity also in the night time hours. We noticed several differences in the behavior of the diurnal and nocturnal events, hence we decided to study these two classes of events separately.

We classified as Class1 the day time events, i.e. the ones that occur between 6 a.m. and 6 p.m., and as Class2 events all the others. The mean between the start and end instant of the saltation interval has been considered as the time of the events.

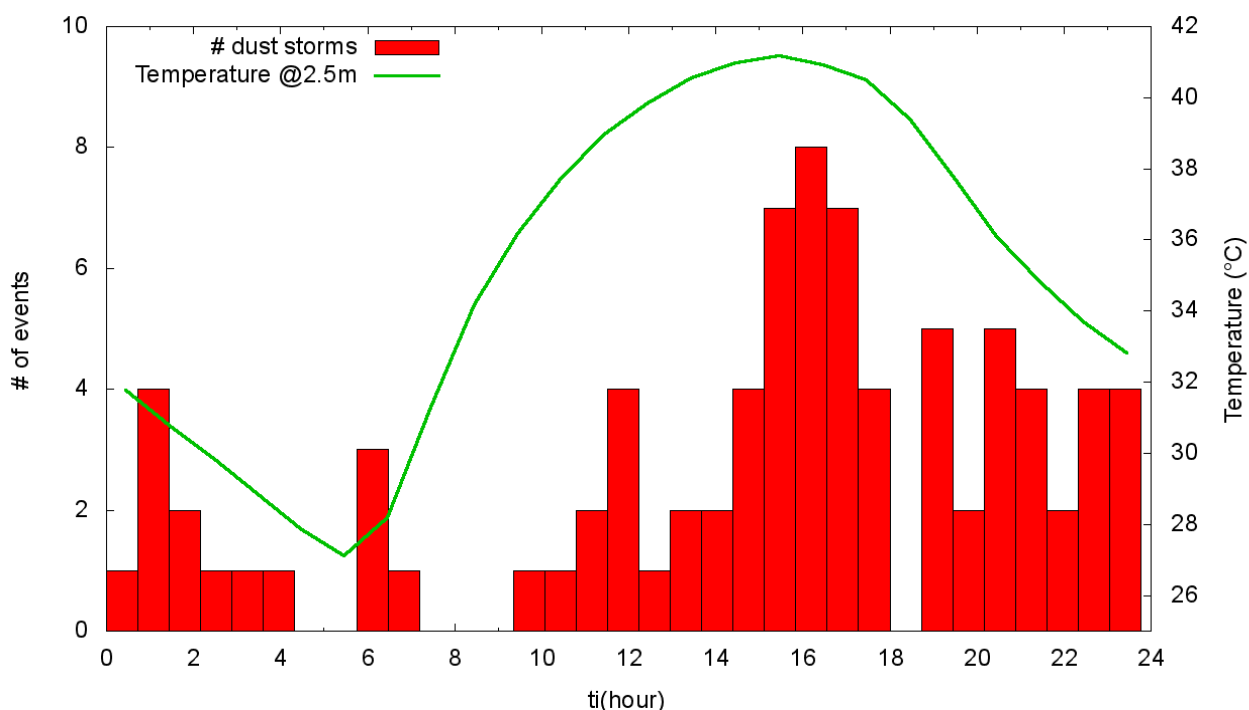

**Fig. 22 Daily distribution of the saltation events observed along the 2013 and 2014 campaign. On the x axis we plotted the mean time of the events. The activity peaks around 4 p.m., however we also observed a moderate night time activity.**

As shown in Fig. 23, Class1 events are mainly characterized by winds blowing from the South and South-Western sectors, while, the Class2 events are mostly associated with North-East winds.

The bimodality of the events distribution is evident also in the air relative humidity, as shown in Fig. 24: Class1 events show a smaller interval of RH variation, with a peak around 8.5%, while, Class2 events have a broader distribution with an larger median value ~23.5%.

Examples of the dusty events of the two classes are depicted in Fig. 25 (Class1) and Fig. 26 (Class2). For both storms we plotted in order from left to right, top-down: the trend of pressure, E-field, saltation and airborne dust activity, temperature, vertical wind speed, horizontal wind direction and speed, solar radiation, air and soil humidity.

The average wind speed, E-field value, saltation level and dust concentration are comparably between the two classes, however, Class2 signals are much shorter and continuous with a duration that usually does not overcome the couple of hours. Moreover, Class2 events are generally characterized by a sudden initial decrease of atmospheric temperature and increase of the relative humidity, that, after the initial step, remain quite stable for the rest of the storm.

The behavior of these events seems to indicate a particular kind of storm known as haboobs. These phenomena are typical of southern borders of Sahara (Williams et al., 2009,), but similar events have been also reported in Morocco, during the SAMUM 2006 campaign (see Emmel et al., 2010; Knippertz et al., 2007; 2009). Haboobs formation arises from to the evaporation of precipitations in dry and desert areas. The resulting cooled air mass gives rise to density currents with strong winds at their leading edge, able to mobilize a large amount of dust. This phenomenon is favored by the presence of mountains, as they help the convection and the acceleration of cooled air along their edges (Knippertz et al., 2007, 2009). The behavior of several Class2 storms is indeed interpretable as connected to the arrive of a moist cold front from the north-est (see Fig. 26),where, around 40 km away from our site, the "Hammada du Guir" is located, the upland of ~1100 m of height that set the border between Morocco and Algeria.

Taking in mind the differences between the two classes of events we analyzed their electric proprieties.

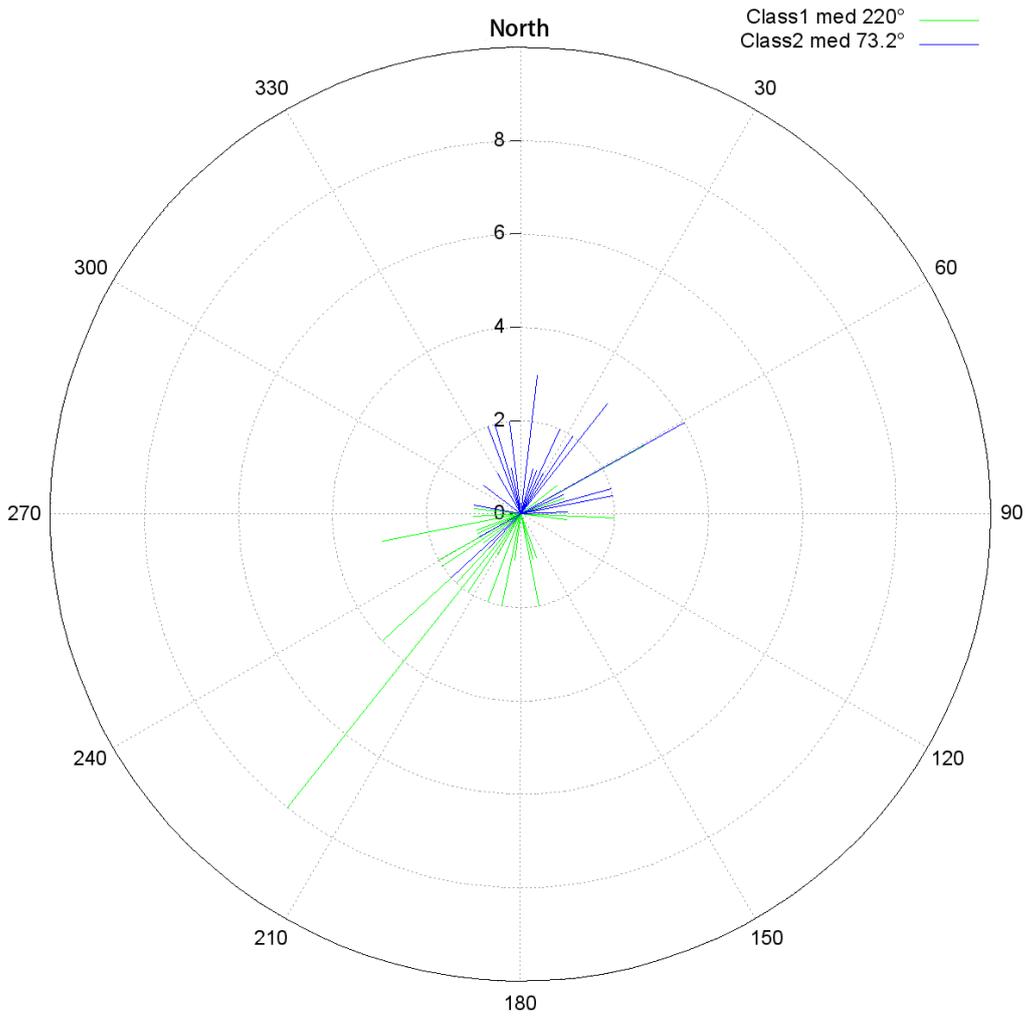

**Fig. 23** Histogram of wind direction that characterizes the dust storm events belonging to Class1 (in green) and Class2 (in blue)

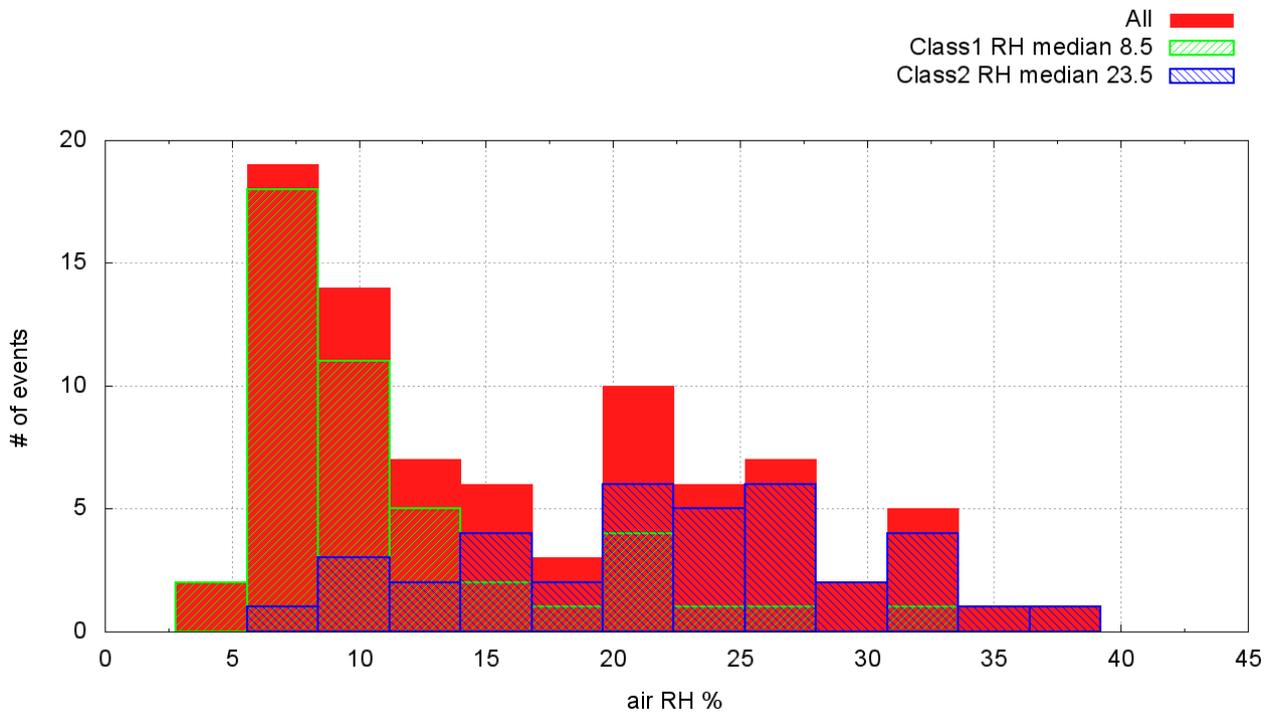

**Fig. 24** Histogram of the relative humidity RH of the events. In red the whole dataset, while in green and blue there are the Class1 and Class2 events.

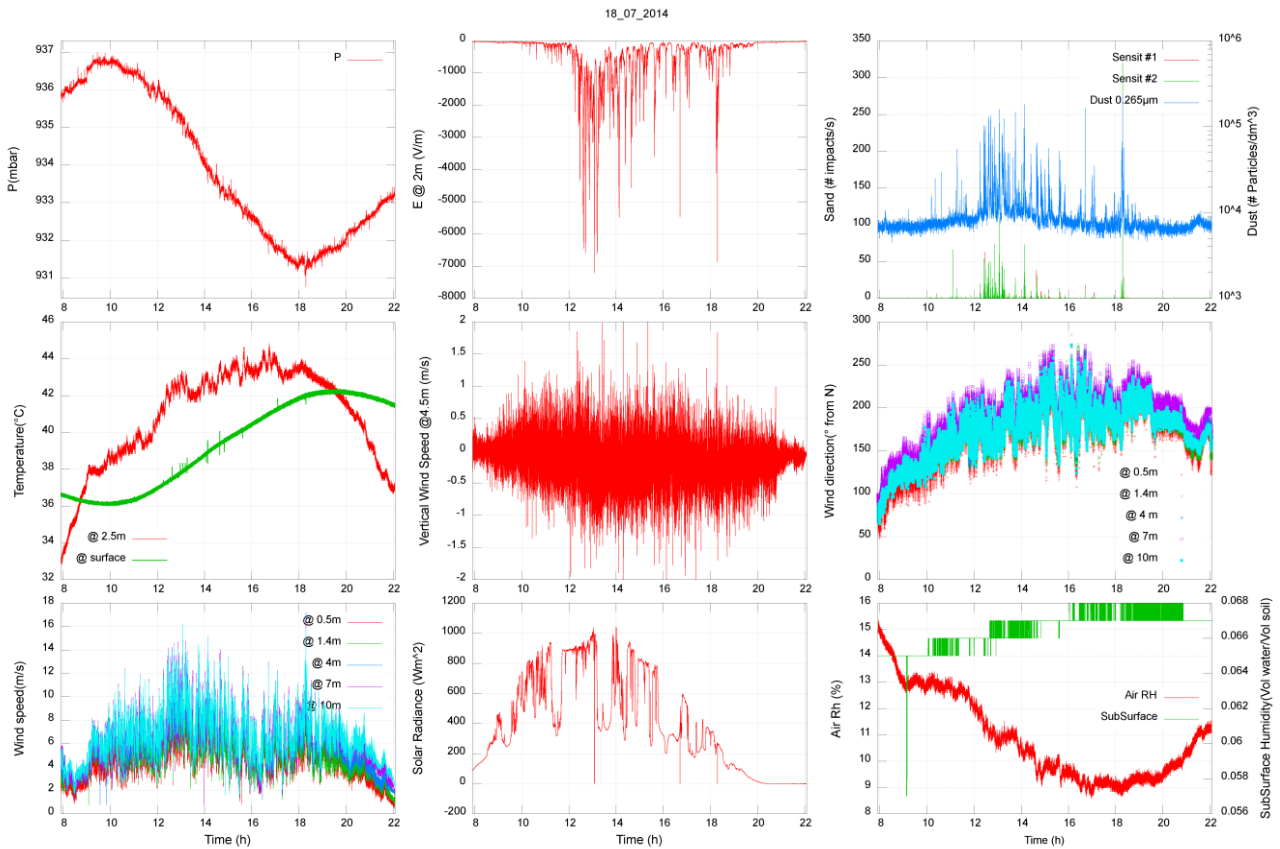

**Fig. 25** Dust storm event belonging to the Class 1. Class 1 usually comes from the south west direction (see sixth plot) and they can last for several hours, up to the whole day, as the one here depicted.

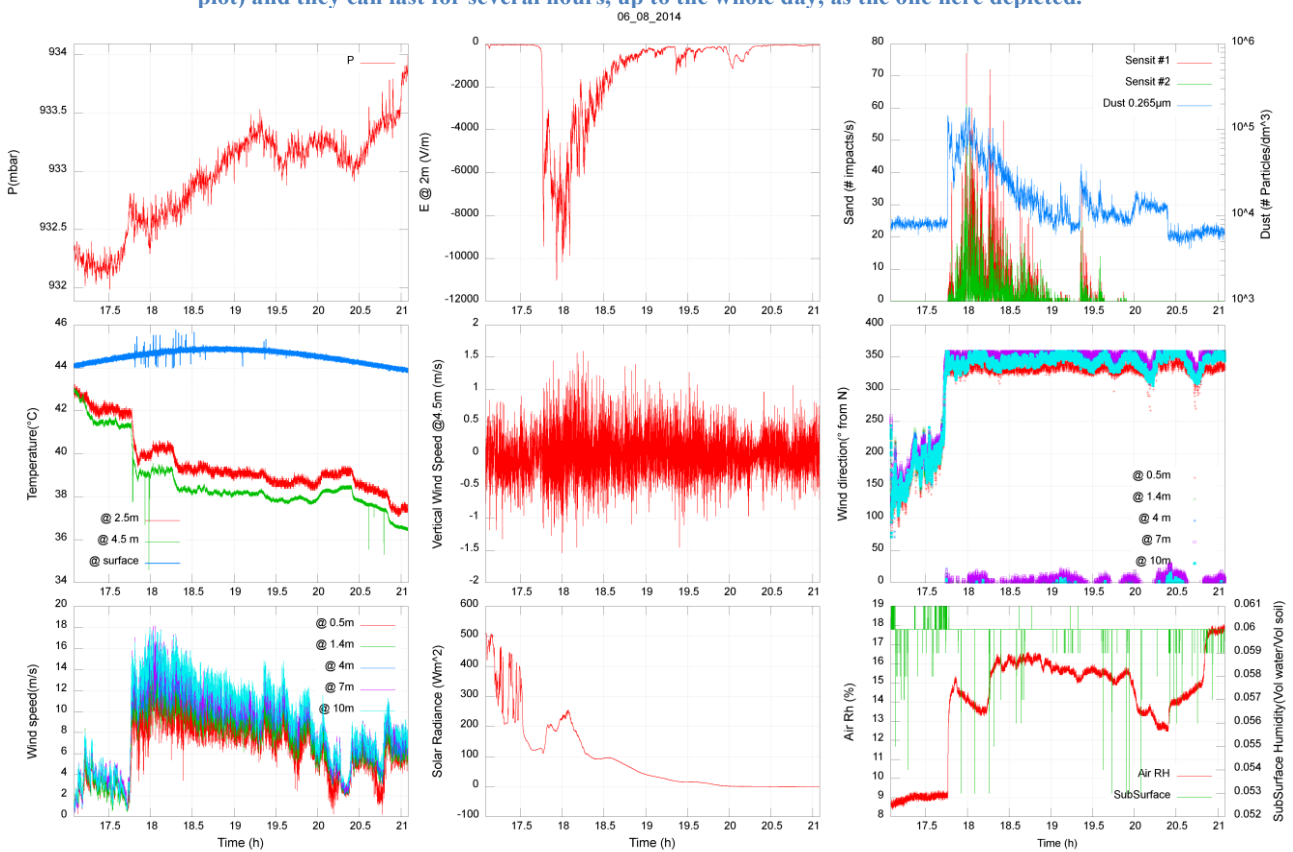

**Fig. 26** Dust storm event belonging to Class 2. The storm is related to the arrival of a moist cold air front from the north. Indeed, it is evident the instantaneous drop of temperature (fourth plot), change of wind direction (sixth plot) and RH increase (ninth plot

## 3.1.2 E-field proprieties

We clearly observed a sudden and strict connection between the increase and behavior of saltation and dust lifting activity and the induced variation of the E-field. The second and third panels of Fig. 25 and Fig. 26 are an example of the trend generally observed during the campaigns, showing how the E-field signal closely mirrors the sand and dust activity. During the dusty events, we observed a generally downward pointing field, directed the same as the fair weather one, ranging from few hundreds of V/m up to ~-18 KV/m (see Fig. 27).

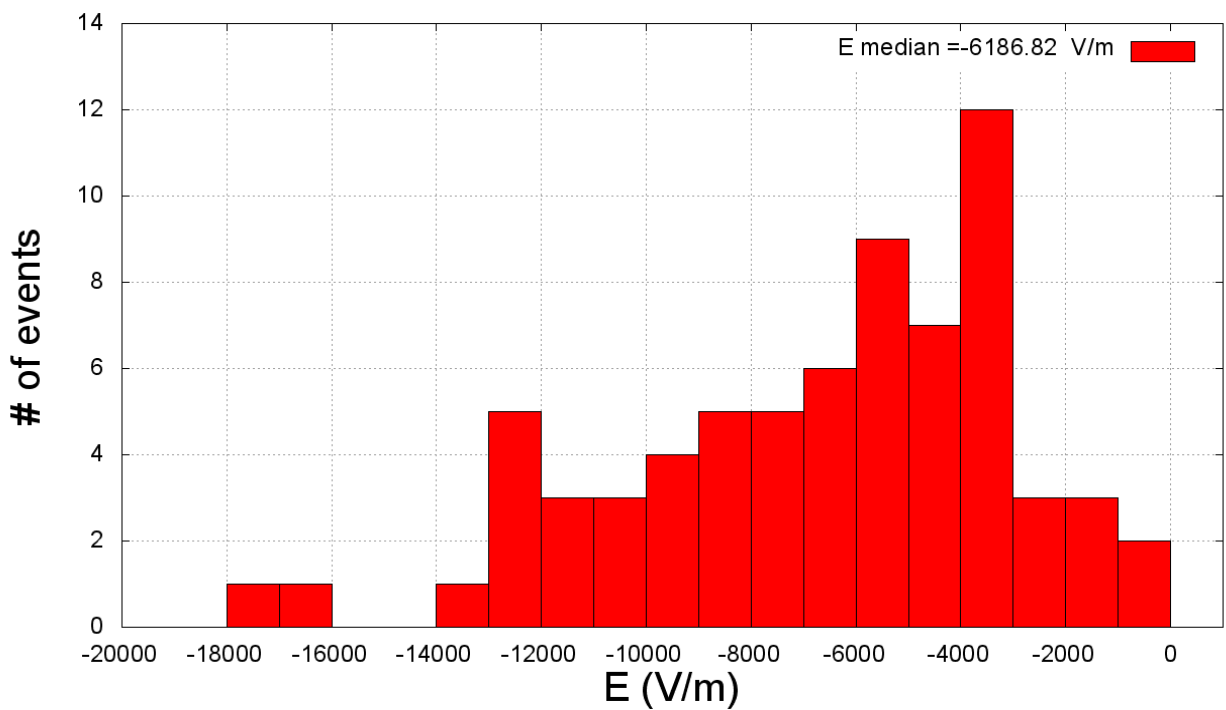

Fig. 27 Histogram of the downward E-field excursions measured during the dust storm events.

We have measured occasional and short inversions in the field sign, especially in the Class2 events, that could be connected to raindrop. As we explained in par. 1.6 the general behavior expected in a dust cloud is that the contact electrification (during the saltation, sandblasting, etc.) leaves the smaller grain negatively charged and the larger one with an opposite sign. The different weight of these grains causes the negative charges to be driven upper in the cloud, leading to the charge separation and the generation of the E-field (upward directed following this description).
However, our observations on the E-field orientation (downward) are in contrast with the diffuse description of the collisional charge transfer as a predominantly size-dependent process, indicating how the composition of the dust cloud could represent a key factor; we will discuss more deeply this topic in par 3.2.4.1.

We analyzed the principal factors that influence the behavior of the E-field, studying also its close connections and feedbacks on the lifted dust concentration.

### *3.1.2.1 Relation with the dust concentration and atmospheric relative humidity*

We observed how the relative humidity (RH) has a significant impact on the E-field intensity. For this reason, we subdivided the dusty events into four *humidity classes*: RH < 10%, 10%-20%, 20%-30%, and > 30%. We considered the whole set of data acquired at 1 Hz rate and performed a binning of the E-field data with 200 V/m steps, comparing this quantity to the total numeric grains concentration (summed over the whole monitored range 0.265-34 μm). The results of this analysis are shown in Fig. 28, separating in two different plots the one regarding the 2013 and 2014 campaign. Indeed, as we mentioned in Chapter 2, the two sites are characterized by a different amount of soil and air humidity that results in a different erodibility level. The moisture content was obtained by averaging the soil humidity data acquired during the campaigns and we obtained a value of 0.49 ± 0.03 $H_2O$ Vol / Soil Vol for the 2013 case, around one order of magnitude higher than the 2014 one: 0.067 ± 0.010 $H_2O$ Vol / Soil Vol.

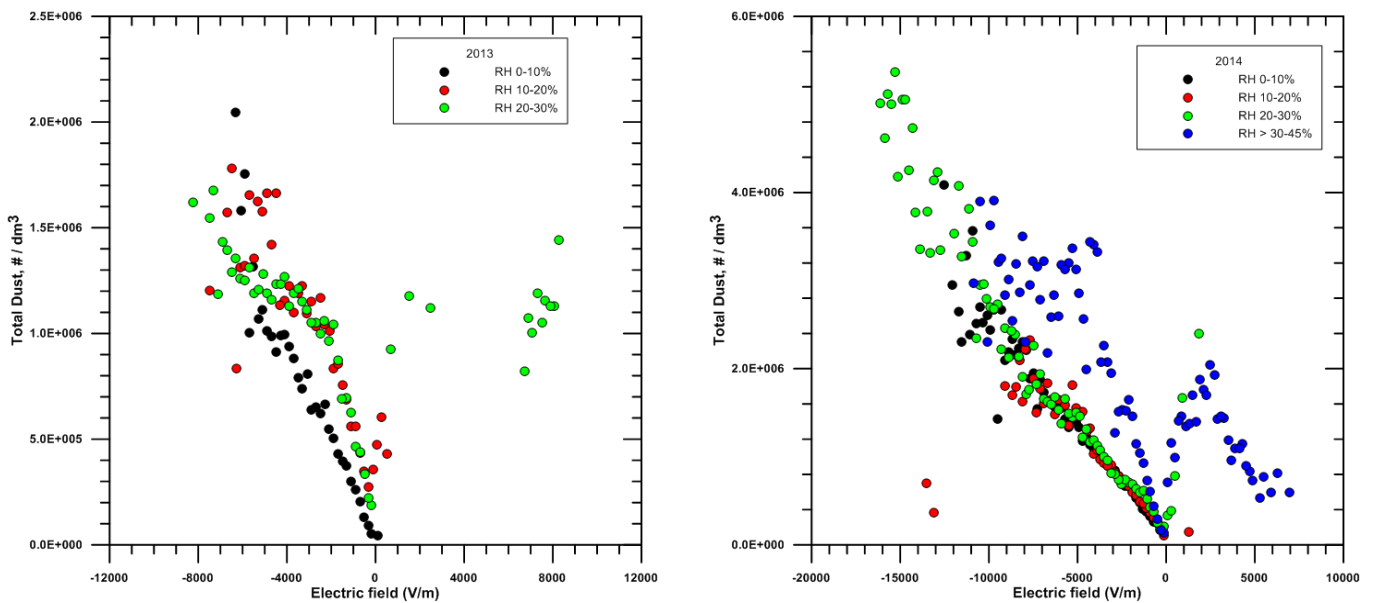

*Fig. 28 Relation between suspended dust concentration and E-field intensity for the 2013 (left) and 2014 (right) campaigns (different colors represent the RH level). In both plots, two main trends can be individuated with different values of the linear slope depending on the RH (black data vs red and green for 2013; black, green and red vs blue for 2014)*

We observed a linear trend between the dust concentration and E-field intensity for each humidity class, with an absolute value of its slope increasing with RH. This means that, at a given dust concentration, the E-field decreases with the increasing RH. However, the dependence of E-field from humidity does not seems to be a continuous relation, but more similar to a step function: we can observe two main trends separated by a critical RH value. Such value is different for the two sites, being the 2013 one around RH ~ 10% and the 2014 one at RH~ 30%, reflecting the different amount of soil humidity observed during the campaigns. Under the RH critical value, the slopes obtained in dry conditions are compatible in both sites, around -240 ± 10 $dm^{-3} \cdot (V/m)^{-1}$ ($R^2$ = 0.90).

Over the RH critical value, the absolute slope value increases up to the higher E-field intensities, when a decrease seems to happen again.

A dependence of E-field behavior on the RH was also observed during wind tunnel experiments by Xie and Han, 2012. The authors measured the E-field in a chamber covered with a sand bed, using a field mill (KDY-IV) varying the RH, wind and temperature conditions. They observed that the E-field linearly increased with the RH up to a critical value and then exponentially decreased. This critical RH value was observed to increase upon increasing the wind speed. They reported values of RH ~20% for a wind speed of 12 m/s and ~40% at 14 m/s. We believe that the re-decrease of the slope for the higher E-field values could be related to the increment of the wind speeds, as supposed by Xie and Han, 2012. Even if the laboratory experiment cannot be fully compared with the conducted Sahara campaigns, due to the different conditions, the obtained results are in good agreement.

Our observations provide strong evidence of how RH represents a critical parameter in aeolian processes, affecting the charging mechanism of dust and sand grains. A likely explanation of this phenomenon is the evaporite deliquescence: absorbing the atmospheric moisture the evaporite may dissolve in a water solution. The hydrated evaporites (Mg and Ca chlorides) present in both the 2013 and 2014 soils (with different abundance) are affected by deliquescence at low RH values (critical RH, C-RH = 28.7% for $CaCl_2 \cdot 2H_2O$ and 32.78% for $MgCl_2 \cdot 6H_2O$).

When the atmospheric RH exceeds the local composition C-RH value, the hydrated salts dissolve in the absorbed water and create a liquid film around the sand and dust grains, increasing their electrical conductivity and impacting the charging process.

A direct relation between the humidity and dust lifting could also exist and, consequently, influence the E-field. In field experiments, Halleaux and Renno 2014, found that the dust lifting increases with the soil humidity. They proposed that this could be due to either an increase in the amount of dust aggregates able to initiate saltation as the humidity increases or an increase in dust lifting due to the formation and breakup of crusts favored by the increase in humidity.

In summary, from our results we can conclude that:
- saltation and dust emission processes produce an enhancement in the atmospheric electric field (due to the exchange of electric charge among grains and successive charge separation by local turbulence);
- the concentration of emitted dust and the E-field intensity are linearly related at fixed RH (Fig. 28);
- humidity affects the conductivity of the surface and evaporite deliquescence could play a key role in this;
- the soil and air humidity affects the slope of the linear relation between the dust concentration and E-field: at a given value of the dust concentration, the E-field decreases when the humidity exceeds a threshold, in agreement with what arises from the laboratory experiment of Xie and Han, 2012.

It is difficult at this stage to discriminate between causes and effects in the phenomena observed in the field, however, it results clearly evident from our data that the concentration of lifted dust, E-field and humidity are very strongly related.

### *3.1.2.2 E-field feedback on dust emission*

In the previous sections, we have shown evidence that the saltation and dust lifting processes cause particles to become electrically charged and generate an atmospheric electric field, which is also affected by the ambient relative humidity and soil moisture. In this paragraph we will study if a feedback mechanism exists and how the generated E-field can in turn influences the grains lifting. Kok and Renno, *2008* and Holstein et al., *2012* supposed that when the E-field intensity exceeds a certain threshold, the electric forces can play a significant role in the lifting, reducing the static friction velocity threshold necessary to initiate the saltation process. We here report in Fig. 29 the original Fig. 4 of Kok and Renno, 2008, where the authors modelled how the density mass of saltating grains depends on the friction velocity u*. In particular, above an E-field threshold value (corresponding to a shear velocity of ∼ 0.6 m/s), their model predicts and abrupt increase of the amount of grain mass in saltation. In turn, this further boosts the E-field magnitude and the saltation activity.

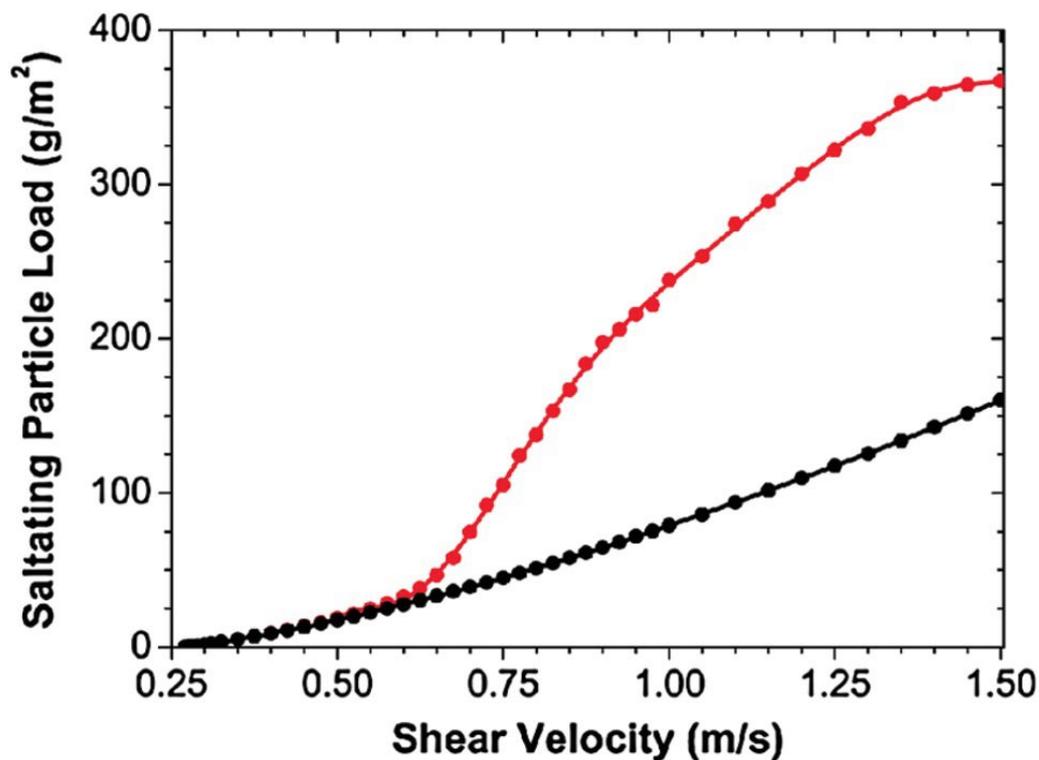

**Fig. 29 Original Figure 4 of Kok and Renno, 2008. The predicted relation between the mass of grains in saltation for two models: in black the one that does not include the electric forces, in red the one in which are included.**

**The models are in agreement up to a critical value of shear velocity ∼ 0.6 m/s, when the red one predict an abrupt increase of the amount of mass in saltation.**

Up to now, this active effect of the electric forces has never been investigated on field, hence we decided to use our data to study this subject. We cannot evaluate directly the saltating mass density, so we used instead the mass density of grains in suspension, evaluated from the numeric lifted grain concentration $\eta$. In order to do so, we considered that the dust size fraction of the soil in the site of measurements is composed mainly of quartz ($\rho$=2.66 g/cm$^3$) and muscovite ($\rho$=2.82 g/cm$^3$), with a minor part of calcite ($\rho$=2.71 g/cm$^3$) and illite ($\rho$=2.79 g/cm$^3$), as measured in our laboratory. Its density can be roughly estimated as $\rho_t$=2.7 g/cm$^3$. We have evaluated the total lifted mass density of dust (with diameter $\leq$ 34 μm) using the direct measurement of $\eta$ for each of the 31 bin size monitored and assuming $\rho_t$ as the mean density.

We derived the friction velocity $u^*$ directly from the high rate 3D wind components ($u$, $v$, $w$) measured by the CSAT3 3D anemometer at 4.5 m (eq. (12): $u^* = (\overline{u'w'}^2 + \overline{v'w'}^2)^{1/4}$ ). Considering that the vertical variation of $u^*$ with height can be assumed to be negligible in the surface layer (Haugen et al., 1971), we use this measured $u^*$ value as representative of the superficial shear stress.

In their model, Kok and Renno, 2008 has assumed a soil made conductive due to the absorption of conducting films of water. Therefore, to compare our results we neglected the driest storms (RH < 10%). Moreover, we have not considered in this analysis the minority of data characterized by an upward pointing E-field and the night-time storms clearly recognizable as haboobs. The physics of these events need to be studied as a separate case, as the haboobs arise in a heavily moist environment taking energy mainly from the latent heat released during the evaporation (Knippertz et al., 2007, 2009).

For the remaining events, we compared the data regarding the concentration with the corresponding shear velocity one, as showed in Fig. 30.

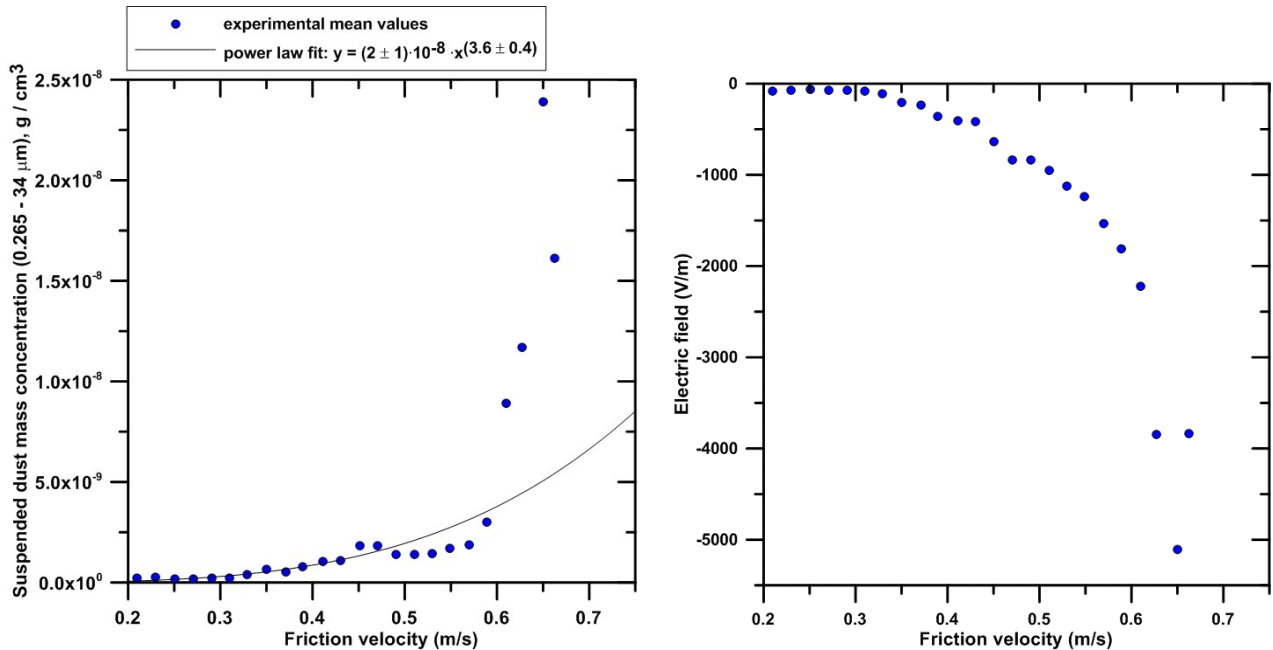

*Fig. 30 Suspended dust mass concentration (left) and atmospheric E-field (right) as a function of the wind friction velocity for dust storm events with RH in the range 10-30%. The data are binned in step of 0.02 m/s.*

In agreement with the theoretical prediction that neglect the E-field contribution (*Kok et al., 2012 ; Merrison, 2012*), for lower shear speed u* the dust mass concentration increases as a power law $\propto u^{*\,(3.6 \pm 0.4)}$ (the gray line in Fig. 30). However, at a threshold value of $u^* \sim$ 0.6 m/s, the mass concentration experiences an abrupt increase coupled to a corresponding increase of E-field (see right panel of Fig. 30).

Kok and Renno 2008 showed that this sudden growth of the particle load is not predicted by lifting models that do not include electric forces. In agreement with their model, by exceeding the threshold, our data show an increase of approximately an order of magnitude in the lifted dust mass concentration.

Kok and Renno, 2008 and Rasmussen et al., *2009* speculated that the E-field should exceed 150 kV/m to significantly affect the particle lifting. However, this value is predicted at short distance from the sand bed (few centimeters) and thus cannot be directly compared with our measurements performed at 2 m from the ground (where E-field was found to be < 20 kV/m).

The effective role of the electric forces on the set in motion of the grains and their trajectory need further studies to be really understood, however, our results provide the first field evidence that the E-field play a key role in the dust lifting process. It suggests the utmost importance of including its contribution in dust entrainment models to improve predictions on dust load and its effect on the atmospheric thermal structure and on the planet climate in general.

## 3.2 Dust Devils

In this section we'll focus on the dust devil's activity observed during the 2014 campaign. Along the 83 days of the campaign, we observed a total of 556 dust devil's events belonging to the Class A (the most probable dust devils candidates, see par. 2.6.1.3 for the description of the detection method and the categorization). For 54 days we have also collected the 3D anemometer data, monitoring 338 Class A events.

For the first time we collected a data set that couples the meteorological measurements of the vortices with the monitoring of the induced saltation and dust lifted activity. In addition, we measured also the E-field signatures of the whirlwinds, a data that has been reported in literature for less than a tens of events. Overall, our survey represents the most comprehensive one available up to now for the dust devils and has allowed the first statistical study of their electric proprieties.

Fig. 31 shows an example of sensors' response to one of these events. In this instance, all the dust devil features described in par. 2.6 are clearly recognizable. In *Fig. 31*a we can see the drop due to the vortex low pressure core, *Fig. 31*b shows the E-field peak due to the passage of the charged grains, *Fig. 31*c shows the increase in saltation activity measured by the Sensit and *Fig. 31*d the increase in temperature related to the convection lifting of the warm air layer near the ground. In Fig. 31e we show the peak in vertical wind speed connected with the upward motion inside the vortex column (negative values indicate a downward directed velocity). In *Fig. 31*f and *Fig. 31*g it is possible to recognize the change in the horizontal wind direction and the related peak in wind speed due to the vortex rotary motion, respectively. Finally, Fig. 31h and i show the peak in the lifted dust concentration and the decrease in the measured solar radiance due to the passage of the dust column and its shadow.

We analyzed the environmental condition that characterize the whirlwinds activity, the distributions of the vortex principal parameters and the correlations arising between them.

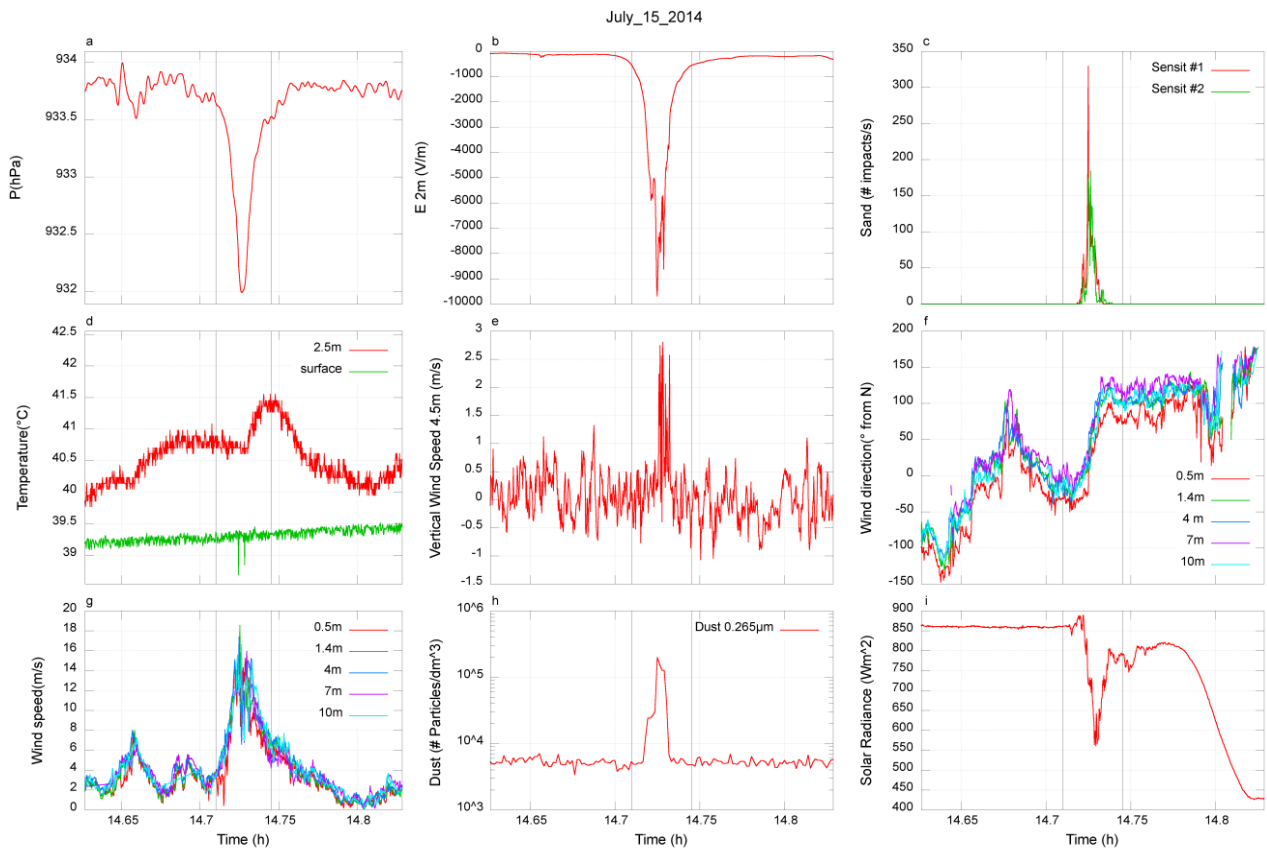

*Fig. 31 A dust devil (Class A) encounter as detected by our measurement station. The plots show the time series of the different monitored parameters. In order from left to right and from top to bottom there are: (a) atmospheric pressure, (b) E-field, (c) counts of near surface saltating grains, (d) temperature, (e) vertical wind speed, (f) horizontal wind direction, (g) horizontal wind speed, (h) lifted dust concentration and (i) solar radiance.*

### 3.2.1 Daily and seasonal activity

The dust devils activity starts around 9 a.m., around 3 hours after the sunrise, increasing up to the early afternoon and then it reaches its minimum around 10 p.m., a couple of hours after the sunset (see Fig. 32).
The daily trend closely follows the temperature trend and this results is in agreement with what reported in the literature and firstly observed by Sinclair 1973. However, unexpectedly, after the first minimum the activity slightly reincreases reaching a maximum around 12 p.m., finally totally stopping around 2 a.m. Despite representing only a small fraction of the total sample, this possible night-time activity is of great interest, potentially opening new scenario for the dust devils formation. Night time dust devils like signatures have been observed also in the Martian data, e.g. in Ellehoj et al. 2010, however these events have still not been considered in the statistic, being not in agreement with the vortex physics as we know it currently. The eventuality of night time activity needs further investigation in order to be validated. This is one of the reasons we decided to equip the station of the 2017 and future campaigns with a camera and a nigh flash system. The results of the 2017 data are described in the next Chapter, but currently they are conclusive on this topic.

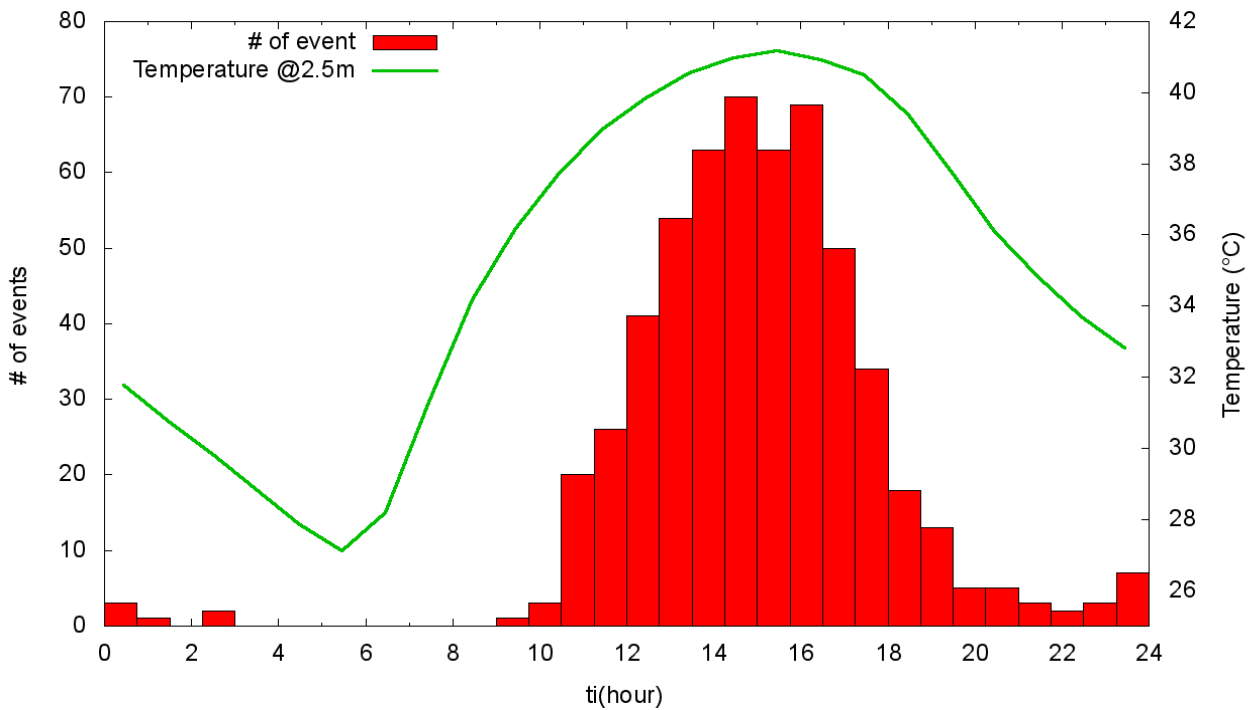

**Fig. 32 Diurnal dust devil's activity observed in the Sahara data, it clearly follows the temperature variation.**

The activity during the entire campaign has been very variable, as shown in Fig. 33, and it seems to weakly correlate with the temperature trend. Therefore, unlike the day-scale activity, the long term one is not mainly driven by the temperature variation. The factors playing a significant role are of difficult individuation and we are still investigating this subject.

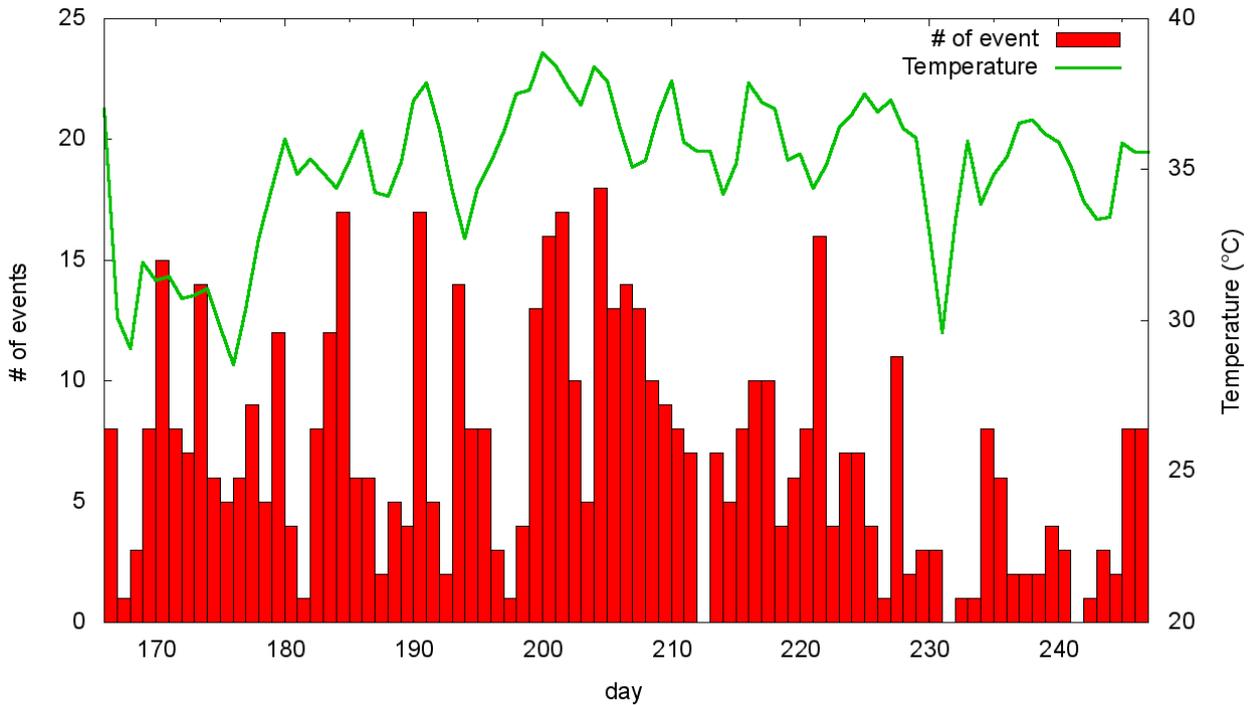

**Fig. 33** Seasonal activity along the entire campaign, on the x axis there are the day of the year, and on the y one the number of events observed that day. It is also shown the mean daily temperature in green, though, it seems to weakly correlate with the activity.

### 3.2.2 Wind speed regime

The wind speed signal ($v_t$) acquired from the station during the encounters is: $\boldsymbol{v_t = v_w + b}$, where $v_w$ represents the vortex wind speed, i.e. the relative horizontal speed of the dust devil in the rest frame of the background wind ($b$).

We analyzed the environmental wind conditions considering a time window of thirty minutes around each event. Most of the dust devils arises from a wind background $b$ of about 6 m/s and comes preferably from the south west direction (see Fig. 34). Overall, this respects the average wind regime faced during the campaign and is in agreement with what we observed for the diurnal dust storms.

A reasonable simplifying assumption is to consider that the vortex travels parallel to $b$. The maximum $v_M$ of the total wind signal $v_t$ is reached in the instant of minimum distance from the station ($d = d_o$). Hence, from the direct measurements of $v_M$, we can evaluate $\boldsymbol{v_w = v_M - b}$ obtaining an estimation of the rotatory vortex wind speed $v_w = v_r(d_o)$. As we explained in par 2.7, we are not able to obtain the intrinsic rotatory wind speed $V_r$ and the impact parameter $d_o$ from this measure. We have evaluated all these quantities using the data acquired by the anemometer placed at 4 m. Moreover, through the 3D anemometer mounted at 4.5 m we were able to monitor also the vertical wind speed of the vortices.

The distribution of maximum vertical wind speed and vortex wind speed $v_w$ are showed in Fig. 35.

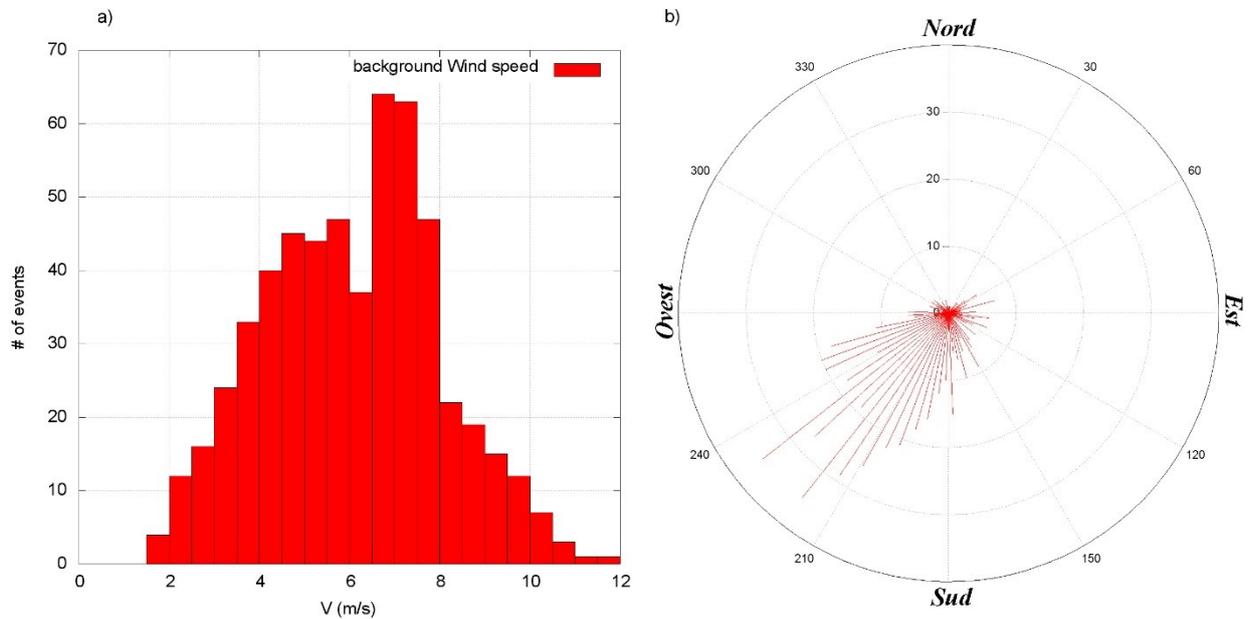

**Fig. 34 The environmental wind regime observed during the dust devils encounters. a) shows the histogram of the background wind speed, while b) shows its direction.**

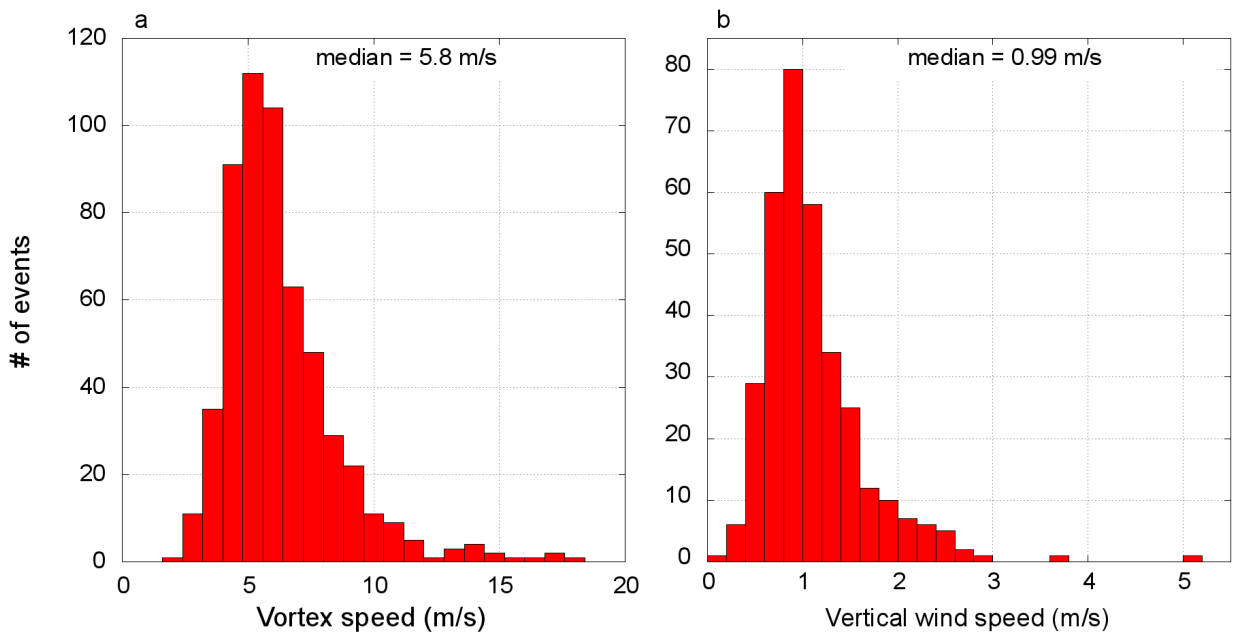

**Fig. 35 The histogram relative to: a) the vortex wind speed $v_w$ and b) the maximum vertical wind speed observed during the dust devils encounters.**

For the rotatory speeds we observed a mean value around 6 m/s with peaks up to 18 m/s, a range of variation in agreement with what previously reported in literature (Sinclair 1973). Regarding the vertical wind speed, it ranges from 0.1 m/s to 5 m/s, with a mean value of ~ 1 m/s. Even when the vortices passed exactly over our station ($d_o < R$) the maximum vertical wind speed remains ~ 1.5 m/s and only during two events it overcomes 3 m/s. Overall, from the collected signatures, we observed how the dust devil passage leaves a signal clearly recognizable in the horizontal wind speed and direction, pressure and E-field time series, with a full duration half maximum of various seconds. Compared to these ones, the trace left in the vertical wind speed is usually more similar to a delta function or a series of deltas, without a clear growing and decreasing trend (see for example the Fig. 31e). The reduced duration of the vertical flow signal coupled with its tiny magnitude makes the vertical wind speed peak less clear to recognize and study than the other dust devils features. This led us to the conclusion that the upward and downward vortex air flows have to decay much faster with the distance than the other monitored parameters.

Reporting the results of the field campaigns he performed in Arizona, Sinclair (1973) observed how the vertical wind speed of the vortices, measured approximatively at 2.1 m and 9.4 m, usually reaches 10 m/s inside the dust column, rapidly decreasing outside it. The vertical speeds we measured at 4.5 m are far slower than the ones observed by Sinclair, even for the vortices that pass exactly over our station (Fig. 35b). We also noted that the measured vortex horizontal wind speeds usually overcome the vertical speeds. This result is in agreement with the observations made by Ryan and Carroll (1970), which reported vertical wind speeds measured at 2 m from ground between 0.2 m/s and 2.2 m/s, approximately one order of magnitude lower than the horizontal wind speed.

Starting from our observed ranges of speed magnitude, we have estimated the contribution to the lifting due to the vortex horizontal (the shear stress force described in par. 1.5.2) and vertical wind flow (the suspension force described in par. 1.5.1 eq. (31)). From our evaluation it results that none of the two contributions can be neglected and, for the dust size particles, they both contribute with a force per grain of the order of $10^{-9}$ N, on average one order of magnitude greater than the gravity force.

### 3.2.3 Pressure drops distribution

The measured pressure drops ($\Delta P$) range from 0.18 (the chosen threshold) to 1.3 hPa (Fig. 36a). Most of the detected vortices lies in the range 0.2-0.8 hPa and in this interval the $\Delta P$ cumulative distribution is well described by a power law ($y = a\, x^b$ with and exponent b = -2.85 ± 0.05, Fig. 36b). For the largest and smallest $\Delta P$ values an appreciable drop off from the power law is evident, as also seen in other terrestrial data (e.g. in Fig. 36b we show a dataset from Lorenz and Lanagan, 2014). The deviation of the distribution of lowest and highest drops values from the main trend could be attributable to the inefficient detection of the smallest vortices and to the under sampling of the largest ones due to the finite sample size (Jackson and Lorenz 2015). However, a reason of physical nature cannot be excluded. Despite the fact we obtained a similar trend distribution, the value obtained for our power law exponent (-2.85) is higher than the ones previously reported for the Terrestrial dust devil surveys (~-1.5).

We observed that in the same range 0.2-0.8 hPa the $\Delta P$ cumulative distribution can be also described by an exponential function with a coefficient of determination comparable to the one of the power law fit ($y = a\, 10^{b\,x}$ with an exponent b = -3.32 ± 0.05). Although there is not a compelling reason to prefer one representation to another, throughout this work we will continue to use the results of the fit power law in order to allow an easier comparison with the literature.

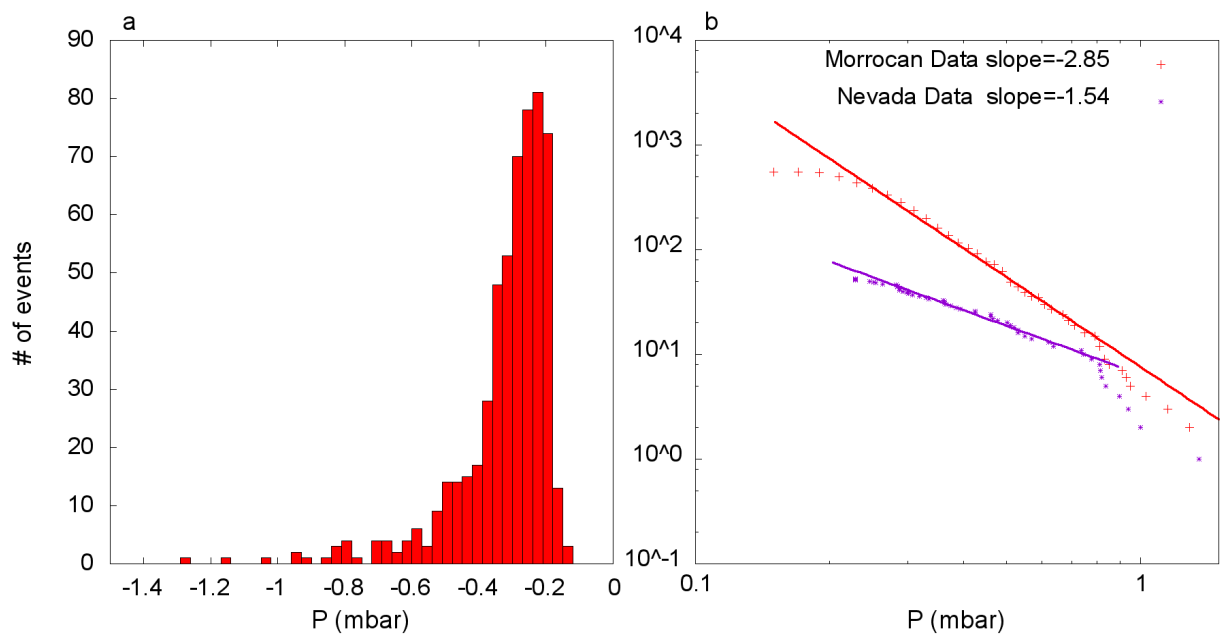

Fig. 36 The distribution of the pressure drop ΔP for our Sahara dust devils. a) shows the histogram of ΔP, while b) shows its cumulative distribution (absolute value): in red for our data and for comparison, in violet another terrestrial survey collected in Nevada by Lorenz and Lanagan, 2014 (Category 1 events in the P28 dataset).

## 3.2.4 E-field proprieties

### 3.2.4.1 Distribution and Orientation

The fair-weather value of the electric field measured during the campaign is almost always negative around −50V/m. During the dust devils passage we observed an excursion of vertical E-field always downward directed from few thousands V/m up to 16000 V/m, with a mean value around -2600 V/m (see Fig. 37a). Up to now, there are no other survey of dust devils electric field with which compare our results, indeed our data have allowed the first statistical study of this parameter. We observed that the cumulative distribution of the E-field excursions (showed in a semi-log plot in Fig. 37b) is well describable in the whole variation range by an exponential law, $y = a\, 10^{bx}$ with an exponent of $-(161 \pm 4) \cdot 10^{-6}$ $(V/m)^{-1}$.

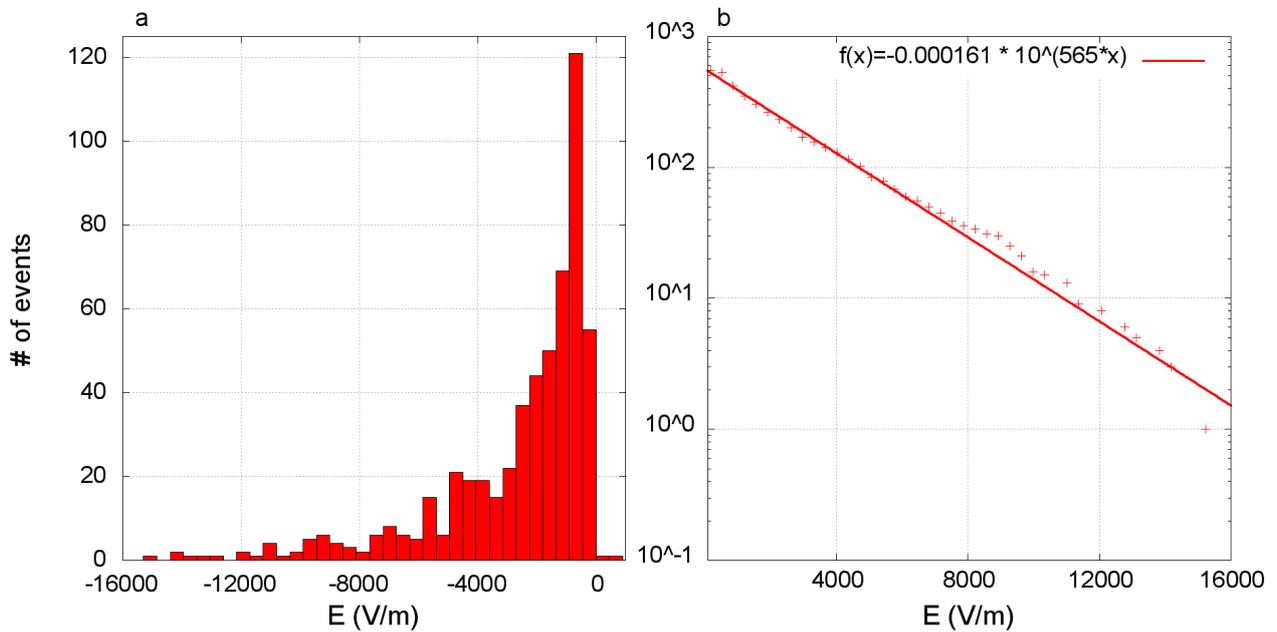

Fig. 37 The distribution of the E-field excursions induced by the dust devils passages. a) shows the histogram with a bin width of 450 V/m, while b) shows the cumulative distribution (absolute value), that is describable by an exponential law.

An important difference between our results and those commonly reported in the dust devil's literature is represented by the orientation of the vertical E-field. Indeed, as we said for the dust storms, in presence of a lifted dust cloud an upward directed E-field is usually reported, corresponding to a size dependent electrification with the smaller grains negatively charged over a base of bigger particles of opposite sign (Forward et al. 2009; Lacks and Levandovsky, 2007; Duff and Lacks, 2008; Desch and Cuzzi, 2000; Melnik and Parrot, 1998; Bo and Zheng, 2013).

Regarding specifically the dust devils, the literature so far reports only cases characterized by upward directed E-field (Frier, 1960; Croizer, 1964 and 1970; Farrell et al., 2004, Jackson et al., 2006). However, these measurements are still few and there is not a physical constrain on the possible direction of the E-field. Indeed, the statement that the charge process of the grains is size dependent is secondary to the hypothesis that the dust cloud is homogenous from the bottom to the top. We don't believe that this is the case, at least for our site of measurement.

For the dust storms, few cases are reported where this electric configuration (smaller grains negatively charged over positive bigger ones) is not reproduced (Trigwell et al., 2003; Sowinski et al., 2010; Kunkel, 1950) and there are even lesser cases where the electric field has been observed with both orientations (Demon et al., 1953; Kamra, 1972). In particular, Kamra 1972 measured both positive and negative electric gradient and space charge during the dust storms, depending on the site and, even at fixed site, depending the atmospheric conditions. He observed that dust storms arising from soil mainly composed by silica show both orientations of the E-field, while the ones arising from soils mainly composed by clays materials usually induce a positive (upward) E-field. Moreover, he speculated that its measurements acquired over the sand dunes suggest a dipole configuration with positive charge up and negative down, like the configuration we are supposing for our acquisitions. As far as we know, our data set is the first example of a dust devils sample entirely characterized by a downward directed E-field, being the first evidence that the orientation of the vortex E-field is not a universal property. We believe that the principal factor that affects the sign of the induced E-field is the composition of the dust cloud, and the relative differences in composition between the smaller and the bigger grains. However, this subject is still matter of controversy and a full comprehension requires further investigation, in particular laboratory analysis on the composition and size distribution of the soil that sustains the whirlwinds.

### *3.2.4.2 Signal trend and connections to the other vortex parameters*

Fig. 38 shows the recorded vertical E-field, pressure drop and lifted dust signals for three different events with a strong, medium and tiny atmospheric E-field variation, respectively. These signatures are always well visible and distinct from the background and the pressure signature appears to last for a longer time than the corresponding E-field variation. Indeed, we observed how the full width half maximum of signal induced by the vortex passage is around 50% longer in the pressure time series than in the E-field one. The E-field signature is usually roughly symmetric, however strongly asymmetric events as the one depicted in Fig. 38c are not totally uncommon.

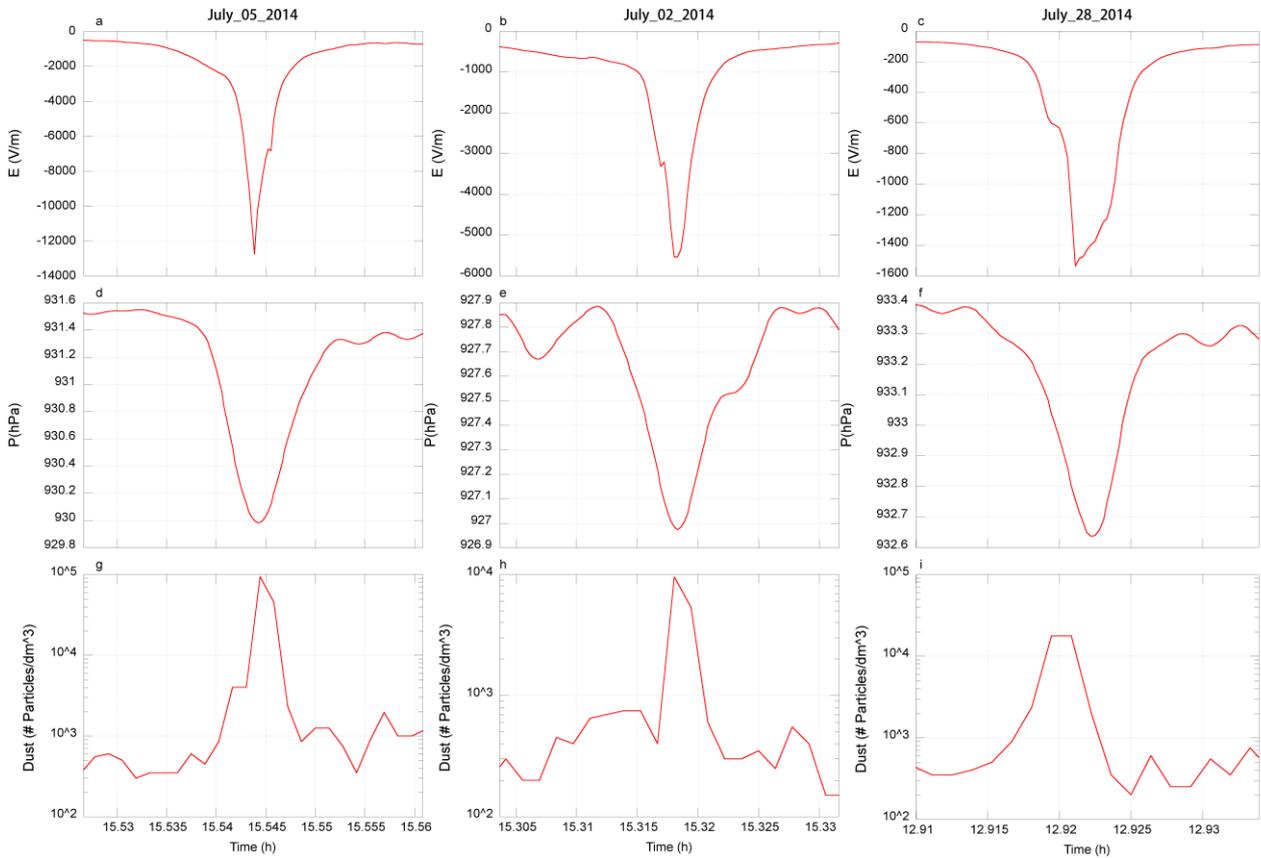

**Fig. 38 The vertical E-field time series for three Class A events of strong (a), medium (b) and tiny (c) magnitude, respectively. For the same events there are also reported the pressure drops (d, e, f) and the dust concentration in the bin-size 1.45μm (g, h, i).**

In particular, during the passages over the station, we observed both single (Fig. 39) and double peaks shape (Fig. 40) as well as irregular and highly noise signals. Unfortunately, the events that directly cross the instrumentation represent only a few percent of the total sample (around 6%). The high number of different observed trends and the low statistics for each case does not allow to proper study the case of dust devils directly passing over the station.

For the whole set of data, we analyzed how the E-field excursions are connected to the other parameters signals (for the discussion regarding the uncertain level of these results see the next paragraph 3.2.4.3).

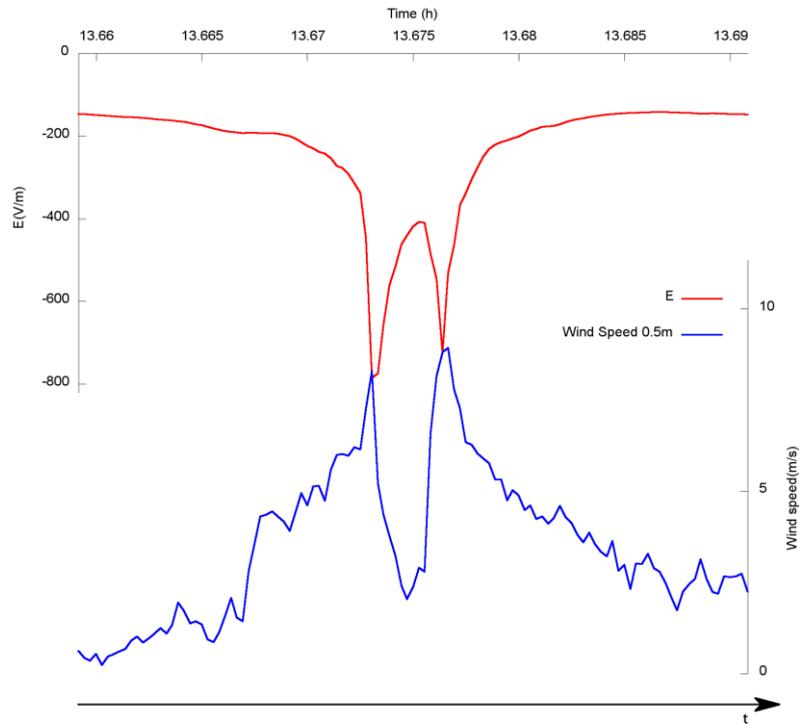

**Fig. 39 An event passing over our station, where it is possible to see the double peaks trend in both the wind speed and the E-field**

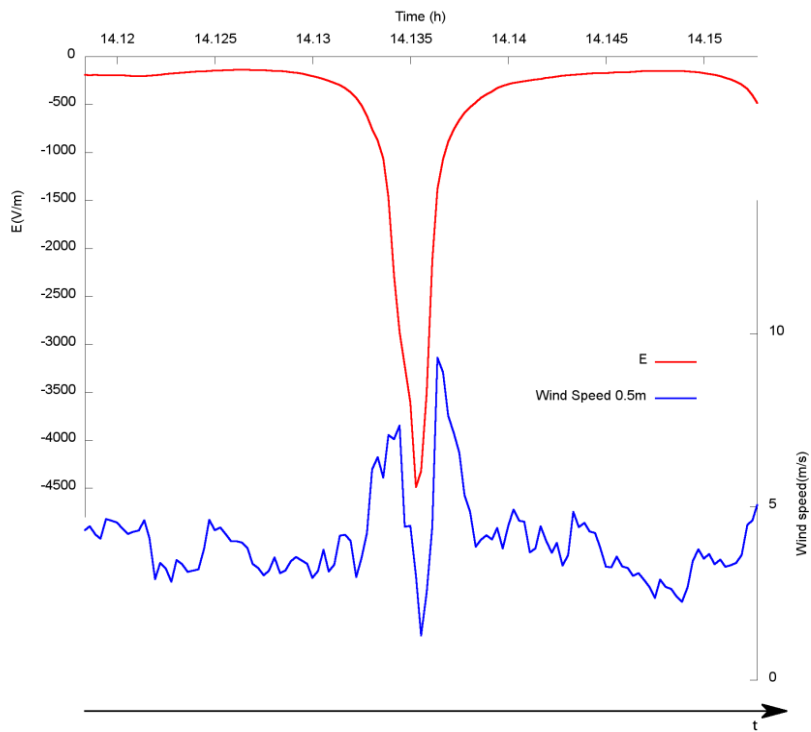

**Fig. 40 Another event passing over our station, where it is possible to see the double peaks trend in the wind speed, but a single peak trend in the E-field**

3.2.4.2.1  E-field vs vortex pressure drop and wind speed

We observed that the induced E-field is linearly related to both the pressure drop (Fig. 41a) and the vortex wind speed of the vortices (Fig. 41b). The physical meaning of these relation has probably to be searched in the dependence of the vortex lifting ability to its horizontal pressure gradient, which in turn is connected to the rotatory speed. Greeley et al. (2003) observed how the dust devils can entrain dust with wind speed up to 80% lower than those required to the simple boundary layer winds. This feature is probably due to the presence of the pressure gradient force generated by the low pressure core, that acts like a lifting force additional to that exerted by the wind (Neakrase et al., 2016). The magnitude of the pressure drop is therefore related to the amount of the lifted dust, which generate the E-field. The theoretical prevision of the relation function is a complex matter, due to the difficulties in quantifying the actual contribution of the pressure gradient force and currently this relation is not investigated by any model of dust devil formation and dynamic. Our data show the first proof that the induced E-field and the vortex ΔP are actually connected, and that the relation is approximatively describable by a linear function (Fig. 41a). The relation between the E-field and the vortex wind speed probably arises from the previous one, considering that $v_w$ could be with good approximation equated to the rotary vortex speed, that is related to the pressure drop through the hydrostatic equilibrium eq. (4).

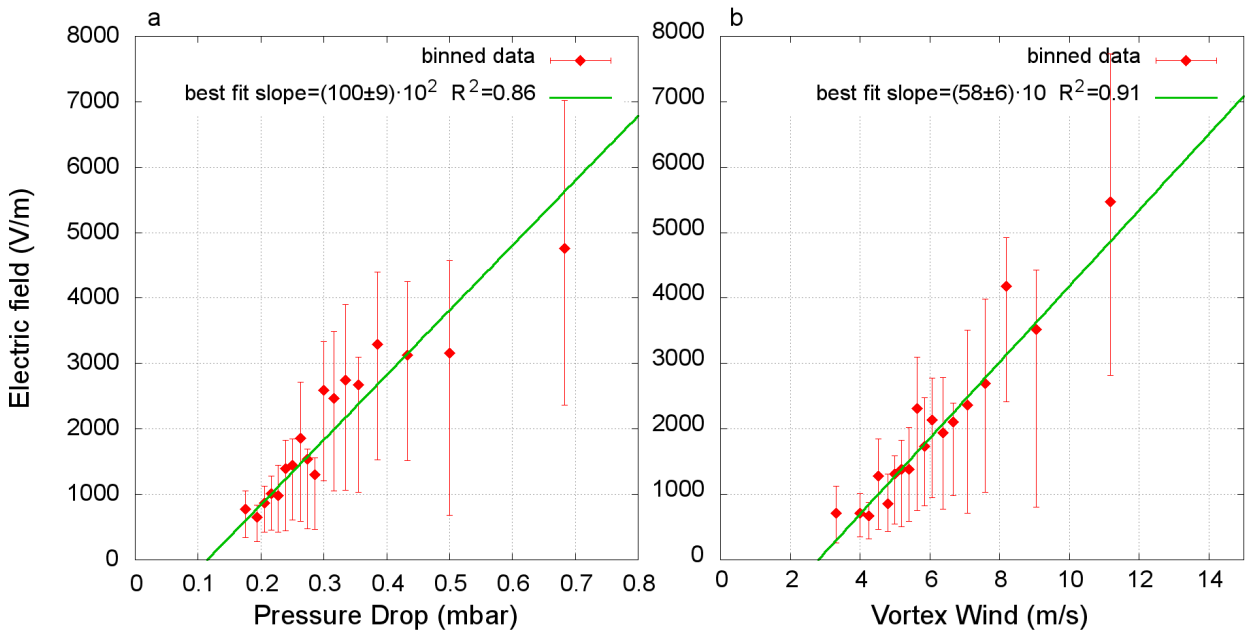

Fig. 41 Relations of the dust devil's induced E-field with: a) the vortex pressure drop and b) the vortex wind speed. Each point represents a bin of 30 dust devils, the error bars represent the 25th and 75th percentile.

3.2.4.2.2 E-field vs vertical flow speed

We also observed that the E-field is linearly related to both the upward (Fig. 42a) and downward (Fig. 42b) directed vertical flow speed measured during the encounters. The vertical wind acts like a driving force for the suspended grains motion, driving their collisions and spatial distribution, hence it is directly linked to the amount of electrification of the system.

Farrell et al. (2006) modelled the electric field generated by a dust devil system simplifying the composition by considering only two species of grains with different sizes. The grains are stratified by gravity and the authors assumed, as it is common in literature, the small negatively charged and the large ones positively (E-field upward directed). Moreover, they assumed the horizontal wind contribution to the lifting process to be negligible compared to the vertical one, basing this hypothesis on the Sinclair results (a vertical wind speed one order of magnitude greater than the horizontal one). In Fig. 7 of their paper, the authors showed how the modelled vertical E-field is linearly related to the upward vertical wind speed. Nonetheless, they pointed out that this relation has to be applied with caution to the case of vertical speed whose magnitude is small enough to be compared with the horizontal one, because this possibility is not contemplated in their model.

According to our data, the E-field and the vertical wind speed seems to be actually linearly related and the relation subsists also for the low upward vertical wind speed (Fig. 42a). Differently to the assumption of Farrell et al. (2006), we have seen that the contribution of the horizontal wind is not negligible (see par. 3.2.2). Indeed, we found a direct relation between and the E-field and $v_w$ (Fig. 41b).

As we said, the induced E-field seems to be also related to the downward directed wind component (Fig. 42b). This downward motion is usually predominant right outside the dust column, where the grains start to sediment by convective flow motion and gravity. This suggests that the electric current related to the descending motion of the grains could play an important role on the behavior of the E field, while, the current dust devils E-field models don't take into account this contribution (Farrell et al., 2006, Barth, 2016).

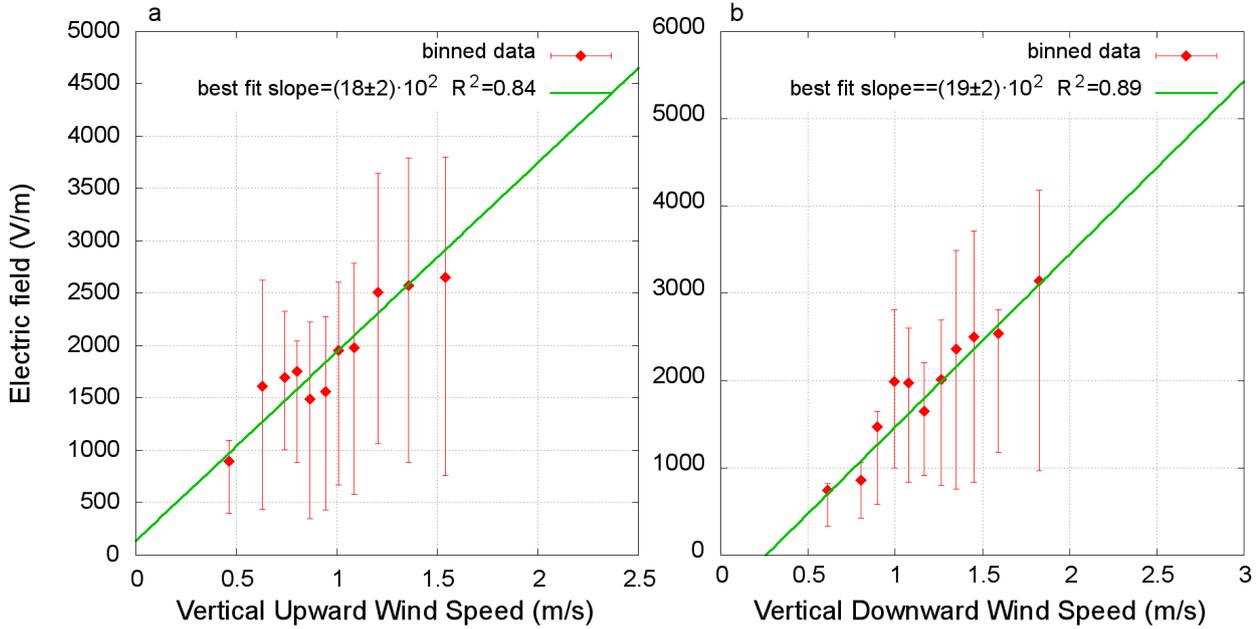

**Fig. 42 Relations of the induced E-field with: a) the upward directed vertical wind speed and b) the downward directed one. Each point represents a bin of 28 dust devils, the error bars represent the 25th and 75th percentile.**

3.2.4.2.3 E-field vs lifted dust

Finally, we studied the correlation between the induced E-field and the amount of lifted dust. We already showed that, during the dust storms, the vertical E-field is linearly related to the numeric lifted grains concentration ($\eta$). Here, in the place of the instant by instant values considered in par. 3.1.2.1, we have analyzed the maximum of grains concentration vs maximum induced E-field intensity for both dust devils and dust storms events (Fig. 43). In Fig. 43a we plotted the lifted dust grains concentration against the corresponding E-field absolute value. The data related to the dust devils are shown in red, while, for comparison the dust storms data are reported in blue.

We have observed that the E-field and the grains numeric concentration ($\eta$) are linearly related also during the dust devil events. Moreover, the slopes of the two relations are compatible: dust devils slope = $(3.7 \pm 0.2) \cdot 10^2$ dm$^{-3}$(V/m)$^{-1}$; dust storms slope = $(4.0 \pm 0.3) \cdot 10^2$ dm$^{-3}$(V/m)$^{-1}$. This correspondence indicates that the linear relation E-field/grain concentration is probably a general law common for the dust lifting phenomena.

From $\eta$ we have also evaluated the total lifted mass concentration as described in par. 3.1.2.1. Fig. 43b shows in a semi-log plot that the relation between the E-field magnitude and the mass concentration can be described by a power law (exponent = $0.53 \pm 0.05$).

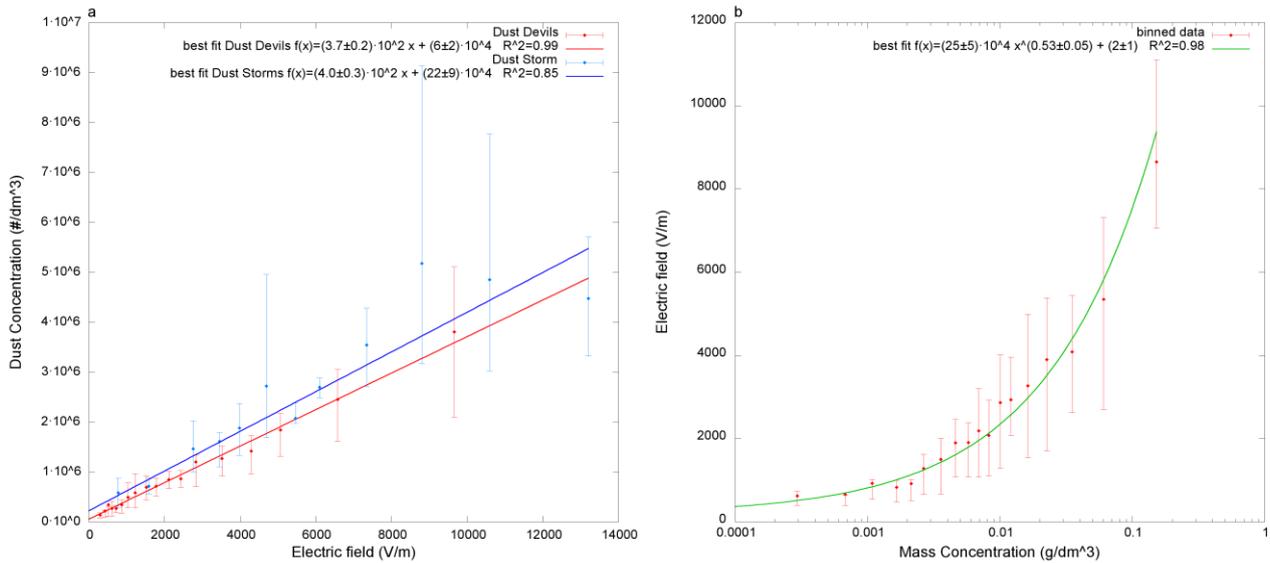

**Fig. 43 a)** the total dust grain concentration is shown as a function of the induced E-field for the dust devils (in red) and the dust storm (in blue) events observed during the field campaign. **(b)** Semi-log plot that shows the relation between the induced E-field and the total dust mass concentration during the dust devils events. Each point represents a bin of 30 events for the dust devils and of 7 events for the dust storms; the error bars represent the 25th and 75th percentile.

For each of the linear regression (y=b+ax) here reported, we have evaluated the test statistic t and the p value of the relations and performed a one tail test, simply assuming as Null hypothesis a=0. In all the cases, choosing a level of significance of 0.05, we can't accept the null hypothesis, confirming that, despite the high uncertain level, the studied quantities are actually related.

Except the for relation between the E-field and the upward vertical wind speed, none of the reported correlation is currently investigated by any model of dust devil formation and dynamic. Our results encourage the acquisition of further data to study this subject and the start of a theoretical study of these relations, also in the optics of the upcoming martian data.

### 3.2.4.3  Sources of uncertain

For all the relations we reported that connect the E-field to the other vortex proprieties, we observed an high level of fluctuation of the data points around the main trend (see Fig. 41, Fig. 42 and Fig. 43).

As we explained in par. 2.7, it is not possible to directly interfere the impact parameter (the minimum distance achieved by the vortex during its passage near the station) from the data acquired from a fixed meteorological station. Therefore, we cannot evaluate the intrinsic parameters of the vortex, as its core pressure drop $\Delta P_o$ and rotatory speed at wall $V_r$, and either its morphological parameters, such as the diameter and the height. This lack of information does not represent a huge problem when we analyze the parameters singularly; indeed, given the large number of observed events, we can suppose that we dust devils are sampled at random vortex radii, height, and intrinsic parameters and no systematic bias is introduced. However, this is no more the case when we have to compare two different quantities.

We here focus in particular on the first relation we reported, E-field vs pressure drop (Fig. 41a), however the general meaning of the following discussion is extendible to all the relations reported in Fig. 41, Fig. 42 and Fig. 43. The large error bars shown are probably mainly due to the different dependence of plotted quantities on the distance between the vortex and the station. Indeed, the $\Delta P$ magnitude decreases like the square of the impact parameter (Ellehoj et al., 2010), while the E-field decreases like the third power of the distance (assuming a dipole configuration). This different dependence is noticeable also in our data; indeed, we observed how the pressure signal lasts significantly longer than the E-field one. Moreover, we have to consider that also the vortex diameter could not be directly inferred from our data. Currently it is not known how the magnitude of the induced E-field depends on the vortex size, hence, a different dependence on the diameter of the two plotted quantities could not be excluded and could lead to a further increase of the uncertain. These last considerations can also be extended the other reported comparisons between the vortex parameters, and we believe that the indetermination on impact parameter and vortex size represents the main source of error for each relation. For these reasons we organized a further field campaign, aimed to acquire synchronous measurements of the electric, meteorological and morphological vortex parameters in order to resolve the degeneration that currently affect the data and further investigate this subject. The results of this campaign are described in the next Chapter.

### 3.2.5 Parallel to the martian case

We compared the results of our survey with those obtained on Mars (Fig. 44). As we mentioned (par. 2.6.2 and par.1.6.1), wind speed and direction data are really sparse and the E-field data have never been collected up to now. Therefore, we focused on the measurements of the pressure drop related to the vortices passage. In particular, we considered the data obtained by the Pathfinder and Phoenix landers and by the Curiosity rover.

NASA Mars Pathfinder lander collected the data of 79 vortices during 83 martian sol in Ares vallis (Murphy and Nelli, 2002), while NASA Phoenix Mars Lander detected 197 vortices with $\Delta P > 0.5$ Pa during the 151 Sol Mission (Ellehoj et al. 2010) at Green Valley.

NASA's Mars Science Laboratory (MSL) Curiosity rover observed one full martian year of dust devil activity in Gale crater, collecting a sample of 245 convective vortices with pressure drops in the range of 0.30–2.86 Pa with a median value of 0.67 Pa (Steakley and Murphy 2016). This represents the martian vortex pressure drop survey with the highest statistics available up to now.

In all these cases a power law cumulative distribution similar to the terrestrial data has been observed and the value of the exponent is close to that of our survey. In particular, for the Curiosity data the similarity with the Sahara data is stronger: also in this case the distribution of the smallest and largest ΔP does not follow the main trend and the exponents (-2.81 ± 0.07 for Curiosity and -2.85 ± 0.05 for Sahara) are compatible inside the uncertainties.

A rigorous comparison of the terrestrial and martian results is not possible due to the different magnitude of pressure drop and the dependence of the regression results on the chose range of variation and binning procedure. However, it appears clear how the dust devils population on the two planets have many points in common. The similarities of the vortices populations on the two planets allow to speculate how our results on the induced E-field on Earth could be representative of the Martian dust devils too.

As we discussed in the previous paragraph, the Saharan data showed that the dust concentration and the induced E-field are linearly related both during the storms and dust devils events, with similar slope. Starting from this result we tried to obtain a rough estimation of the possible E-field inside a martian dust lifting event.

To a first approximation, dust devils can be outlined as a dipole distribution, hence, the magnitude of the field is directly proportional to the total charge. We assume a dipole model for the vortices, with the charges uniformly distributed inside the walls of a cave cylinder. In order to take into account that the particles are mainly distributed inside the vortex walls, we considered only 30% of the cylinder volume as filled by the grains; moreover, we considered the vortex height to be five times the radius, a ratio that is on average indicated by the surveys.

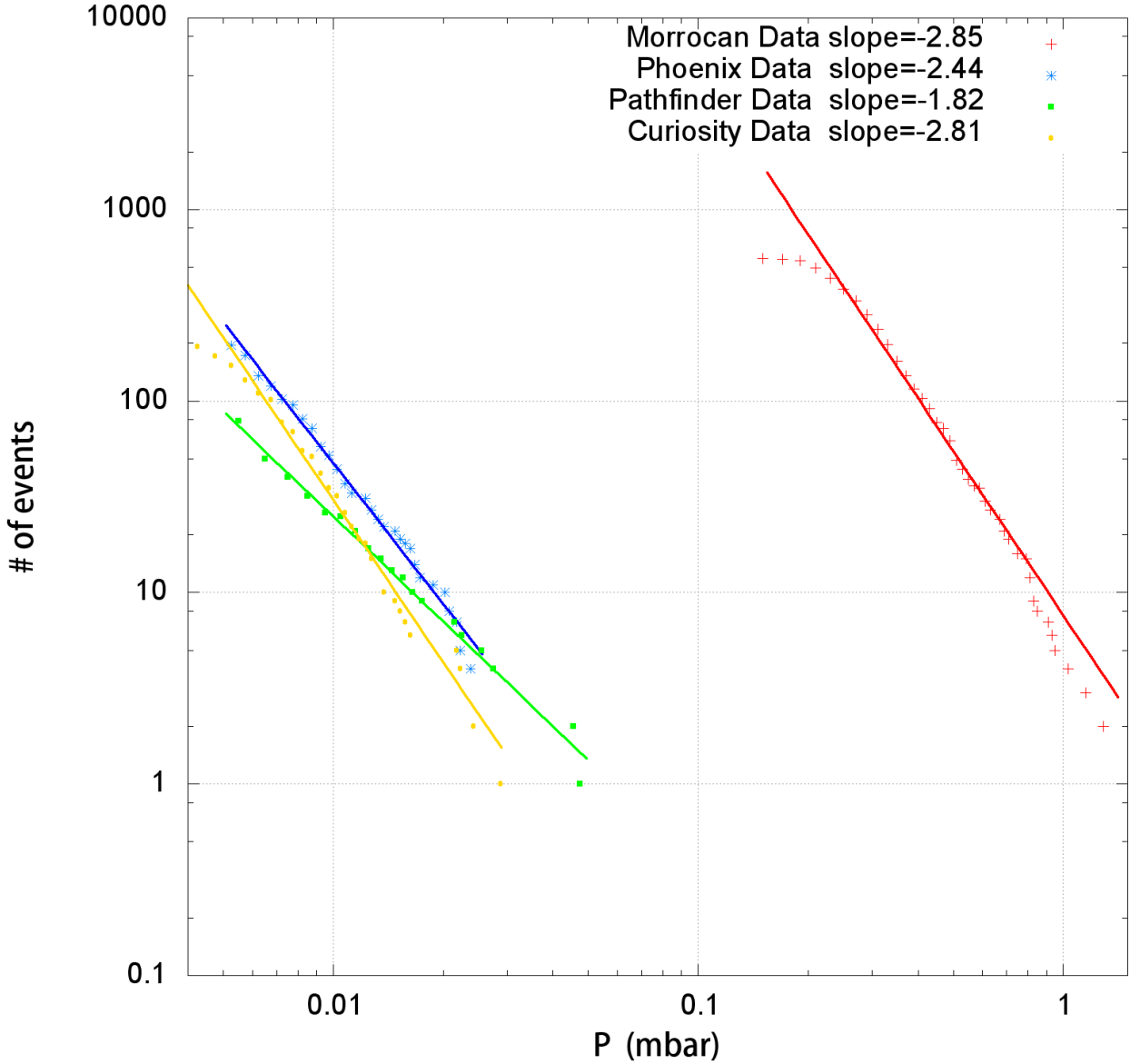

Fig. 44 Comparison of the cumulative distribution of dust devils pressure drop for the martian survey and our Sahara data.

In these hypotheses we obtain that the E-field magnitude is directly proportional to the charge density $\rho$, where the density is:

$$\rho = \sum_i q_i \eta_i \qquad (52)$$

the index $i$ labels the different grains species that compose our dust cloud and $\eta_i$ and $q_i$ are the numeric concentration and charge per grains of that specie. From our data we found an $E/\eta$ linear relation, where $\eta = \sum_i \eta_i$. For our model the $E$-field and the $\rho$ are linearly related, therefore, we obtain a linear relation between the space charge density $\rho$ and $\eta$. Hence, on a first approximation the charge per grain $q_i$ have to be similar for the different species, equal to an average $q$. The E-field at vortex wall of our distribution can be simplified as:

$$\begin{cases} E_{xy} = \dfrac{45\,\eta\,q\,R^5\left(\frac{h}{2}-z\right)}{16\left(R^2+\left(\frac{h}{2}-z\right)^2\right)^{5/2}\epsilon_0} \\ \\ E_z = \dfrac{15\,\eta\,q\,R^4\left(-R^2+2\left(\frac{h}{2}-z\right)^2\right)}{16\left(R^2+\left(\frac{h}{2}-z\right)^2\right)^{5/2}\epsilon_0} \end{cases} \quad (53)$$

where we made explicit the horizontal $E_{xy}$ and vertical contribution $E_z$, $R$ and $h$ are the radius and the height of the vortex respectively, $\eta$ is the numeric density of particles, while $z$ is the altitude where we monitor the field.

We applied this model to the vortices passing over the station, roughly estimating their radius from the passage duration and traveling speed. With these parameters, we have estimated the average charge per grain ($q$) and the charge density $\rho$ of the events. The found value of $\rho$ ranges between $3\cdot 10^{-9}$ and $4\cdot 10^{-8}$ C/m$^3$, while, $q$ ranges between $8\cdot 10^{-18}$ and $5\cdot 10^{-17}$ C, in agreement with what reported by Kunkel (1950).

We have considered this range of $q$ variation scaling the other model parameters to the ones expected on Mars. As we said, there are not direct measurements of the Martian near surface dust concentration (a gap will be filled up right by the MicroMED sensor of the ExoMars 2020 mission). However, from the opacity measurements reported by Montabone et al., 2014 we can estimate a numeric grain concentration ranging from few tens up to several hundreds of grains per cm$^3$. Regarding the sizes of the martian dust devils we know from the work of Greeley et al., 2010 that they range from 2 to 276 m. Hence, supposing a similar value of charge per grains on Earth and Mars, we obtain an *E-field* from few hundreds of V/m up to over 35 kV/m. Despite the fact that our estimations do not take into account the different atmospheric composition of the two planets, a so high E-field values strongly indicate the possibility to overcome the martian electric breakdown voltage, as firstly supposed using laboratory measurements by Eden and Vonnegut (1973) and recently repurposed by the work of Farrell et al.(2017), see par. 1.6.1.

Overall, the results presented in this Chapter represent some of the most advanced ones available for the dust lifting processes. In particular, we have performed the first statistical study of the electric proprieties of the dust clouds. This work is fundamental in order to understand the dust physical processes taking place on Earth, and also as preparation for the study of the martian environment, in anticipation of the ExoMars campaign

# Chapter 4
# Evaluation of the intrinsic vortex parameters

In this Chapter we will discuss about the last campaign we performed in the Sahara desert. The survey has been mainly aimed to the study of the dust devil activity, in order to improve and extend the data set acquired in the previous campaigns. The purpose was to fill some gaps in the previous data, such as the evaluation of vortices diameters, heights and passage distances. We will expose the method we developed to estimate the impact parameters of the encounters using the tracks leaved in wind speed and direction signals. We tested this method on the field data, comparing the results with the ones obtained by using the high rate images acquired.

## 4.1 2017 field campaign

In the last Chapter we saw how the E-field induced by the dust lifting phenomena is directly linked to the amount of entrained grains. The relation seems to be consistent for both dust storm and devil events. During the storms, we observed how E-field seems to play a feedback that can further increase the lifting. While, during the devils we found how the induced E-field correlates with the principal vortex parameters, as its pressure core drop and its rotatory speed. Another peculiar aspect that we observed is the presence of various night-time dust lifting events. The ones with duration of about one hour are identifiable as possible haboobs (microbursts), while the ones with a length of few seconds could be related to an unexpected nocturnal dust devil activity. Unfortunately, the obtained results regarding the dust devils are affected by high uncertainty, related to the impossibility to resolve the degeneration distance-radius of the acquired data set.
For this reason, we worked on the development of a new technique to evaluate the distance between the acquiring station and the passing whirlwind by analyzing the induced wind speed and direction variation, we will present this in par.4.2. The method has to be tested and verified on a data set for which it is already known the passage distances.
Therefore, the aforementioned points arisen from the previous campaigns that need further investigation have motivated the execution of a further field mission, where the monitoring of the meteorological parameters of the dust devils was coupled also with the observation of vortices morphological characteristics (radius and height) and impact parameters on the station.
With these purposes in mind, I personally led the planning and organization of a new Saharan campaign. The proposal of the mission has been submitted to the first call of the EuroPLANET 2020 research infrastructure, that finances planetary science projects receiving funding from the European Union's Horizon 2020 research and innovation programme. The mission has been approved and performed with the collaboration of the IBN Battuta centre, an EuroPLANET facility with headquarters in Marrakech (Morocco). The simplest way to evaluate the sizes and the impact parameter of the encountered dust events is by catching the images of their passages. Therefore, as mentioned in par.2.3.3, we planned to equip the station with a high acquisition rate camera. This was intended to acquire an image every 2 sec during the day, and to wake up during the night together with a nigh lighting system in the eventuality of passing dust phenomena.

The campaign foresaw a shorter duration than the 2014 one, with a reduced deployment of instruments, being the whole set up more similar to the 2013 mission (see par.2.3.3).

We arrived by flight to Marrakech, from there we have continued the journey by car until the south part of the Tafilalt region, near Merzouga and the boarder with Algeria (see Fig. 10). The site chosen for the measurements was few km away from the ones of the previous campaigns, along the dry bed of Ziz river, with a soil composition similar to the 2014 campaign (see Fig. 45).

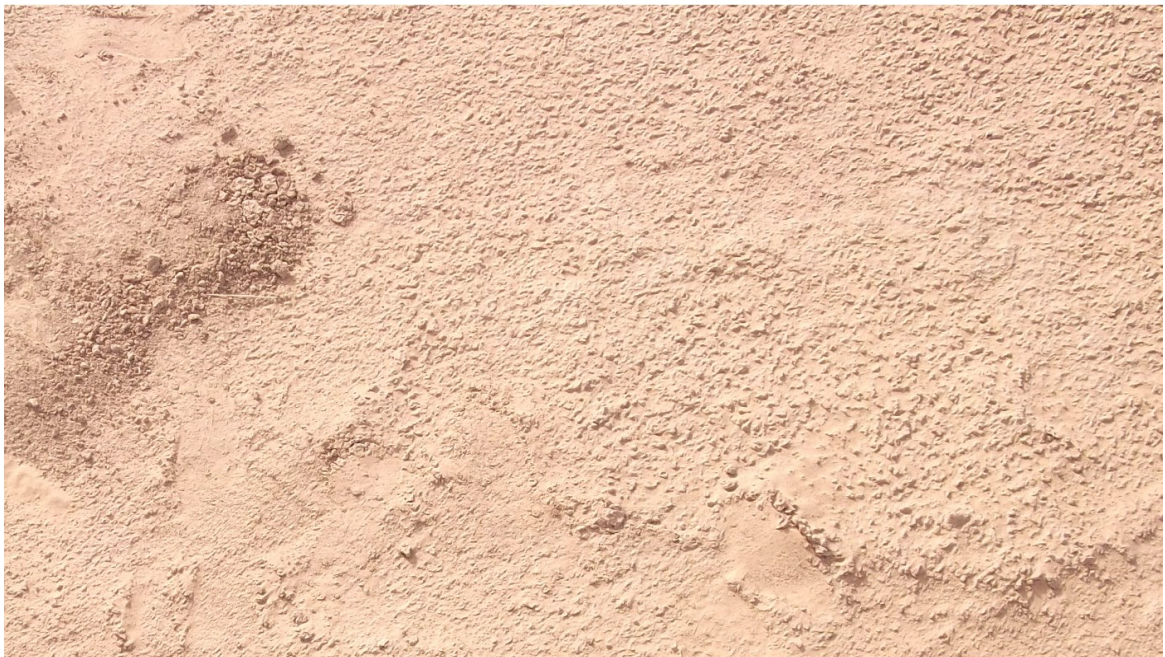

**Fig. 45** *The 2017 soil site*

Summarizing, the three principal aims of the survey were:
- the acquisition of synchronous measurements of dust devils induced E-field coupled with the images of their passages, in order to study the intrinsic variation of the E-field and its trend with the distance;
- the measurements of the horizontal wind speed signal of the vortices, in order to test our model for evaluation of the impact parameter, comparing the results with the ones obtained from the images;
- the test of the possible presence of night time dust devils by catching their images using the nigh flash system built on purpose for the camera.

Unfortunately, during the flight Italy-Morocco, the air company (Royal Air Maroc) managed to lose almost a quarter of our instrumentation. In particular, they lost two key pieces such as the electric field sensor and the camera, in addition to the most part of station mounting tools and part of instruments holders and cables. Both the missed instruments were fundamental to achieve our objectives.

After ten days, the air company has been able to recover the lost stuff, regrettably, not soon enough for the campaign. Therefore, we had to give up the planned station set up in favor of an emergency solution.

We tried to fix the lack of tools buying on field everything we manage to recover. Regarding the instruments, we replaced at least the camera with a GoPro camera. However, the GoPro was powered by its internal battery, providing a coverage of only four or five hours during the day. Moreover, the GoPro was not compatible with our night flash system, so we bought a second battery to power the lights, mounting them separately from the camera.

The issues with luggage caused a series of delays, reducing the days of measurement from eleven to seven. During the first day we tested the new set up, acquiring only meteorological data. For the other ones we acquired also diurnal camera images, managing during the last two days to find the best position for the new camera in order to evaluate the distances. Finally, for the last two nights we acquired also nocturnal images.

We had also to tune on field the data logger acquisition code for the new set up. This has led to a problem with the data acquisition that we recognized only after the end of the campaign and that had shifted the effective sampling rate from 1 to 2 Hz.

The final station set up is showed in Fig. 46, to sum up we mounted:
- three 2d anemometers at 0.7 m, 2.0 m and 3.35 m;
- three thermometers, one comprehensive also of the air humidity sensor at 1.70 m, a second one placed right under the surface and a last one at 0.1 m under the surface;
- a solar irradiance sensor at 0.3 m from the ground;
- an atmospheric pressure sensor at the soil level;
- one Sensit sensor placed at 2.70 m from the principal mast;
- a camera placed on a secondary mast at about 6 m from the first.

The whole station was powered by a battery recharged by the solar panel, only the camera and the night light system used separate batteries.

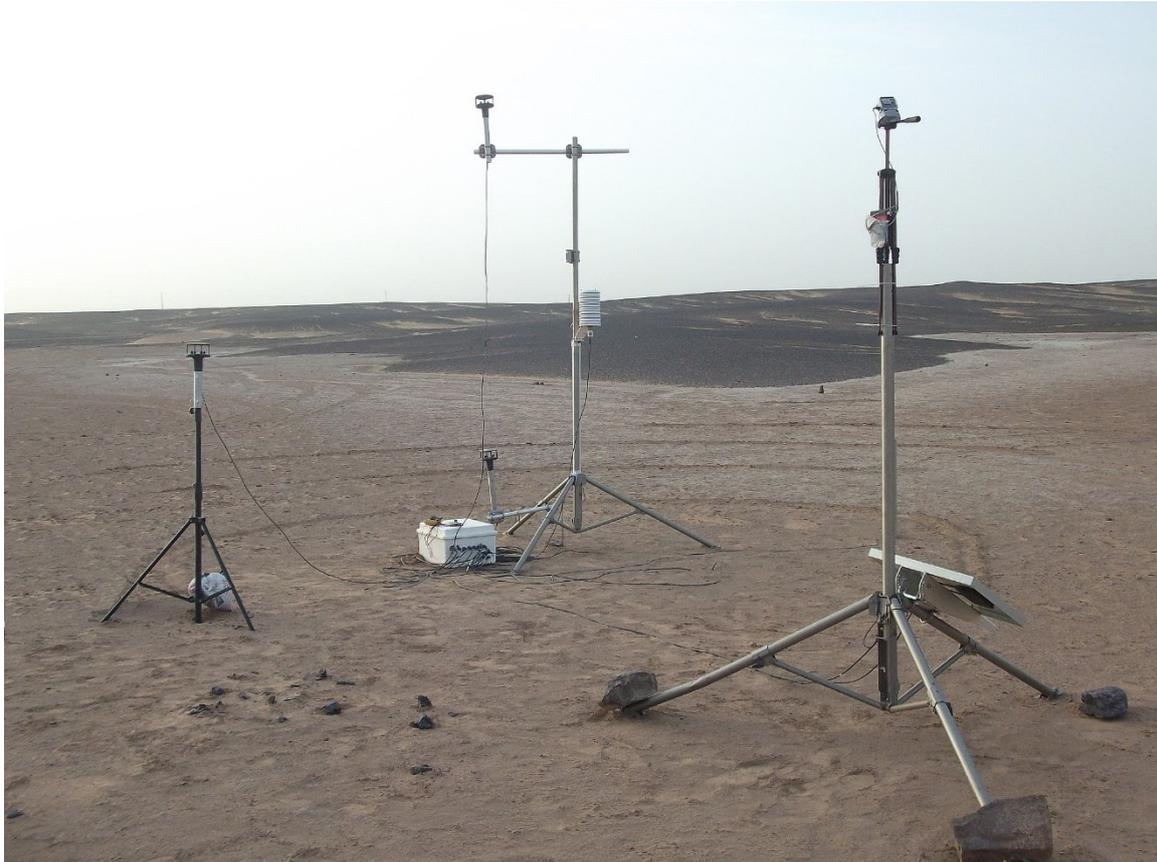

**Fig. 46 The station deployed during the 2017 campaign. On the principal mast there are two anemometers, a thermometer, a pressure sensor and a solar irradiance sensor. A third anemometer is placed on a separate tripod. The secondary mast houses the solar panel and the camera. Near the principal mast we placed other two thermometers, for the superficial and sub-superficial temperature, and a Sensit impact sensor.**

Overall, we could not achieve the first aim of the campaign due to the lack of the E-field sensor, and we acquired too few nigh-time data to proper study the last point. However, despite all the problems we faced, we managed to find a good set up to measure the distances and we observed a sufficient number of dust devils in order to achieve our second goal, i.e. the test of our model for the evaluation of the impact parameters.

## 4.2 Impact parameter evaluation model

As we saw in par. 2.7, monitoring the signal of one particular parameter $q(t)$, the most easily quantity to measure during the dust devils passage is the induced variation $\Delta q_{Max}$. This variation depends (see eq.51) both on the intrinsic vortex variation $q_o$ (the one that we would measure if the event passes exactly over the station) and on the impact parameter $d_o$ (the effective passage distance). Due to this dual dependence, we cannot separate the two factors using the measure of the signal variation, i.e. dust devils of different sizes passing at different distances could leave similar tracks in our data. Therefore, we need to found a measurable quantity related to the signal $q(t)$ that depends only on one of the two factors.

In order to do this, we started from the study of the wind speed signal dividing the different cases that could occur. Summarizing what we said in par.1.3.2 the rotary wind speed ($v_r$) inside the vortices can be reasonably described by the Rankine model:

$$v_r(d) = \begin{cases} V_r \dfrac{R}{d} & \text{if } d > R \\ V_r \dfrac{d}{R} & \text{if } d < R \end{cases} \quad \begin{array}{l}(54a)\\ b)\end{array}$$

The maximum value ($V_r$) of the rotary speed is reached at distance $R$ from the center of the vortex. $R$ is the radius of the vortex and the value of $v_r$ decreases below $V_r$ for distances larger and smaller than $R$. The parameters $V_r$ and $R$ are intrinsic of the dust devils, here we will consider a vortex fully formed in a stable state where these two parameters are fixed in time. The parameters $d$ and $v_r$ are instead dynamic and they depend on the point of observation and on the relative motion between the observer and the vortex.

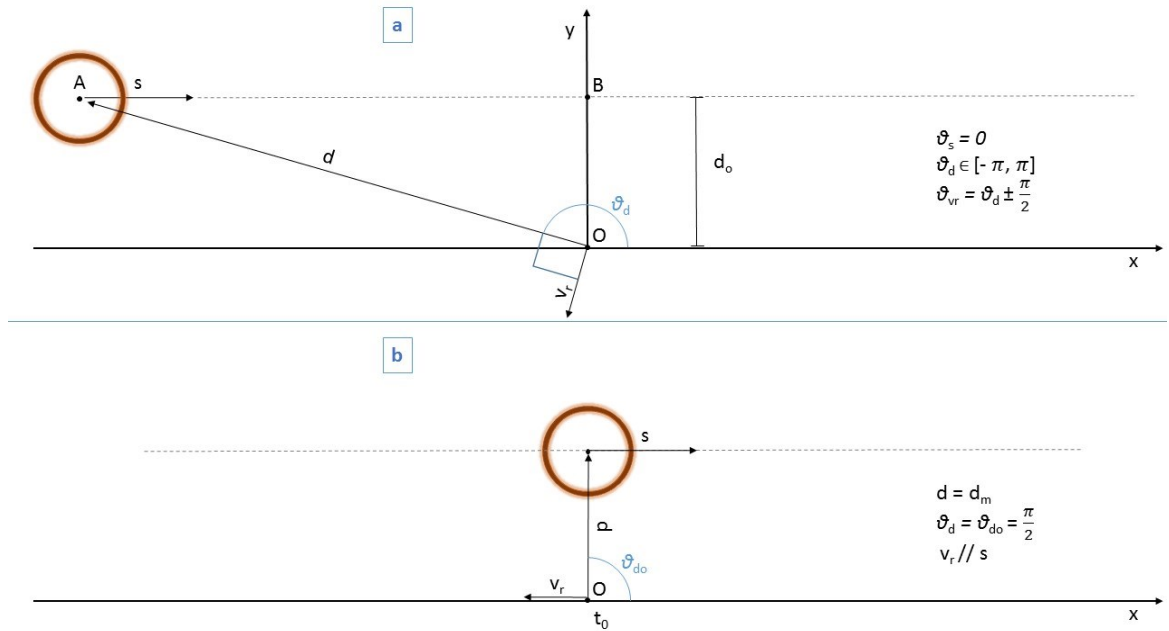

**Fig. 47** The frame that we use to study the passage of the dust devil. It is centered on the fixed meteorological station and it has the x-axis oriented in the direction of the vector *s*. Here, it is depicted the case of a dust devil that does not pass directly over the point O. a) shows the instant when the vortex is still far from the station. The rotational speed $v_r$ is always orthogonal to $d$, we have $\theta_{vr} = \theta_d \pm \frac{\pi}{2}$ depending on the sense of rotation of the vortex. In this case the dust devil rotates discordantly. b) shows the same dust devil in the instant $t_o$ of its minimum approach from the station.

Fig. 47 shows the frame that we use to study the dust devil passage near the fixed meteorological station. The station is placed in *O*, the center of the frame. The dust devil moves at speed ***s*** in the direction $\theta_s$ and ***d*** represents the vector position of the center of the vortex from *O*, while $d_o$ is the impact parameter, the minimum approach that the vortex reaches from *O* at the instant $t_o=0$. The time evolution of ***d*** is given by the equations:

$$\begin{cases} d_x = s\cos(\theta_s)\, t + d_{ox} \\ d_y = s\,\text{sen}(\theta_s)\, t + d_{oy} \end{cases} \quad (55)$$

the suffix x and y indicates the projection of the vector along the axis, the time $t\in[-\infty,+\infty]$. As showed in Fig. 47b), when the dust devil reaches the point of minimum approach, **d** and **s** are orthogonal, while **s** and **v**$_r$ are parallel or antiparallel depending on the sense of rotation of the vortex.

Let's us define the rotation as discordantly to the translational motion if, at the time $t_o$, **s** and **v**$_r$ are antiparallel; otherwise, if they are parallel the rotation will be concordantly. From the literature we know that there is not a generally preferred sense of rotation for the dust devils, for this reason we will consider both cases in our analysis. We have to take into account that the sense of rotation is another parameter not directly inferable using the measure of the signal variation.

We make the simplifying assumption that the dust devil is totally advected by the wind background, this means that **s** coincides with background speed **b** (**b**≡**s**). Hence, the wind speed measured in $O$ is the sum of **b** and of the value of the vortex rotary wind speed in $O$, we call this sum "total wind speed" ($v_t$). Supposing that **b** does not evolve in time, we have that the dust devil moves along a straight line and that $v_t$ depends on its distance from $O$:

$$\mathbf{v}_t(d) = \mathbf{s} + \mathbf{v}_r(d) \tag{56}$$

Here, to avoid to overload the notation, we have indicated **v**$_r$ as a function of only d, considering for now only the geometric parameters. However, even if not made explicit, we have to take into account that there is still the dependence from $R$ (eq.(54)).

Decomposing $v_t$ the along $x$ and $y$, we have that:

$$v_{tx}(d) = s\,cos(\theta_s) + v_r(d)\,cos(\theta_{vr}) \tag{57}$$

$$v_{ty}(d) = s\,sen(\theta_s) + v_r(d)\,sen(\theta_{vr}) \tag{58}$$

where $\theta_{vr}$ is the angle between the *x*-axis and the velocity **v**$_r$.

The direction $\theta_{vt}$ of the vector $v_t$ will be simply:

$$\theta_{vt}(d) = \text{arctg}\left(\frac{v_{ty}(d)}{v_{tx}(d)}\right) \tag{59}$$

Let us now derive an expression that will be useful in the next paragraphs. Considering the triangle OAB of Fig. 47a) we have:

$$d_m = d\,\cos(\theta_d - \theta_{do}) \tag{60}$$

and hence:

$$\cos(\theta_d - \theta_{dm}) = \frac{d_m}{d} \tag{61}$$

$$sen(\theta_d - \theta_{dm}) = \sqrt{1 - \left(\frac{d_m}{d}\right)^2} \tag{62}$$

To simplify the equations, without losing any generality, we can rotate the frame in order to have the x-axis parallel to the direction $\theta_s$. We recall for simplicity also the eq.(48) that connect the distance to the time:

$$d^2 = s^2\,t^2 + d_o^{\,2} \tag{63}$$

In the following section we will study separately the two cases of a dust devil rotating concordantly or discordantly using the here described frame.

## 4.2.1 Concordant Rotation

The vector **d** and **$v_r$** are perpendicular in every instant t and in case of concordant rotation we have:

$$\theta_{vr} = \theta_d - \frac{\pi}{2} \tag{64}$$

We want to focus on the axial components of the measured speed $v_{tx}$ and $v_{ty}$. Taking into account that in our frame $\theta_{dm}$ equals to $\frac{\pi}{2}$ (see Fig. 47b), we can substitute the equations (61) and (64) in eq(57):

$$v_{tx}(d) = s + v_r(d)\frac{d_m}{d} \tag{65}$$

and the equations (62) and (64) in eq. (58):

$$v_{ty}(d) = v_r(d)\sqrt{1 - \left(\frac{d_m}{d}\right)^2} \tag{66}$$

Let's us subdivide the analysis in other two cases: the case of direct passage, where the dust devils pass directly over the meteorological station ($d_o \leq R$) and the case of not-direct passage ($d_o > R$).

We define also the quantity $w$ as the maximum value of the rotational speed measured in O. In the case of not-direct passage we have that $w = v_r(d_m) = V_r \frac{R}{d_m}$, while, in case of direct passage, $w$ simply coincides with $V_r$.

### *4.2.1.1 Not-direct passage*

We start from the case of a dust devils that passes outside the meteorological station ($d_m > R$). Substituting the eq.(1a) in eq.(65) and eq.(66), the functions $v_{tx}(d)$ and $v_{ty}(d)$ became:

$$v_{tx}(d) = s + V_r\frac{R\, d_o}{d^2} \tag{67}$$

$$v_{ty}(d) = V_r\frac{R}{d}\sqrt{1 - \left(\frac{d_o}{d}\right)^2} \tag{68}$$

$v_{tx}(d)$ has a single maximum for $d = d_o$, which corresponds to the instant $t = t_o$. Instead, $v_{ty}(d)$ has four stationary point located at $d = \pm\infty$ and $d = \pm\sqrt{2}\,d_o$. Changing variable from d to t using eq.(63), we have that the maximum and the minimum of $v_{ty}$ are located in a specular way to the center, at $t = \pm\frac{d_o}{s}$. This means that the position of the maximum and of the minimum does not depends on the intrinsic parameters of the vortex (its radius $R$ and vortex speed $V_r$) and that these points will be located at the same instants for every dust devils that translates at speed $s$ passing with the same impact parameter $d_o$.

This is the principal point of our methods, i.e. we have found a quantity measurable from the acquired signal that does not depend on the intrinsic vortex characteristics, in order to resolve the degeneration radius-distance.

As for $v_{ty}$, also $\theta_{vt}(d)$ (eq.(59)) has 4 stationary point: $d = \pm\infty$ and $d = \pm\frac{d_o\sqrt{2s+w}}{\sqrt{s}}$, which correspond to the instants $t = \pm\infty$ and $t = \pm\frac{d_o\sqrt{s+w}}{\sqrt{s^3}}$. Hence, the position of the maximum and minimum of $\theta_{vt}(t)$ depends on the intrinsic vortex parameters, $V_r$ and $R$, through the variable $w$. The points at $t = \pm\infty$ are not interesting for our analysis and we will avoid to consider them from now on.

Fig. 48 shows the trend of $\mathbf{v}_t(t)$, $\mathbf{v}_{tx}(t)$, $\mathbf{v}_{ty}(t)$ and of the direction $\theta_{vt}(t)$ for a dust devil of radius $R = 3$ m, that travels parallel to the x-axis with a speed $s = 5$ m/s ,rotating concordantly with $V_r = 25$ m/s , passing at distance $d_o = 4.5$ m from the station. As the dust devil approach the station, $v_t(t)$ growths from the background value $s$ to its maximum value $s + w$ in $t_o$. The same trend is visible also for the component $v_{tx}(t)$, but with a shorter width of the signal. Instead, as we said, $v_{ty}(t)$ and $\theta_{vt}(t)$ have two stationary points and Fig. 48c) clearly shows how the maximum and minimum have the same position also for a dust devil of different radius and intrinsic vortex wind speed. Moreover, in case of concordant rotation, we can also notice how both for $v_{ty}(t)$ and $\theta_{vt}(t)$ the maximum precedes the minimum.

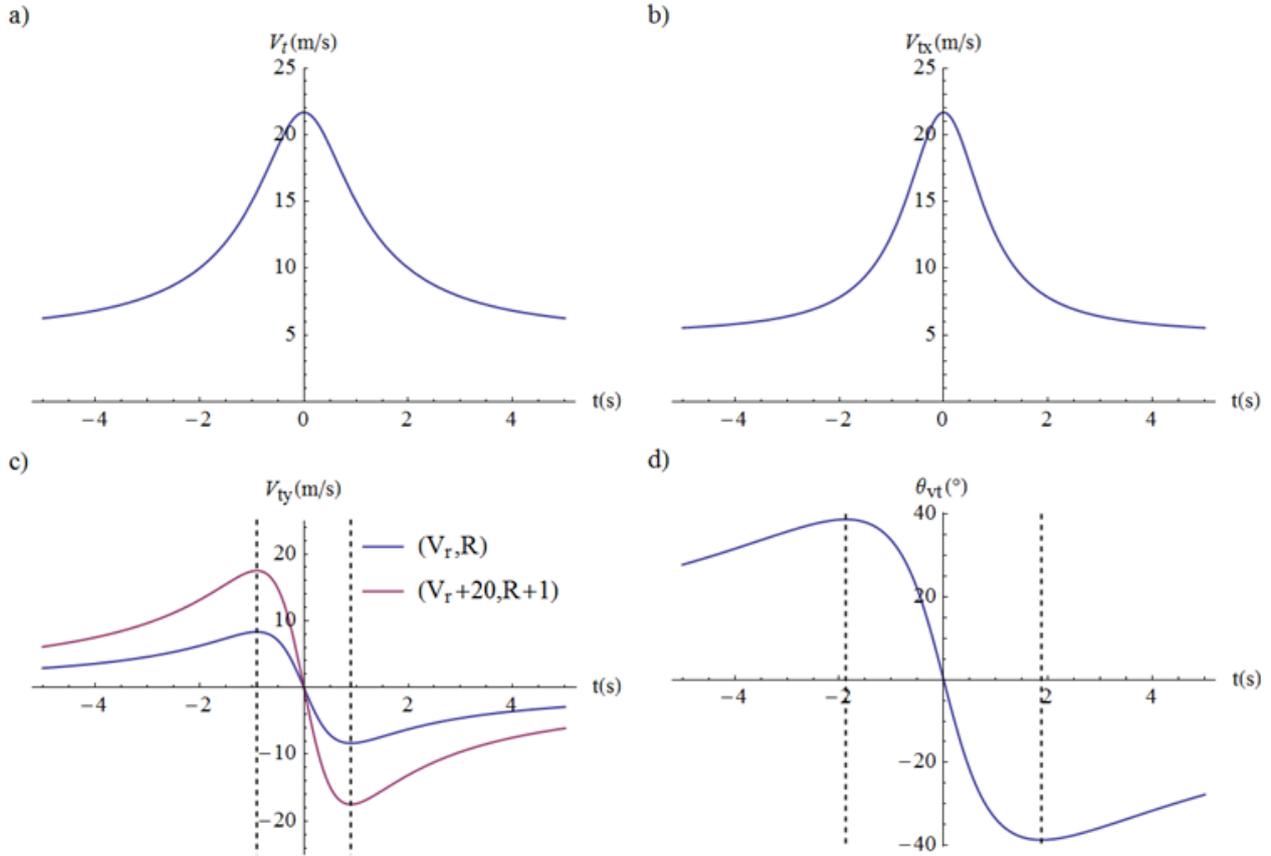

**Fig. 48** *The time trend of the wind velocity measured by the fixed meteorological station during a dust devil encounter concordantly rotating. The dust devil depicted in blue has the following parameters: s = 5 m/s, $\theta_s$ = 0, $V_r$ = 25 m/s, R = 3 m, $d_o$ = 4.5 m . The four plots are:*
*a) the total wind speed $v_t$; b) the x-axis component of $v_t$; c) the y-axis component of $v_t$; d) the direction of $v_t$. In c) and d) the dashed lines indicate the position of the maximum and of the minimum. In c) we also show how these points are located in the same position for dust devils of different intrinsic rotational speed and diameter. Indeed, the dust devil in red rotates 20 m/s faster and it has 1m bigger radius than the blue one.*

### 4.2.1.2 Direct passage

The dust devils starts to pass over the station when $|d| = R$, this happens for instants $t \in [-\frac{\sqrt{R^2-d_o^2}}{s}, +\frac{\sqrt{R^2-d_o^2}}{s}]$. In this interval, substituting eq.(1b) in eq.(65) and eq.(66) we have:

$$v_{tx}(d) = s + V_r \frac{d_o}{R} \tag{69}$$

$$v_{ty}(d) = V_r \frac{d}{R} \sqrt{1 - \left(\frac{d_o}{d}\right)^2} \tag{70}$$

This means that during the passage of the dust devil over the station the quantity $v_{tx}$ remains constant. Instead, $v_{ty}$ has still a maximum and a minimum, located at $d = \pm\sqrt{2}\,d_o$ ($t = \pm\frac{d_o}{s}$) if $R < \sqrt{2}d_o$, and at $d = \pm R$ ($t = \pm\frac{\sqrt{R^2-d_o^2}}{s}$) if $R > \sqrt{2}d_o$.

Hence, when $R < \sqrt{2}d_o$ we have still that the position of the points do not depend on the intrinsic vortex parameters.

The total velocity $v_t(d)$ has not a single peak in $d = d_m$ ($t = t_o$), like in the previous case, but two peaks at $d = \pm R$ ($t = \pm\frac{\sqrt{R^2-d_o^2}}{s}$). Regarding the direction $\theta_{vt}(d)$, it has two stationary point at $d = \pm\frac{d_o\sqrt{2s+w}}{\sqrt{s}}$ ($t = \pm\frac{d_o\sqrt{s+w}}{\sqrt{s^2}}$) if $R < \frac{d_o\sqrt{2s+w}}{\sqrt{s}}$, while, the stationary points will be located at $d = \pm R$ ($t = \pm\frac{\sqrt{R^2-d_o^2}}{s}$) if $R > \frac{d_o\sqrt{2s+w}}{\sqrt{s}}$. Fig. 49 shows the function $v_t(t)$ and its components and direction for the same dust devil of Fig. 48 passing at $d_m = 1\,m$ from the station.

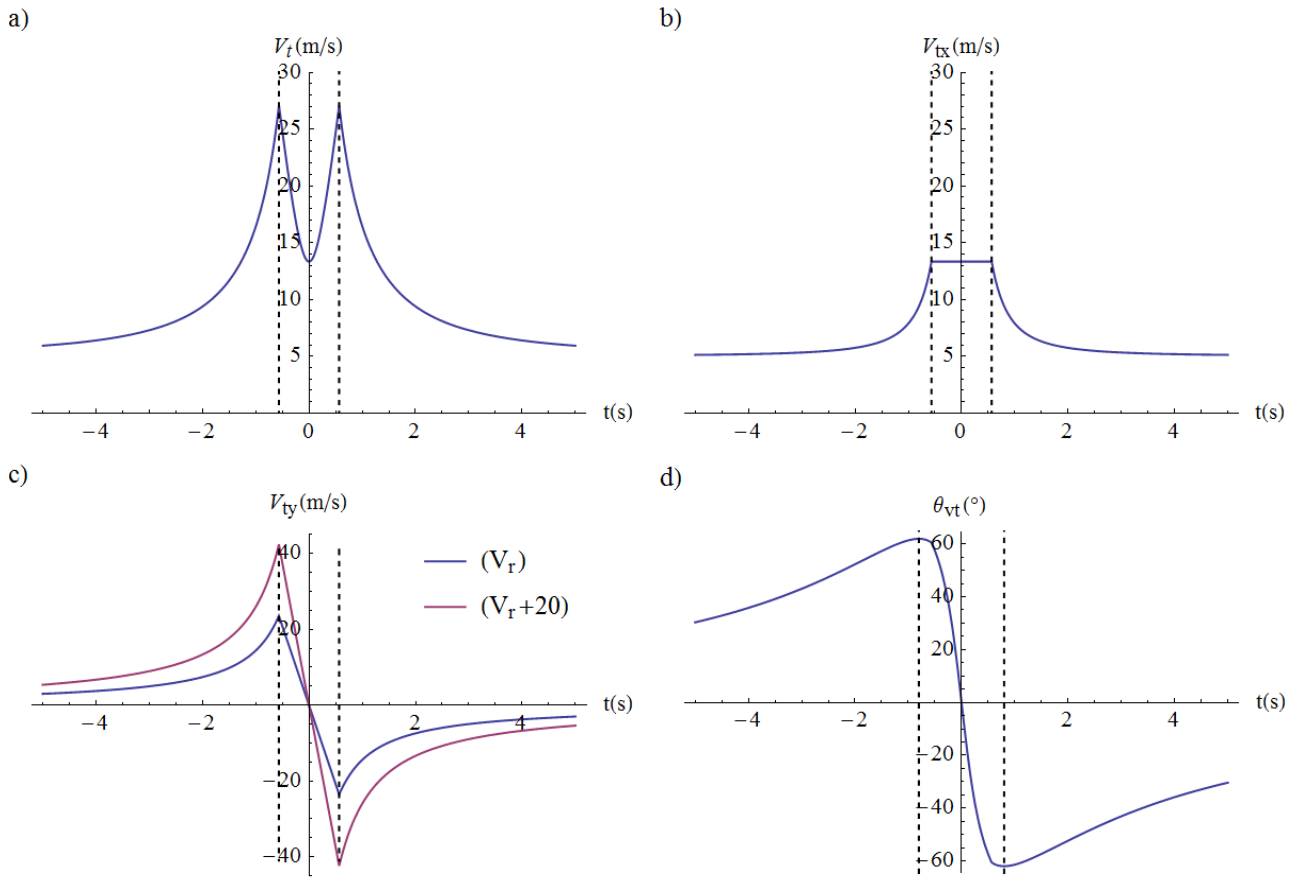

Fig. 49 *The trend in time of the wind velocity measured by the fixed meteorological station during a dust devil encounter in the case of concordant rotation. The dust devil depicted in blue has the following parameters: s = 5 m/s , θ$_s$ = 0, V$_r$ = 25 m/s, R = 3 m , d$_o$ = 1 m. The four plots are:*
*a) the total wind speed v$_t$; b) the x-axis component of v$_t$; c) the y-axis component of v$_t$; d) the direction of v$_t$. The dashed lines indicate the position of the maximum and of the minimum of the functions. In c) we also show*

## 4.2.2 Discordant Rotation

In case of discordant rotation, we have that:

$$\theta_{vr} = \theta_d + \frac{\pi}{2} \tag{71}$$

Adding half period to the function sin and cos and taking into account that $\theta_{do} = \frac{\pi}{2}$ : we obtain:

$$\cos\theta_{vr} = -\cos\left(\theta_d + \frac{\pi}{2} - \pi\right) = -\cos(\theta_d - \theta_{do}) \tag{72}$$

$$sen\,\theta_{vr} = -sen\left(\theta_d + \frac{\pi}{2} - \pi\right) = -sen(\theta_d - \theta_{do}) \tag{73}$$

Hence, substituting the equations (61) and (72) in eq.(57) and the equations (62) and (73) in eq.(58), we obtain:

$$\mathrm{v}_{tx}(d) = s - \mathrm{v}_r(d)\frac{d_o}{d} \tag{74}$$

$$\mathrm{v}_{ty}(d) = -\mathrm{v}_r(d)\sqrt{1 - \left(\frac{d_o}{d}\right)^2} \tag{75}$$

### 4.2.2.1 Not-direct passage

The trend of $\mathrm{v}_{ty}(d)$ is simply opposite than the one of concordantly rotation (see eq.(66)). So, its stationary points have the same magnitude and position in $d = \pm\sqrt{2}\,d_o$, which correspond to $t = \pm\frac{d_o}{s}$. For the function $\theta_{vt}(d)$ the maximum and minimum are now placed at $d = \pm\frac{d_o\sqrt{|2s-w|}}{\sqrt{s}}$, which correspond to $t = \pm\frac{d_o\sqrt{|s-w|}}{\sqrt{s^3}}$. Fig. 50 shows the trend of $\mathrm{v}_t(t)$ , $\mathrm{v}_{tx}(t)$ , $\mathrm{v}_{ty}(t)$ and of the direction $\theta_{vt}(t)$ for the same dust devil of Fig. 48, but, in the case of discordant rotation. The function $\mathrm{v}_t(t)$ reaches a smaller maximum value, because now s and v$_r$ are antiparallel in t$_o$. For this reason and for the fact that we chose $V_r>s$ (as it usually happens), the vector $\mathrm{v}_{tx}(t)$ has a direction opposite to *s* near the point $t_o$, hence it appears negative in this representation. Looking at the trends of $\mathrm{v}_{ty}(t)$ and $\theta_{vt}(t)$, the most recognizable difference from the concordant case is that now the minimum precedes the maximum.
Hence we can simply deduct the sense of vortex rotation looking at the chronological sequence of the stationary points.

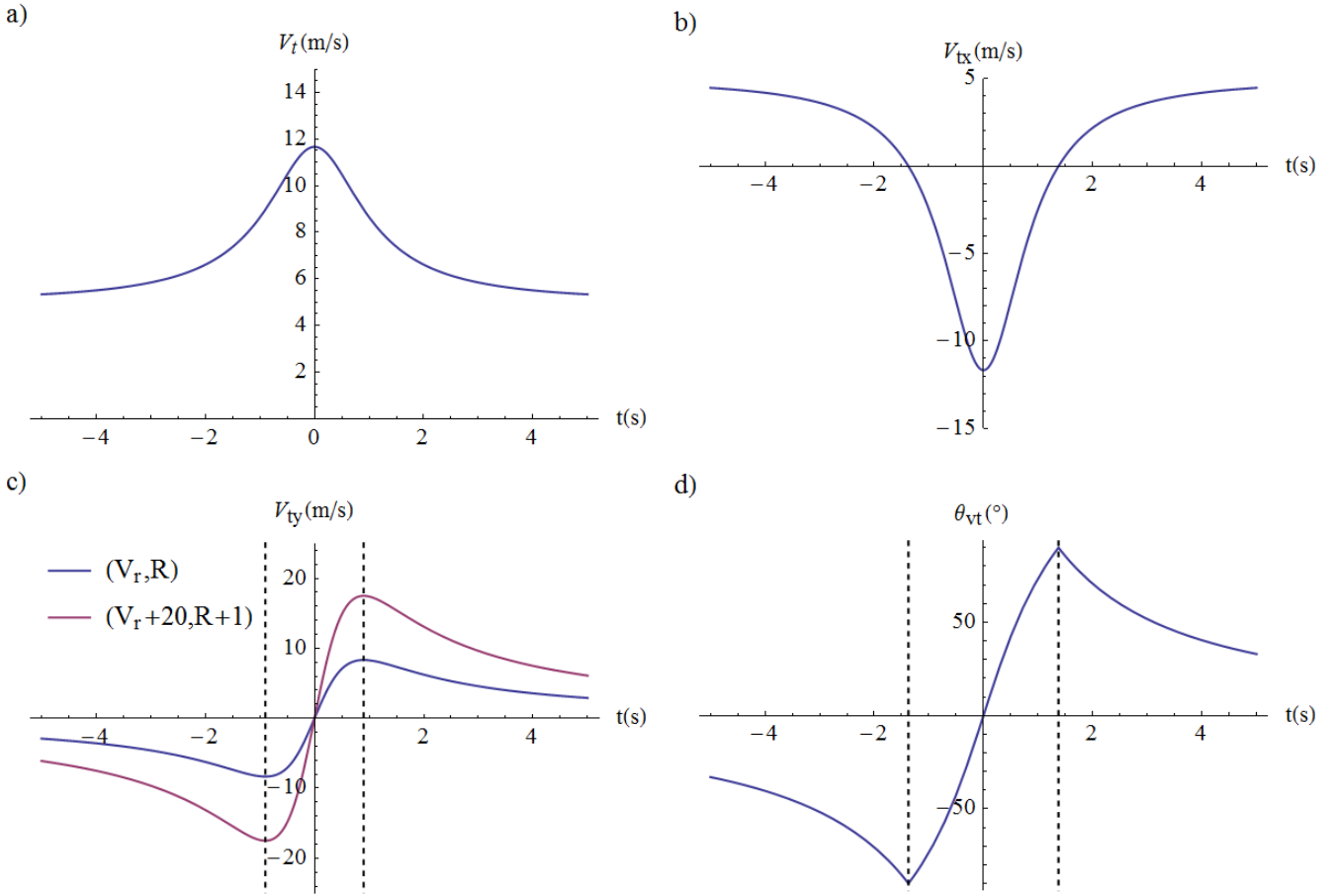

**Fig. 50** *The trend in time of the wind velocity measured by the fixed meteorological station during a dust devil encounter in the case of discordant rotation. The dust devil depicted in blue has the same parameters of the one of Fig. 48: s = 5 m , $\theta_s$ = 0, $V_r$ = 25 m/s, R = 3 m , $d_o$ = 4.5 m . The four plots are:*
*a) the total wind speed $v_t$; b) the x-axis component of $v_t$; c) the y-axis component of $v_t$; d) the direction of $v_t$.*
*In c) and d) the dashed lines indicate the position of the maximum and of the minimum and in c) we also show how these points are located in the same position for dust devils of different intrinsic rotational speed and diameter. Indeed, the dust devil in red rotates 20 m/s faster and it is 1m bigger in radius than the one in blue.*

#### 4.2.2.2 Direct passage

In order to obtain the trends of $v_{tx}(d)$ and $v_{ty}(d)$ during the vortex passage over the station, we have to substitute eq.(1b) in eq.(74) and eq.(75):

$$v_{tx}(d) = s - V_r \frac{d_o}{R} \tag{76}$$

$$v_{ty}(d) = -V_r \frac{d}{R}\sqrt{1 - \left(\frac{d_o}{d}\right)^2} \tag{77}$$

Like the case of concordant rotation the function $v_t(d)$ have two peaks at $d = \pm R$ ($t = \pm \frac{\sqrt{R^2 - d_o^2}}{s}$), while $v_{tx}$ remains constant during the passage. The maximum and minimum of $v_{ty}$ will be located at $d = \pm \sqrt{2}\, d_o$ ($t = \pm \frac{d_o}{s}$) if $R < \sqrt{2} d_o$, and at $d = \pm R$ ($t = \pm \frac{\sqrt{R^2 - d_o^2}}{s}$) if $R > \sqrt{2} d_o$. Instead, the maximum and the minimum of $\theta_{vt}(d)$ will be located at $d = \pm \frac{d_o \sqrt{|2s + w|}}{\sqrt{s}}$ ($t = \pm \frac{d_o \sqrt{|s + w|}}{\sqrt{s^2}}$) if $R < \frac{d_o \sqrt{|2s + w|}}{\sqrt{s}}$, and at $d = \pm R$ ($t = \pm \frac{\sqrt{R^2 - d_o^2}}{s}$) if $R > \frac{d_o \sqrt{|2s + w|}}{\sqrt{s}}$. Fig. 51 shows the function $v_t(t)$ and its components and direction for the same dust devil of Fig. 48 passing at $d_o = 1\ m$ from the station.

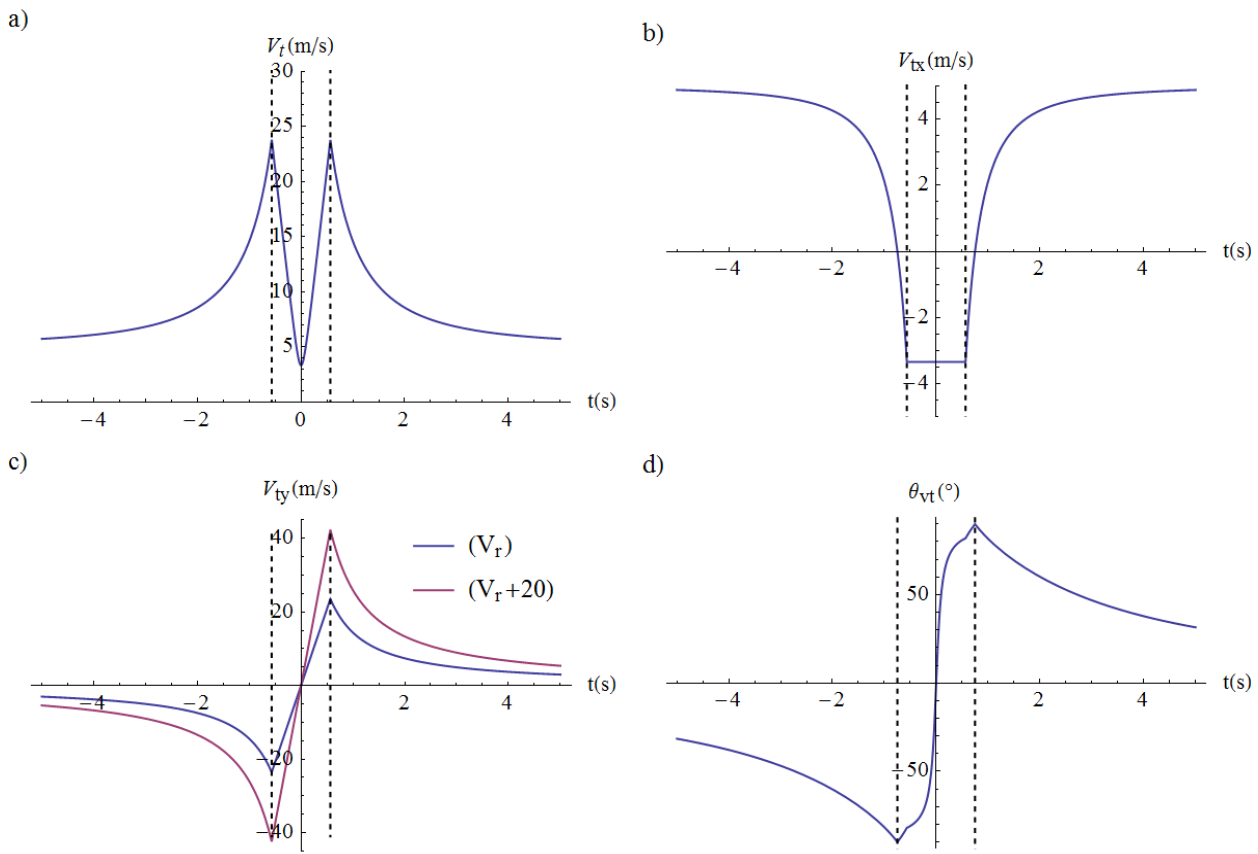

**Fig. 51** *The time-trend of the measured wind velocity during the encounter of a discordantly rotating dust devil that passes over the fixed meteorological station. The dust devil depicted in blue has the following parameters:*
*$s = 5\ m$, $\theta_s = 0$, $V_r = 25\ m/s$, $R = 3\ m$, $d_o = 1\ m$. The four plots are:*
*a) the total wind speed $v_t$; b) the x-axis component of $v_t$; c) the y-axis component of $v_t$; d) the direction of $v_t$.*
*The dashed lines indicate the position of the maximum and of the minimum of the functions. In c) we also show how the position of the maximum and minimum is the same for dust devils of different intrinsic rotational speed (20 m/s faster rotation).*

### 4.2.3 Reconstruction of the vortex intrinsic parameters

Now that we have found a way to evaluate the impact parameters $d_o$ of the dust devils encounters, we can also try to estimate the vortex intrinsic parameters.
From the measure of the observed pressure drop $\Delta P$ and the value of $d_o$ we can reconstruct the value of the intrinsic pressure core drop $\Delta P_o$ (eq.(2)).
Taking into account the equation of the cyclostrophic balance eq.(3), and the equation of Rankine vortex rotation model eq.(1) we obtain a system in the variables $V_r$ and $R$:

$$\begin{cases} \dfrac{dP_o}{\bar{P}} \bar{T} = \dfrac{V_r^2}{R} \\ v_r(d_o) = V_r \dfrac{R}{d_o} \end{cases} \qquad (78)$$

Taking into account that $v_r(d_o)$ is measurable from the time signal as they are the environmental temperature $\bar{T}$ and pressure $\bar{P}$, we can resolve the system to calculate the intrinsic rotatory speed $V_r$ and the radius $R$ of the dust devil.
Depending on which quantity is the easiest to measure, we can substitute the second equation of the system with the value of $v_y$ in the stationary points:

$$v_y(\sqrt{2}d_o) = \frac{1}{2} V_r \frac{R}{d_o} \qquad (79)$$

In any cases, once the impact parameter is known, we can evaluate the full set of vortex intrinsic characteristics $\Delta P_o$, $V_r$ and $R$ from the acquired pressure and wind speed signatures.

### 4.2.4 Application of the model

In this paragraph, we are going to take stock on the things we saw from the model and how we applied it to the real data.
On field, the most common situation is the passage of the dust devil outside the station ($d_m > R$). As we have seen, in this case we are able to connect directly the impact parameter $d_m$ and the translational velocity $s$ to the position of the stationary point of the function $v_{ty}(t)$. In addition, the degeneration due to the sense of rotation of the vortex does not affect the position of these points.
In this study, we used the median value of $v_t(t)$ and $\theta_{vt}(t)$, evaluated on a time interval of thirty minutes around the vortex encounter, to estimate the background wind speed $b$ and direction $\theta_b$. The trajectory of the dust devil can differ from a straight line, approaching sometimes a cycloidal motion with a speed $s$ that can moderately differ from the background wind velocity b (Lorenz, 2013). However, usually the total duration of the recorded encounter is of the order of few dozens of seconds (Franzese et al., 2018). Therefore, for a so short time interval, the approximation of a straight line motion at speed $s = b$ and $\theta_s = \theta_b$ is reasonable.

Knowing $\theta_b$, we rotated the acquired speed data to have $\theta_s = 0$ in the new frame and we decomposed the velocity along the two axis.

We have studied the trend of $v_{ty}(t)$ and evaluated the total time distance $\Delta t$ between the maximum and the minimum position, $\Delta t = 2\frac{d_o}{s}$. Therefore, from the measure of $\Delta t$ and $s$ we easily deduced the impact parameter $d_o$.

A similar way to perform the analysis is to study $\theta_{vt}$ instead $v_{ty}$. We can start identifying the vortex rotation sense from the trend of $\theta_{vt}$ and measuring the magnitude of $w$ from the relation $w = v_t(t_o) \pm s$, where the $\pm$ depends on the rotation sense. Then we can study the stationary points and measure their time distance $\Delta t_\theta = 2\frac{d_o\sqrt{|s-w|}}{\sqrt{s^3}}$ to evaluate $d_o$. This method introduces a slightly greater uncertainty in the evaluation of $d_o$ due to the necessity to evaluate also the quantity $w$, so we based our analysis on the study of $v_{ty}(t)$.

We have evaluated all the quantities using the data acquired by the anemometer placed at 3.35 m. In order to validate our technique, we compared the impact parameters estimated with the ones obtained by the acquired images. In the next paragraph, we briefly describe the method used to analyze the images.

## 4.3 Camera images distances evaluation

In order to test the dust devil distances resulting by analyzing the wind speed and direction data, we evaluated the distances using also the acquired images. The procedure adopted for this purpose is briefly described in this paragraph.

We marked out on the soil a circle and various arches of 3, 6, 9, 12 and 18 m of radius, centered on the position of our station. In addition, we placed three recognizable group of rocks at 21, 27 and 34 meters from the station. The entire set-up of our marked points (circle arches and rocks) is showed in Fig. 52.

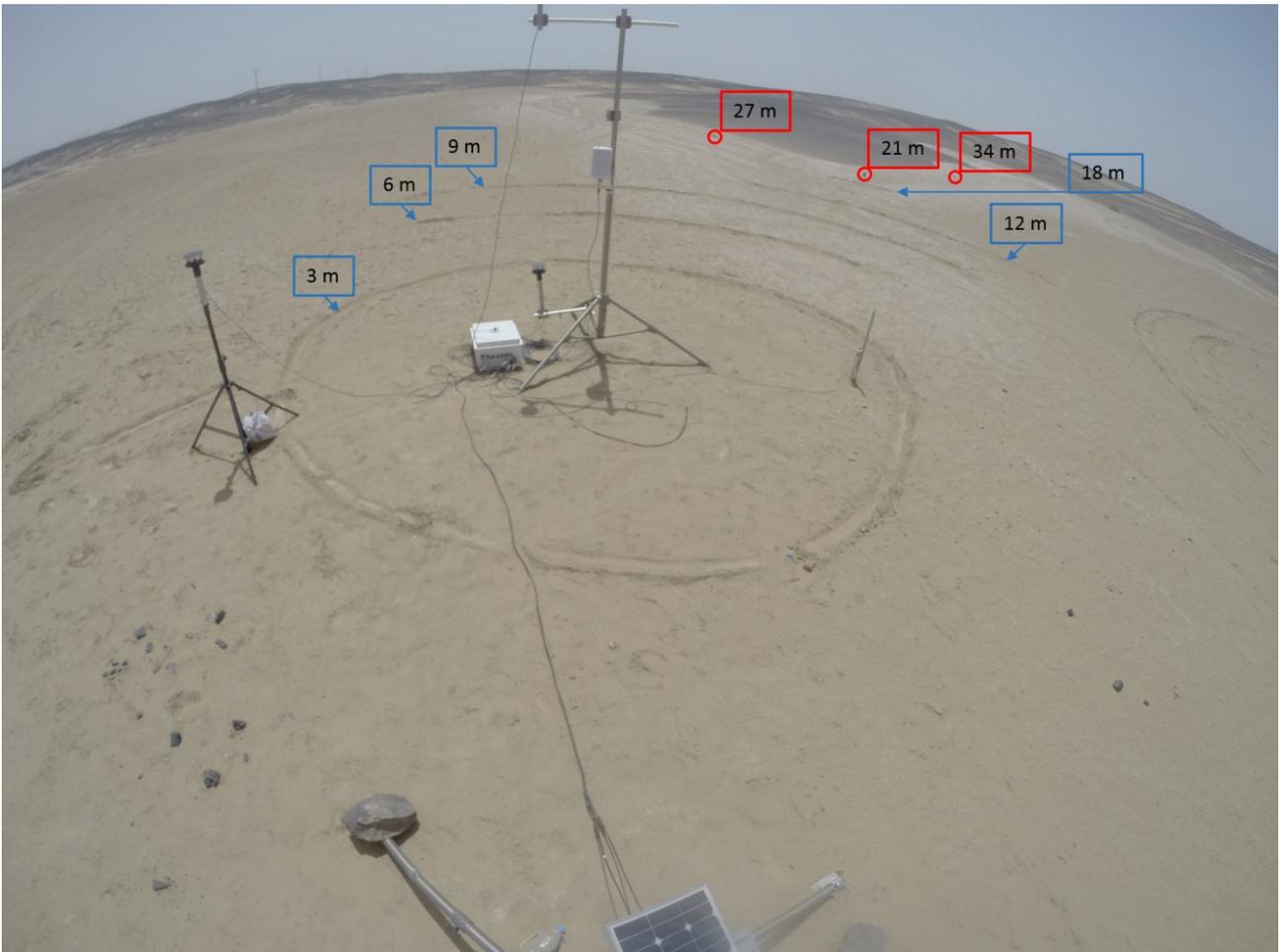

**Fig. 52** *In order to reconstruct the distances from the images, we drew a series of circle arches (blue marks) and we placed three groups of rocks (red marks) at know distances from the station. This image has been acquired by the mounted camera at noon on July 24th.*

We started defining a first 2D orthogonal coordinate system *Oxy* centered on the meteorological station that represent the bird's-eye projection of our site. Then, for each image we corrected the fish-eye effect due to the grand angle camera lens and defined a second orthogonal coordinate system *O'x'y'* on the pictures, centered on their left lower edge (Fig. 53). This system represents the projection of the site view as seen by our camera. We want to find the transformation of coordinates that maps *Oxy* into *O'x'y'*.

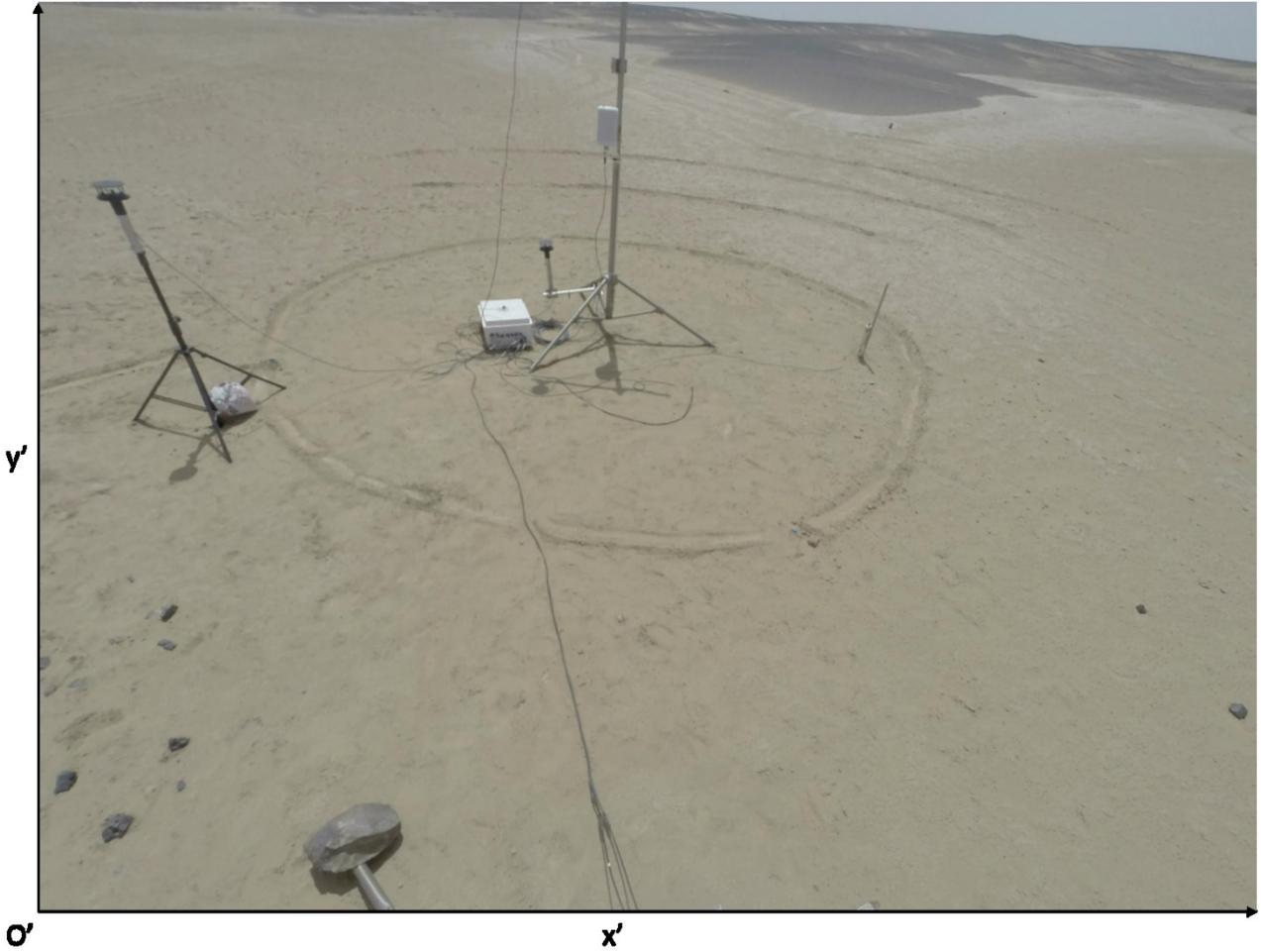

**Fig. 53** *The same picture of Fig. 52 after the correction of the fish-eye effect, it is shown also the coordinate system O'x'y' that we used to map the image pixels.*

To simplify the calculation, we resized our images to 1024x768 pixels, therefore, $x' \in [0,1024]$, while, $y' \in [0,768]$. The projection $Oxy \rightarrow O'x'y'$ is a linear fractional transformation, also known as Möbius transformation or bilinear transformation. It is the composition of scaling, translation, rotation and skew. For a two-dimensional system the transformation matrix has the form:

$$M = \begin{pmatrix} a_{11} & a_{12} & b_1 \\ a_{21} & a_{22} & b_2 \\ c_1 & c_2 & d \end{pmatrix} \quad (80)$$

that correspond to the map $T(x,y)$:

$$\begin{cases} x' = T_x(x,y) = \dfrac{a_{11}\, x + a_{12}\, y + b_1}{c_1\, x + c_2\, y + d} \\ y' = T_y(x,y) = \dfrac{a_{22}\, y + a_{21}\, x + b_2}{c_1\, x + c_2\, y + d} \end{cases} \quad (81)$$

Hence, we have to find the 9 parameters $(a_{ij}, b_i, c_i, d)$ of the transformation. Considering the high number of parameters and the fact that we have still no constraint on their magnitude and range of variation, we decided to preliminary search for a reasonable approximation of their values, before starting any kind of fitting procedure on the whole set of marked points. We performed the whole images processing using a routine that we developed using Wolfram Mathematica.

### 4.3.1 Evaluation of the starting values

We selected the marked points relative to the circle of radius 3 m, fitting an ellipse on them. In order to simplify the calculation we performed the fit in polar coordinates, then, we returned to the Cartesian coordinate $O'x'y'$ writing the equation of the ellipse in the form:
$$A_o x'^2 + B_o x'y' + C_o y'^2 + D_o x' + E_o y' + F_o = 0 \tag{82}$$
The values $A_o, B_o, C_o, D_o, E_o, F_o$ are the results of this preliminary fit.
The equation of the corresponding curve in the system $Oxy$ is simply:
$$x^2 + y^2 - R^2 = 0 \tag{83}$$
If we apply the transformation $T$ to the eq.(82), we have to obtain again the form of eq.(83). Hence, substituting the map of eq.(81) into the eq.(82) and collecting the factors of $(x^2, y^2, xy, x, y)$ we obtain the form:
$$A x^2 + B xy + C y^2 + D x + E y + F = 0 \tag{84}$$
Where $A,B,C,D,E,F$ are coefficients that depend only on the parameters $(a_{ij}, b_i, c_i, d)$ of the transformation $T$ and from the coefficients $A_o, B_o, C_o, D_o, E_o, F_o$ found with the previous fit.
By equating the coefficients in eq.(84) to the ones of eq.(83), we obtain the system of equations:
$$\begin{cases} A = 1 \\ B = 0 \\ C = 1 \\ D = 0 \\ E = 0 \\ F = -R^2 \end{cases} \tag{85}$$
Using the pictures showed in Fig. 53, taken at noon, and the shadow of the station, we can estimate the station center position $(x'_c, y'_c)$ in the system $O'x'y'$. This point is the transformed of the origin of $Oxy$, then from the eq. (81):
$$\begin{cases} x'_c = T_x(0,0) = \dfrac{b_1}{d} \\ y'_c = T_y(0,0) = \dfrac{b_2}{d} \end{cases} \tag{86}$$
Joining this last two equations to the previous ones we found a system of 8 equations in the 9 parameters $(a_{ij}, b_i, c_i, d)$. We decided to solve the system using the simplifying assumption of an equal scaling of the axes x and y: $a_{11} = a_{22}$. We call $(a_{ij}{}^V, b_i{}^V, c_i{}^V, d^V)$ the solutions obtained in this way. We will use them as starting points for the fitting procedure described in the next paragraph. Fig. 54 shows the preliminary image mapping of parameters ($a_{ij}^V, b_i^V, c_i^V, d^V$), as it was easy to predict, it still does not represent a good fit for the points farther than 3 meters from the station.

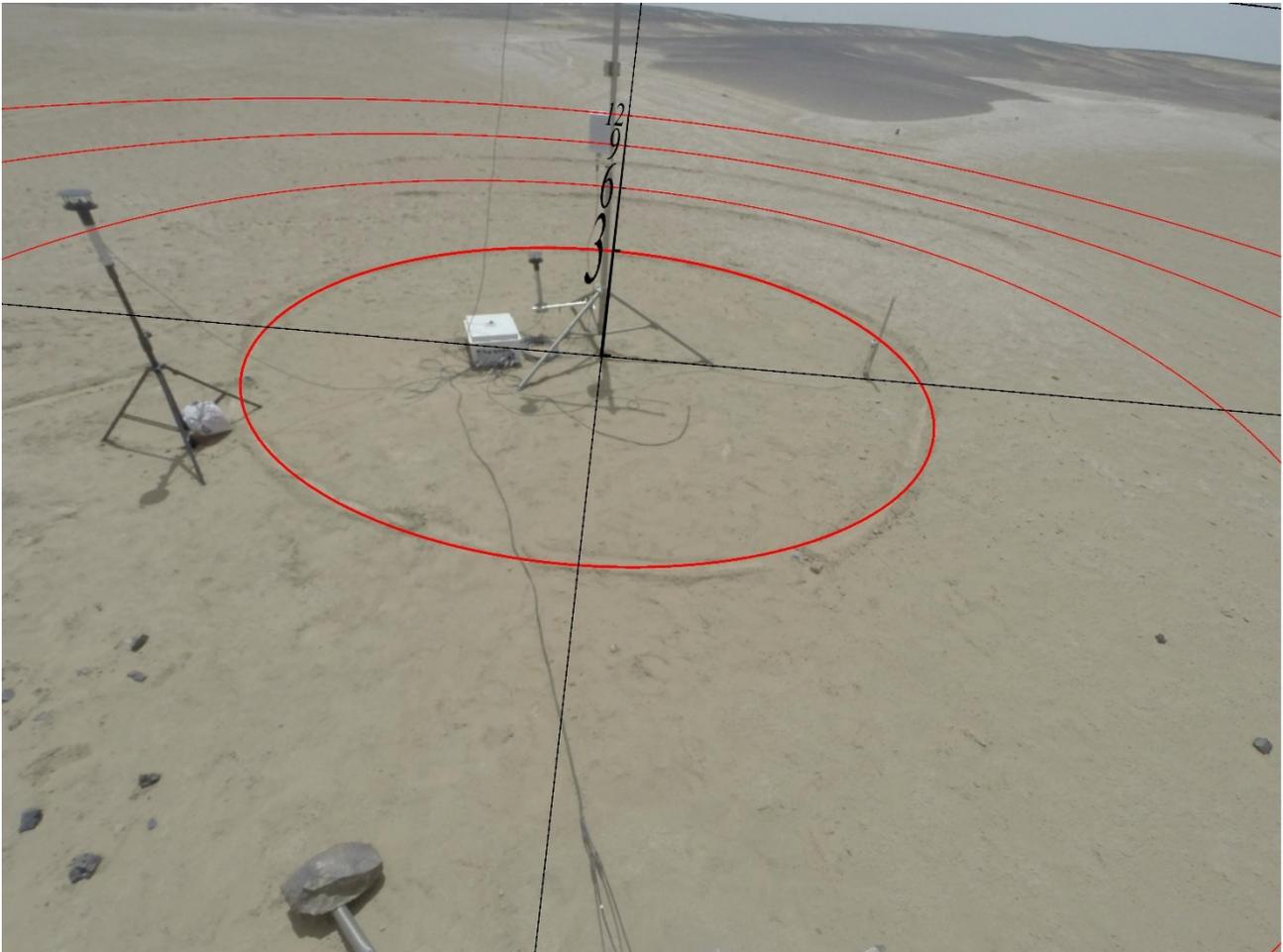

**Fig. 54** *The preliminary result of our image mapping. In red we plotted the circles of radius of 3,6,9 and 12 meters. In black we plotted the axes x and y relative to the original system of coordinates Oxy. In the left corner of the figure it is possible to notice the horizon line of our model. It is evident how the marked points farther than 3 meters and the horizon are not well fitted from this map.*

### 4.3.2 Fitting procedure

For each marked point we know the distance from the station $D_M$, $D_M = \sqrt{x_M^2 + y_M^2}$, but we do not know their original coordinates $(x_M, y_M)$ in Oxy. We can write $D_M$ in function of the coordinates in O'x'y': $D_M = \sqrt{T_x^{-1}(x_M', y_M')^2 + T_y^{-1}(x_M', y_M')^2}$. Making explicit the inverse transformation $T^{-1}$ we obtain the relation:

$$D(x', y') = \sqrt{\frac{(a_{21}b_1 - a_{11}b_2 + (b_2c_1 - a_{21}d)x' + (-b_1c_1 + a_{11}d)y')^2 + (-a_{22}b_1 + a_{12}b_2 + (-b_2c_2 + a_{22}d)x' + (b_1c_2 - a_{12}d)y')^2}{(-a_{12}a_{21} + a_{11}a_{22} + (-a_{22}c_1 + a_{21}c_2)x' + (a_{12}c_1 - a_{11}c_2)y')^2}} \quad (87)$$

We fitted this function on the whole ensemble of marked points showed in Fig. 55, performing a least-squares estimation using the Levenberg-Marquardt algorithm (Levenberg, 1944; Marquardt, 1963), choosing the values $(a_{ij}{}^V, b_i{}^V, c_i{}^V, d^V)$ previously selected as starting points for the transformation parameters. Here, we relaxed the assumption $a_{11} = a_{22}$.

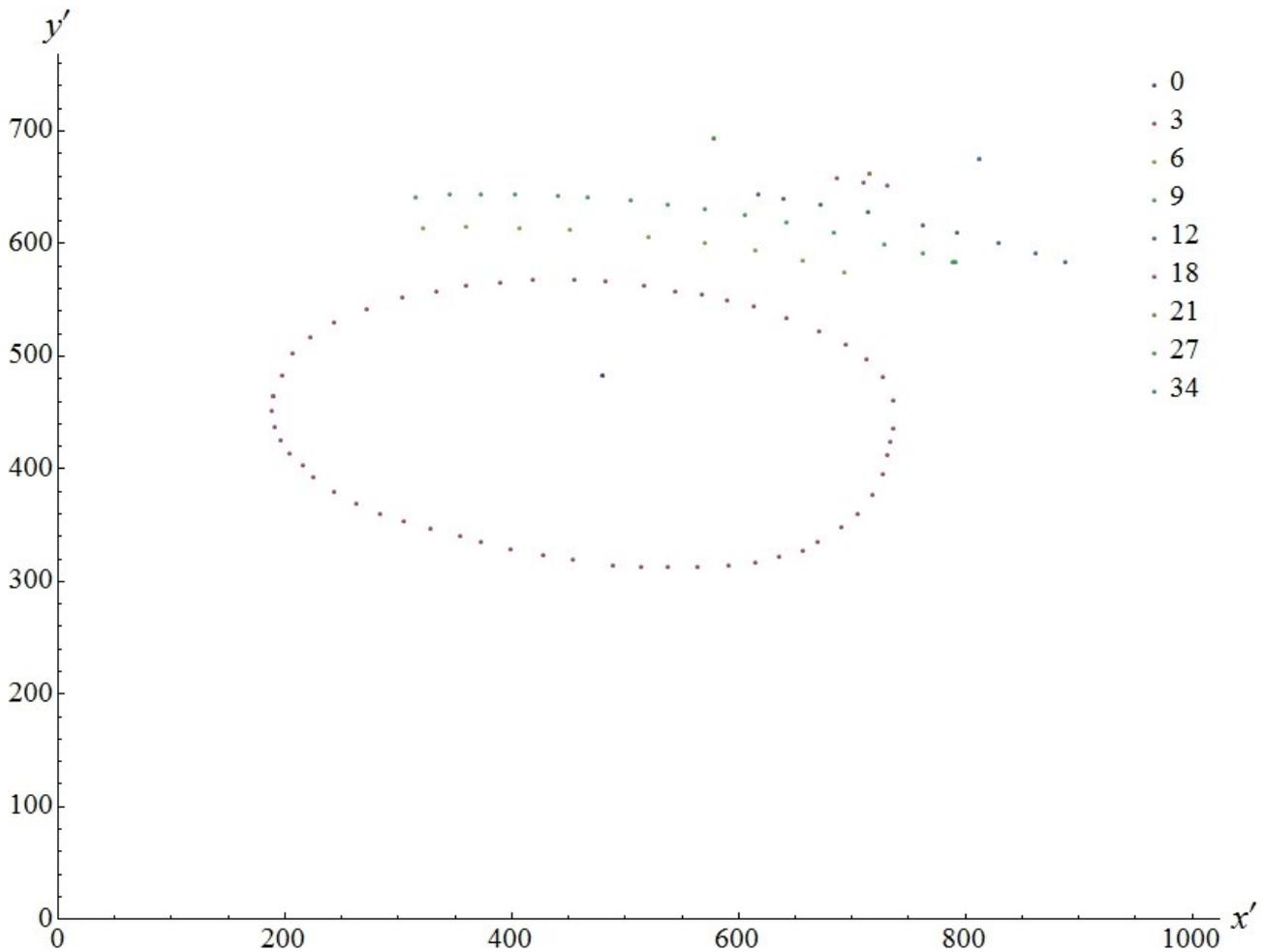

**Fig. 55** *The whole set of marked points used for the map fitting. Different colors indicate points at different distance (in meters) from the station. The unit of measure on the axis are in pixels.*

Fig. 56 shows our best fit model ($R^2 = 0.9994$), the agreement of the map with the picture is clearly evident.

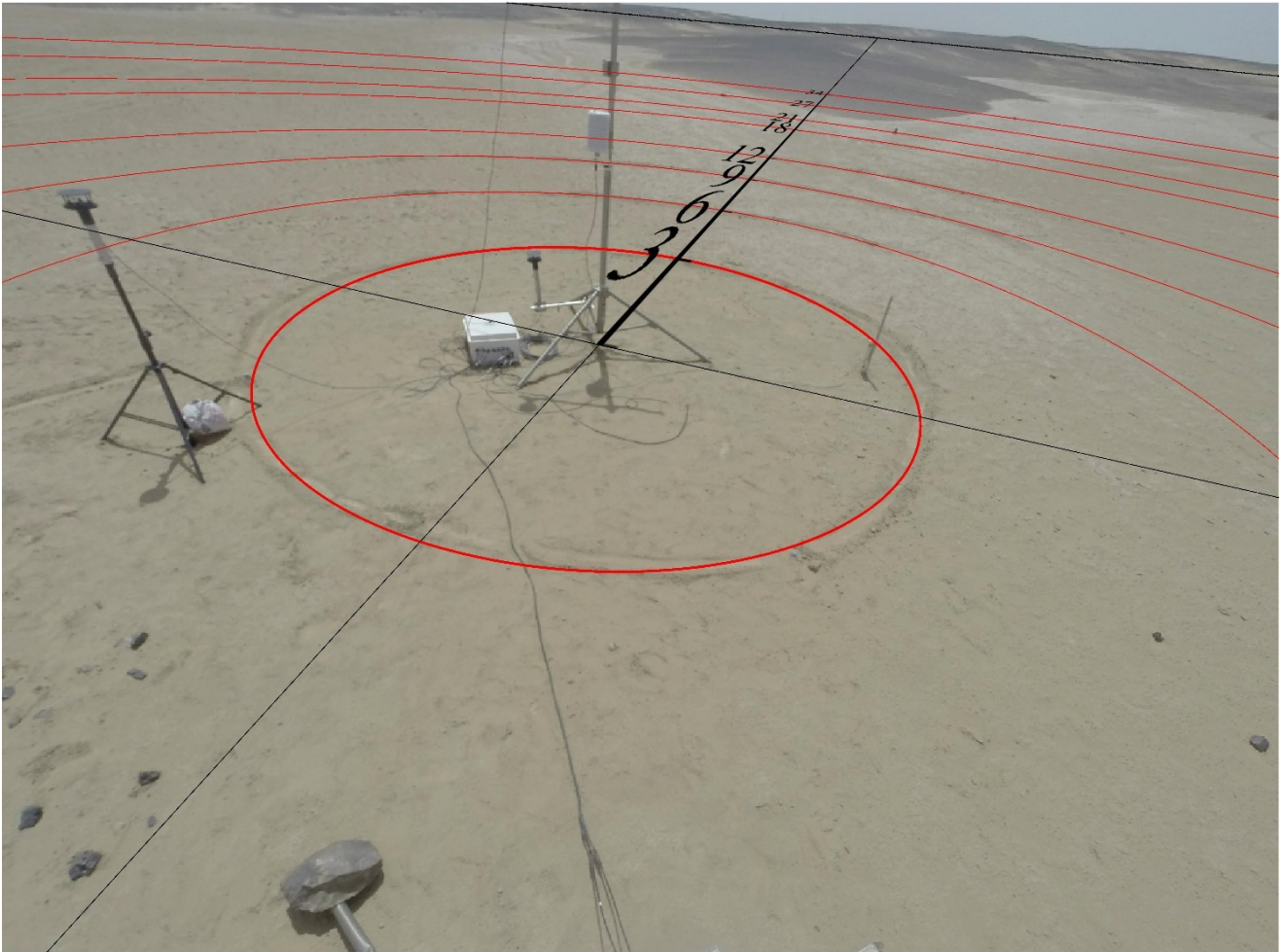

*Fig. 56 The result of our mapping of the horizontal plane on the images taken by the mounted camera. In red we plotted the circles passing for each marked points (radius of 3,6,9,12,18,21,27 and 34 meters as shown). In black we plotted the axes x and y relative to the original system of coordinates Oxy and the evaluated horizon line. Each marked point and the horizon are well fitted from this map.*

We repeated the fitting procedure by varying the starting points around the values $(a_{ij}^V, b_i^V, c_i^V, d^V)$, in order to confirm that the results remain consistent, e.g Fig. 57 shows two of the obtained results. Our mapping technique does not impose constrains on the direction of the axis *x,y* of *Oxy*, because we are only interested in distances. Hence, changing the starting parameters of the fit it is possible to obtain different systems *Oxy* with the axis rotated. However, in all cases we have tested, the distances from the system center remained always compatible within the uncertainties.

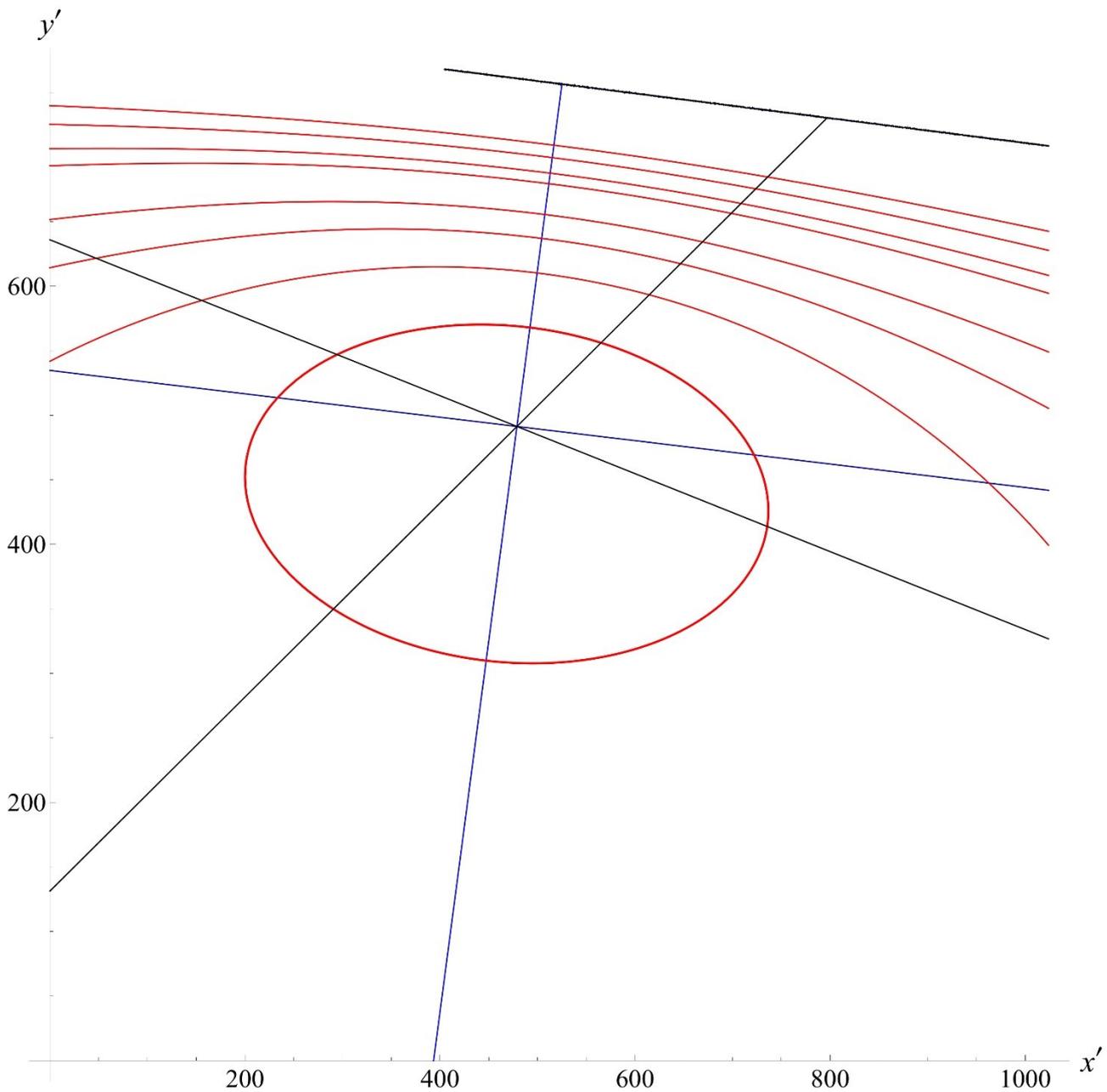

**Fig. 57** *Two results of the map fitting obtained by starting from two different initial points. The axis of the two systems obtained are plotted in black and blue, while we plotted in both cases the circles of radius 3,6,9,12,18,21,27,34 meter in red, and they result coincidence, as the horizon lines.*

By knowing the map $T(x,y)$, we can simply estimate the passage distance $d$ of the dust devils from our station. We have to measure the position $(x'_D, y'_D)$ of the dust devil center from the images in the system $O'x'y'$ and reconstruct its coordinate $(x_D, y_D)$ in the system $Oxy$ by using the relations: $x_D = T_x^{-1}(x'_D, y'_D)$ and $y_D = T_y^{-1}(x'_D, y'_D)$. Finally, the distance will be simply: $d = x_D^2 + y_D^2$.

## 4.4 Comparison of the techniques

We have not acquired the E-field measurements, therefore, we could not directly use the detection method developed during the 2014 campaign. Despite an optimization of the detection technique to work only on pressure and wind data does not represent a big issue, we decided to follow another way. Since we had to analyze only few days of data and not a long campaign, and we are limited in any case to the only events captured by the camera, we decided to check directly the acquired images, looking for the passing dust devils.
The individuation of the whirlwinds in the images can result tricky because they are clearly visible only against a highly contrast area. Both the surface and the sky don't represent a good background.
The site was partially surrounded by black low hills, mainly covered by pebbles that originate their color. These hills represent the optimal background, so we tried various placement of the camera, variating the altitude and the orientation to maximize the portion of image covered by hills. The best set up for dust devils and distances individuation has been reached in the last two days of acquisition, therefore we have focused on these data.
Fig. 58 shows the closest approach image of one of the vortices encounters.
For the individuated events we studied the acquire wind data, evaluating the background wind speed and direction on a time interval of 30 minutes around the center of the encounter. We used these values to evaluate the vortex travel speed $s$ and rotate our frame to obtain $\theta_s = 0$. In this way we can evaluate the axial velocity components $v_x$, $v_y$. Fig. 59 shows the wind speed signal $v_t$ acquired during one of the dust devils passages. The Fig shows also the wind direction already normalized around $\theta_b = 0$ and the two retrieved component of the speed along $x$ and $y$ axis.

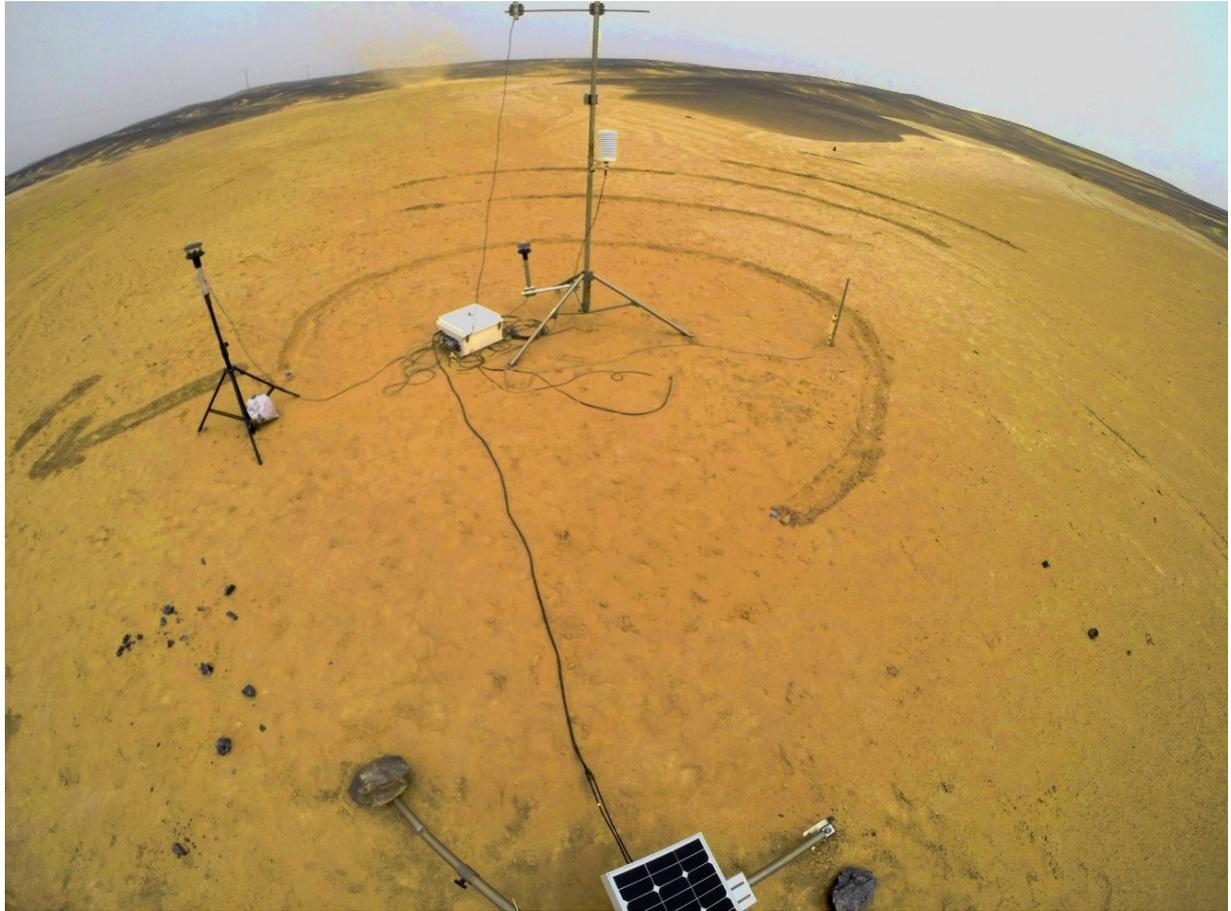
**Fig. 58** *The passage of a dust devil caught by our camera at July 23 around 3 p.m.. The vortex is particularly clear against the black hills background; the colors have been stretched to make it visible also against the sky background.*

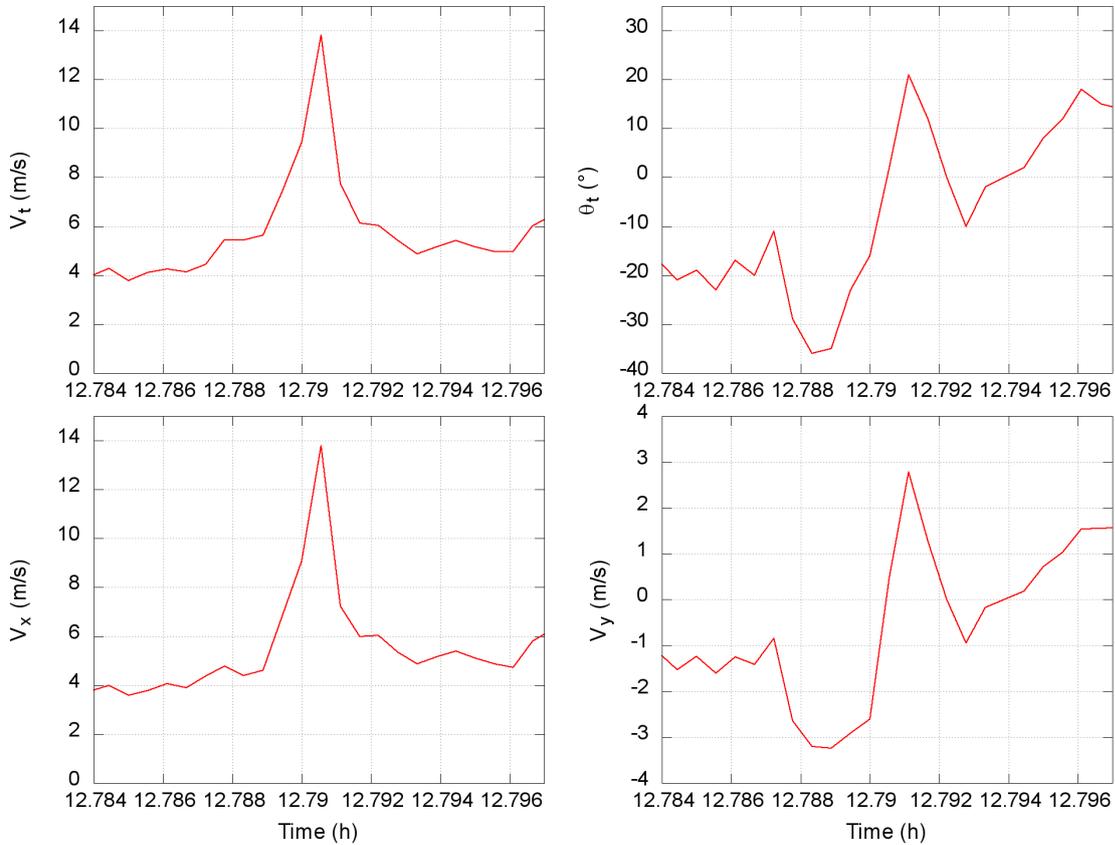

**Fig. 59** *The wind speed signal acquired during a dust devil passage at July 23$^{th}$ around 12:40 p.m. . The first plot represents the total measured wind speed $v_t$, the second one is its direction $\theta_t$, while the third and the fourth one are the two axial component $v_x$ and $v_y$ of the speed, along x and y respectively.*

We monitored the passage of eight dust devils evaluating their impact parameter with both techniques. Taking into account that the measurement rate is 2 Hz, the main contributor to the uncertain of the evaluated impact parameter is given by the distance travelled by the vortex between two consecutive acquisition. Hence, we have estimated the uncertain level using the measured value of *s* (the vortex travelling speed). The only way to reduce this error is to increase the acquisition rate. Table 6 reports the obtained results for the whole dust devils set, the uncertain on the distance evaluated by the wind signal analysis is a factor of two larger than the one obtained from the images. Because we have to evaluate two points from the signal analysis: the maximum and minimum positions.

| date | $t_i$ | $d_{o\ wind}$ | $d_{o\ images}$ |
|---|---|---|---|
| 23_07_2017 | 13.76 | 53 ± 10 | 57 ± 5 |
| 23_07_2017 | 14.67 | 54 ± 8 | 47 ± 4 |
| 23_07_2017 | 14.72 | 56 ± 8 | 62 ± 4 |
| 24_07_2017 | 10.38 | 30 ± 10 | 29 ± 5 |
| 24_07_2017 | 10.545 | 50 ± 10 | 45 ± 5 |
| 24_07_2017 | 11.145 | 74 ± 10 | 75 ± 5 |
| 24_07_2017 | 11.3 | 156 ± 10 | 149 ± 5 |
| 24_07_2017 | 11.857 | 4 ± 8 | 3 ± 4 |

**Table 6** *The impact parameters $d_o$ obtained from the wind signal analysis and from the images*

For comparison, the results are plotted one vs the other in Fig. 60. It is possible to see how all the points are compatible with the bisector inside the uncertainties. Therefore, despite the large error bars caused by the not optimal acquisition rate, the agreement of the results is evident.

The low acquisition rate and the short campaign length have also prevented the study of the case of dust devils passing over the station. This eventuality is indeed quite rare and even if we have observed an event with a $d_o$ compatible with 0, the vortex passed too fast to catch the double peak trend of the signal $v_t$.

Overall, the technique is obviously limited also by the quality of the acquired signal: a high noise level caused both by the internal instrument noise or by a high flow turbulence level can mask the trend of $v_y$ preventing the individuation of the stationary points.

This is especially true for small vortices passing far from the station, however our method seems to be suitable also for dust devils with very high impact parameter. Indeed, the farther whirlwind observed passed at around 150 m from the station, with a diameter that we estimated from images around 30 m.

In conclusion, the method has proven to be very useful to study a dust devils data set acquired by a fixed meteorological station, being able to resolve the degeneration size/distance that affect the measurements and that heavily affect the analysis.

Indeed, the technique retrieves not only the impact parameter of the encounters, but it also estimates the intrinsic vortex parameters, such as its radius and rotatory wind speed at vortex wall $V_r$. Moreover, by knowing these parameters it is possible to directly estimate the pressure core drop $\Delta P_o$ from the acquired pressure signal. Hence, we have found a method able to fully characterize the passing events which greatly enhance our possibilities to study the physics of dust devils.

As a next step, we intend to use this method to re-analyze the data acquired in the 2013 and 2014 campaigns, in order to retrieve the intrinsic parameters of the vortices improving the achieved results. Moreover, we are planning another Saharan campaign to complete the study of those features that we were not able to investigate during the 2017, due to the various issues we faced.

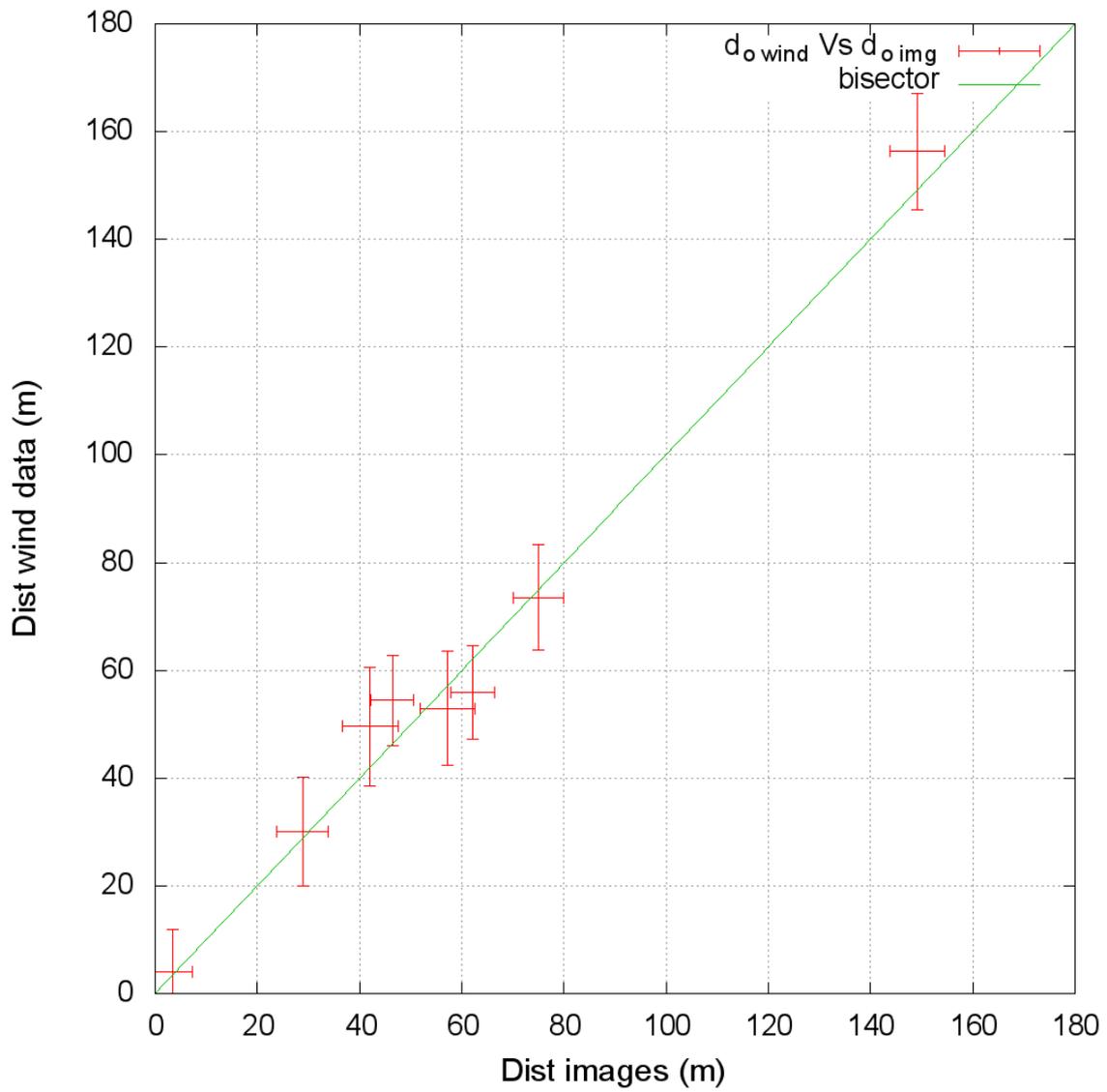

**Fig. 60** *The impact parameter results from the images on x axis, compared with the ones obtained by the study of the wind speed signal. All points are compatible within error bars with the bisector, proving the agreement of the two methods.*

# Conclusion and Future Works

The main topic of my PhD research has been the analysis of the dust lifting phenomena observed during three field campaigns developed in the Sahara desert, in the Moroccan region of Tafilalt, on the border with Algeria. The campaigns have been performed in the frame of the development and test of the DREAMS and MicroMED instruments, which are part of the scientific payload of the ExoMars missions.

Among the goals of the ExoMars programme, those related to this work are:
- the investigation of the possible presence of an atmospheric E-field at the martian surface; this was initially planned through the measurements of the MicroARES sensor included in the DREAMS station of the ExoMars 2016 mission, and now postponed to 2020 mission through the measurements of the Dust Complex on board of the ExoMars Surface Platform;
- the direct measurement of the concentration and size distribution of the lifted dust, this will be achieved by the MicroMED particle optical counter included in the Dust Complex on board of the ExoMars 2020 lander.

In preparation to the analysis of the Martian data related to dust events, we have deployed a meteorological station in the Sahara desert, analogous under many points of view to the DREAMS station of the ExoMars 2016 mission, and to the Meteo Package and Dust Complex of the ExoMars 2020 mission. Our station was able to monitor the dust lifting events by acquiring synchronous measurement of meteorological data (vertical and horizontal wind speed, pressure, air and soil humidity and temperature) and the atmospheric electric field, coupling also the observations of the saltation activity and suspended dust concentration. Currently, we have acquired the most complete data set available for the dust storms and dust devils.

Usually, the signatures of the dust devil passages last only few tens of seconds and, depending on the meteorological parameter considered, it can be partially masked by the noise level. Therefore, the individuation of the dust devils is not a straightforward procedure and the adoption of an automatic detection algorithm is almost obligatory. This is especially true in case of long surveys with a great amount of data, as the ones we performed in the Sahara or the one that will be performed by the ExoMars 2020 mission.

For this reason, we developed two new detection techniques. The first one is based on the "phase piker" analysis, a standard technique for the whirlwind detection, that we modified in order to use for the first time the dust induced E-field as principal detection parameter. Indeed, from our field data, we have observed how the induced atmospheric electric field variation represents the clearest and easiest to recognize feature associated with the dust devil passage. In addition to the E-field data, our algorithm analyses in a second step also the other acquired meteorological parameters, such as the pressure and the wind speed and direction, to crosscheck the results and eliminate the false positive detections.

In view of potential issues that could affect the acquisitions, we developed also a second technique specifically aimed to detect the dust devils by using only one parameter, eliminating most of the false positive detections. We applied a tomographic analysis for the first time to the vortex detection. The technique analyses the pressure signals but can be tuned also for other parameters, such as the electric field (E-field) or the wind speed.

We directly tested the reliability of the tomography by comparing its results to the ones of a standard phase piker detection. Tomography proved to have a good detection efficiency and an optimal ability to distinguish between true and false detections. Indeed, the method provides filtering, separation and characterization of the dust devil signal components even in presence of strong noise. The tomography results have been published in Aguirre et al. 2017.

We analyzed the detected dust devils and dust storms focusing on the study of the induced electric field and the dependences of this quantity to the environmental conditions and the amount of lifted dust. Indeed, the mobilized grains acquire electric charge during the lifting process by colliding between them and the soil, but currently the electric proprieties of the dust are still poorly understood. So far, the literature describes the triboelectrification in the dust clouds as a predominantly size-dependent process, where the smaller grains acquire on average a negative charge, opposite to the one of the largest grains. The gravity stratification of the particles leads to the charge separation and the consequent generation of an upward directed electric field.

On the contrary, we observed how the monitored dust events are almost always associated with a downward pointing E-field, corresponding to a dipole like electric configuration with the smaller grains positively charged on a bed of larger negative particles. Our results strongly indicate how the different composition of the colliders have to play a key role in the electrification process.

We found that during both dust storms and dust devils the induced E-field is linearly related to the numeric concentration of lifted grains ($\eta$) with a slope that is compatible for the two dust lifting processes. The observed development of the electric field is heavily influenced by the air and soil humidity (RH): the slope of the E-field/$\eta$ relation increases upon exceeding a critical air relative humidity value. Moreover, we observed that the grains ability to acquire and hold the electric charge is probably affected by the deliquescence of the evaporites.

In addition, we found how, under a critical E-field value, the relation between the lifted dust concentration and the friction speed follows the same trend predicted by the theoretical models that neglect the electric forces. However, when the critical E-field is reached, the amount of grains in suspension abruptly increases, indicating a probable positive feedback of the E-field on the lifting process. These results have been published in Esposito et al. 2016.

Regarding the dust devils activity, our survey has allowed the first statistical study of the electric proprieties of the phenomenon. We observed how the induced E-field is linearly related to the vortex pressure drop, rotatory speed and the vertical air flow speed.

We compared the pressure drop cumulative distribution of the Saharan events with the martian data available in literature. The distributions are very similar and can be described in all cases by a power law. In particular, we observed how the best fit exponent of the Saharan distribution is compatible with the one of the NASA Curiosity rover survey. Supposing that the dust devils populations are similar also from an electric point of view, it is possible to speculate on the possible magnitude of the dust induced E-field on Mars. These results have been published in Franzese et al. 2018.

The study of the signals that the dust devil passage leaves in the meteorological measurements does not allow to directly evaluate the whirlwind size, its distance of passage from the sensors and its sense of rotation. Usually, the direct images of a camera are needed to retrieve these parameters.

We developed a simple method to evaluate the distance of passage of the events analysing the wind speed and direction signal. The method is able to fully characterize the monitored dust devils, resolving the degeneration of the data, allowing to evaluate the vortex diameter, rotatory wind speed at vortex wall, pressure core drop value and sense of rotation. We successfully tested our method during a Saharan campaign performed in 2017, financed by the EuroPLANET 2020 research infrastructure that approved my mission proposal.

A further Saharan field campaign has been proposed in order to study the dependence of the dust devil induced E-field on the distance and investigate the possibility of night-time vortices activity. This survey will allow the first development of a proper model of the whirlwinds E-field. The mission proposal is currently under evaluation of the EuroPLANET offices.

One the greatest issues that prevents the quantification of the dust role on the global terrestrial and martian climate is represented by our lack of knowledge of the dust injection rate and size distribution. The Saharan data represent one of the few measurements of these parameters on Earth and anticipate the first martian acquisitions that will be achieved by MicroMED. Therefore, our dust injection rate and size distribution data will greatly aid to improve the predictions of the climatic models on both planets.

In conclusion, the acquisition and the analysis of the Saharan field campaigns data represent a fundamental step forward in the understanding of the physics of the dust lifting processes and the proprieties of the airborne dust in general. The acquired data are unique in literature and have allowed the first investigation of the electric proprieties of dust processes. We have observed how the behavior of the dust lifting activity is strictly mirrored by the development of strong electric fields. Our analysis sheds new light on the electrification process, showing as the current size-dependent description need to be improved in order to take into account also the differences in composition of the moving grains. We found also the first evidence that the electric forces are not negligible, but they can probably play an active role in the lifting, significantly increasing the amount of raised dust.

Moreover, our dust devils study shows a series of results not predicted by any model of vortex formation and dynamics. This proves the need of further theoretical and on field studies of the whirlwinds, in order to proper understand their physics. This is a key goal considering also the huge contribution given by dust devils to the global dust budget on Mars.

The general importance of this thesis work is further enhanced by its relevance for the ExoMars programme. Indeed, the developed terrestrial campaigns have allowed to test and tuned some of the sensors of the ExoMars scientific payload. In addition, the obtained scientific results will greatly help to shed light on the martian dust processes and the study initiated with the terrestrial surveys will culminate with the analysis the of the martian data.